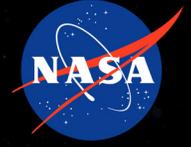

PLANETARY MISSION CONCEPT STUDY FOR THE 2023–2032 DECADAL SURVEY

# Mercury Lander

Transformative science from the surface of the innermost planet

August 08, 2020

**Carolyn M. Ernst**
Principal Investigator
Johns Hopkins University Applied Physics Laboratory
carolyn.ernst@jhuapl.edu

**Sanae Kubota**
Design Study Lead
Johns Hopkins University Applied Physics Laboratory
sanae.kubota@jhuapl.edu

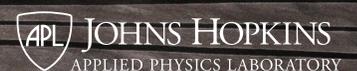

# DATA RELEASE, DISTRIBUTION & COST INTERPRETATION STATEMENTS

This document is intended to support the 2023–2032 Planetary Science and Astrobiology Decadal Survey.

The data contained in this document may not be modified in any way.

Cost estimates described or summarized in this document were generated as part of a preliminary concept study, are model-based, assume an APL in-house build, and do not constitute a commitment on the part of APL.

Cost reserves for development and operations were included as prescribed by the NASA ground rules for the Planetary Mission Concept Studies program. Unadjusted estimate totals and cost reserve allocations would be revised as needed in future more-detailed studies as appropriate for the specific cost-risks for a given mission concept.

# LIST OF CORRECTIONS TO ORIGINAL FILE

This report was updated on 12 July 2021:

- Pages 8, 49, 70: The Orsini et al. 2020 (in review) reference was updated to Orsini et al. 2021.
- Exhibit 15: A typo was corrected to "Diode Heat Pipes".
- Exhibit 27: The spacecraft velocity was corrected to 3.395 km/s.
- Exhibit 37: The costs of previous New Frontiers missions were corrected.

# ACKNOWLEDGEMENTS

The Johns Hopkins Applied Physics Laboratory would like to thank all of the Mercury Lander team members and the NASA Planetary Mission Concept Study Program for supporting this study. Special thanks are due to Shoshana Weider, the NASA Point of Contact, for her contributions.

| ROLE | NAME | AFFILIATION |
|------|------|-------------|
| **Science Team** | | |
| Leadership | Carolyn Ernst, Principal Investigator | APL |
| | Nancy Chabot, Deputy Principal Investigator | APL |
| | Rachel Klima, Project Scientist | APL |
| Geochemistry | Kathleen Vander Kaaden, Group Lead | Jacobs/NASA JSC |
| | Stephen Indyk | Honeybee Robotics |
| | Patrick Peplowski | APL |
| | Elizabeth Rampe | NASA JSC |
| Geophysics | Steven A. Hauck, II, Group Lead | Case Western Reserve University |
| | Sander Goossens | University of Maryland, Baltimore County |
| | Catherine Johnson | Planetary Science Institute |
| | Haje Korth | APL |
| Mercury Environment | Ronald J. Vervack, Jr., Group Lead | APL |
| | David Blewett | APL |
| | Jim Raines | University of Michigan |
| | Michelle Thompson | Purdue University |
| Geology | Paul Byrne, Group Lead | North Carolina State University |
| | Brett Denevi | APL |
| | Noam Izenberg | APL |
| | Lauren Jozwiak | APL |
| Programmatic Expertise | Sebastien Besse, BepiColombo Liaison | European Space Agency |
| | Ralph McNutt, Jr. | APL |
| | Scott Murchie | APL |
| **Engineering Team** | | |
| Avionics | Norm Adams / Justin Kelman | APL |
| Cost | Kathy Kha / Meagan Hahn | APL |
| Flight Software | Chris Krupiarz | APL |
| G&C | Gabe Rogers | APL |
| Mechanical | Deva Ponnusamy / Derick Fuller | APL |
| Mission Design | Justin Atchison / Jackson Shannon | APL |
| Mission Operations | Don Mackey | APL |
| Landing | Benjamin Villac | APL |
| Payload | Rachel Klima / David Gibson | APL |
| Power | Dan Gallagher / Doug Crowley | APL |
| Propulsion | Stewart Bushman | APL |
| Management | Dave Grant | APL |
| Systems Engineering | Sanae Kubota / Gabe Rogers | APL |
| Telecomm | Brian Bubnash | APL |
| Thermal | Jack Ercol / Allan Holtzman | APL |
| **Report Development Team** | | |
| Editing | Marcie Steerman | APL |
| Graphics | Gloria Crites / Christine Fink / Ben C. Smith / Matt Wallace | APL |

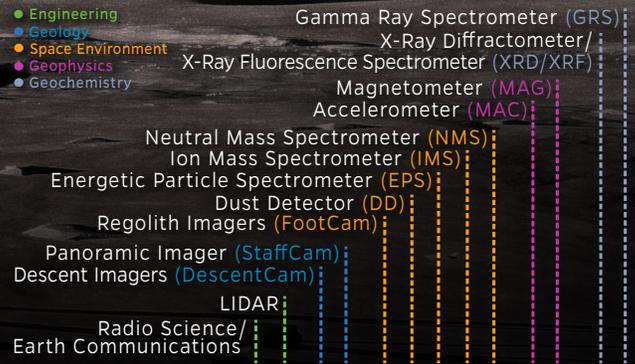

# Mercury Lander

PLANETARY MISSION CONCEPT STUDY FOR
THE 2023–2032 DECADAL SURVEY

**Overview:** The only inner planet unexplored by a landed spacecraft, Mercury is an extreme end-member of planet formation with a unique mineralogy and interior structure. Mercury is also a natural laboratory to investigate fundamental planetary processes—including dynamo generation, crustal magnetization, particle–surface interactions, and exosphere production.

**Study Objective:** Evaluate the feasibility of a landed mission to Mercury in the next decade to accomplish four fundamental science goals.

**Science Goal 1 (Geochemistry):** Investigate the mineralogy and chemistry of Mercury's surface.

**Science Goal 2 (Geophysics):** Characterize Mercury's interior structure and magnetic field.

**Science Goal 3 (Space Environment):** Determine the active processes that produce Mercury's exosphere and alter its regolith.

**Science Goal 4 (Geology):** Characterize the landing site at a variety of scales and provide context for landed measurements.

### A Full Mercury Year On The Surface:

The Mercury Lander touches down at dusk, permitting ~30 hours of sunlit measurements. Surface operations continue through the Mercury night (88 Earth days)—one full trip of Mercury about the Sun—providing unprecedented landed measurements of seasonal variations in Mercury's space environment, a long baseline for geophysical investigations, and time for multiple geochemical sampling measurements. Sunrise brings an end to mission operations.

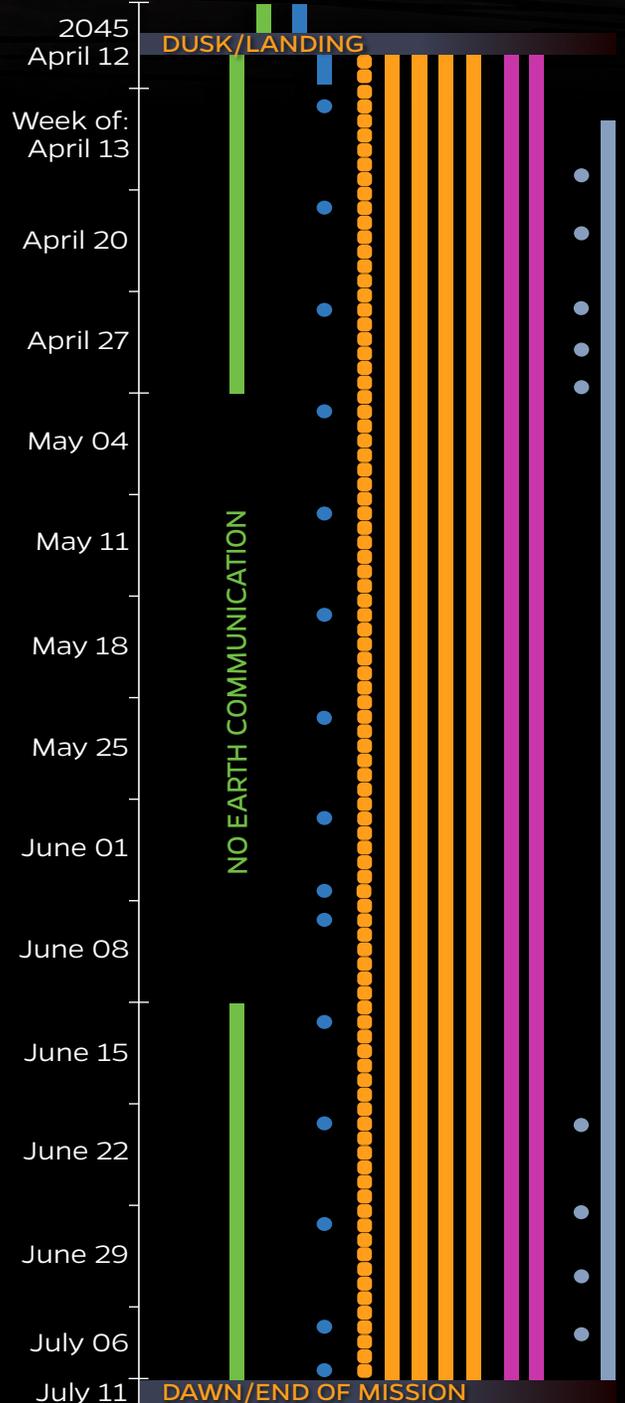

# Mercury Lander

PLANETARY MISSION CONCEPT STUDY FOR THE 2023–2032 DECADAL SURVEY

The Mercury Lander flight system maximizes use of heritage components and leverages major NASA investments (e.g. ion propulsion, NextGen RTG) to enable a New-Frontiers-class landed mission to Mercury.

| Key Mission Characteristics | |
|---|---|
| Launch | March 2035, expendable Falcon Heavy |
| C3, Launch Mass (MEV) | 14.7 km$^2$/m$^2$, 9410 kg (wet), 3680 kg (dry) |
| Design Life | 10.5 years |
| Propulsion | Cruise stage: Solar Electric, Xenon<br>Orbital stage: Bi-propellant, MMH and MON-3<br>Descent stage: Solid Rocket Motor, TP-H-3340<br>Lander: Bi-propellant, MMH and MON-3 |
| Power | Cruise stage solar array<br>· 9.3 kW BOL @ 0.99 AU for SEP<br>· 1.4 kW BOL @ 0.99 AU for spacecraft<br>Orbital stage solar array: 1.1 kW BOL @ 0.46 AU<br>Orbital stage battery: 60 Ah BOL<br>Lander battery: 4.5 Ah BOL<br>Lander RTG: 16 GPHS NextGen RTG, 373W BOL |
| Telecomm | X-band and Ka-band, direct to Earth |

### Cruise Phase
10 Year solar electric propulsion cruise

### Orbital Phase
Jettison cruise stage
Mercury orbit insertion
2.5 Months in 100 x 6000 km orbit

### Descent Phase
Jettison orbital stage
Initiate braking burn

Jettison descent stage
Hazard detection and avoidance
Final landing

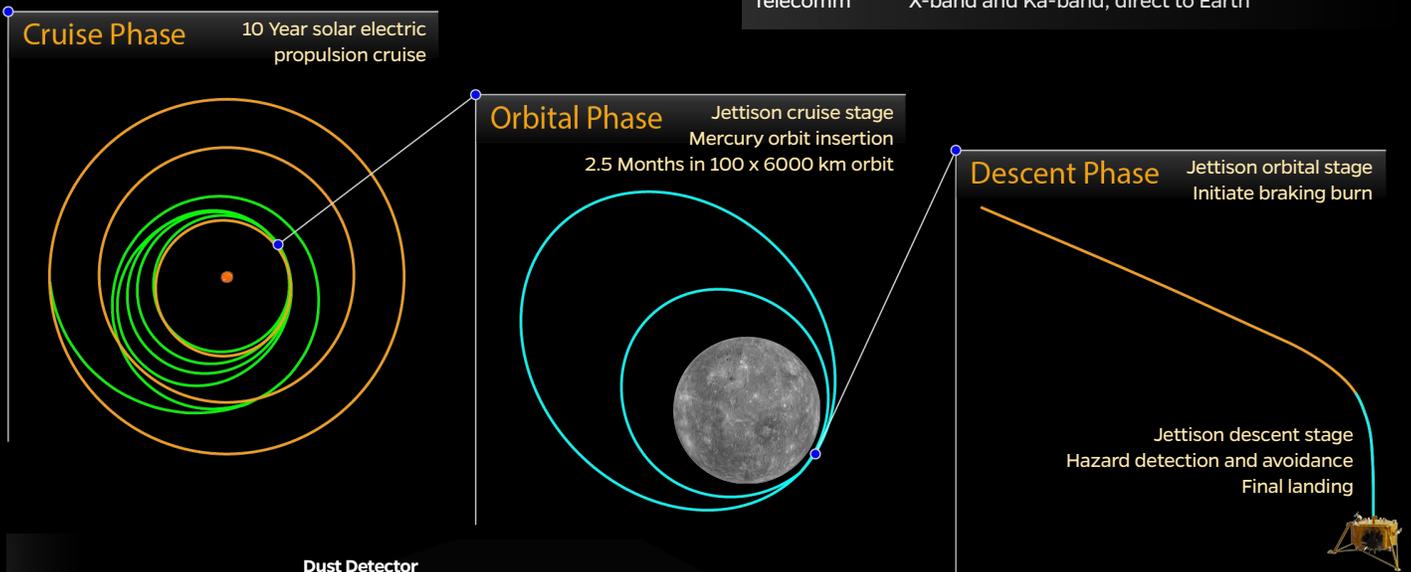

**Dust Detector**
Influx of micrometeoroids

**Neutral Mass Spectrometer**
Composition and density of the near-surface neutral exosphere

**High-Gain Antenna and StaffCam**
Ka-band data return, radio science, and panoramic landing site characterization

**FootCam**
Regolith characterization

**Magnetometer**
Magnetic field as a function of time

**Gamma-Ray Spectrometer**
(not shown)
Elemental composition

**NextGen RTG**
Enables continuous operations through the Mercury night

**Ion Mass Spectrometer**
Fluxes of low-energy charged particles

**Energetic Particle Spectrometer**
Fluxes of high-energy charged particles

**DescentCam**
Landing site characterization

**PlanetVac sample transfer to X-Ray Diffractometer / X-Ray Fluorescence Spectrometer**
Mineralogical composition

**Accelerometer / Short Period Seismometer**
Gravitational acceleration and seismic activity

| PI-Managed Cost | |
|---|---|
| Phase A-D w/o LV | $1192M |
| Phase E-F | $316M |
| Total w/o LV | $1508M |

# Planetary Science Decadal Survey

## Mercury Lander Mission Concept Design Study Final Report

# EXECUTIVE SUMMARY

As an end-member of terrestrial planet formation, Mercury holds unique clues about the original distribution of elements in the earliest stages of solar system development and how planets and exoplanets form and evolve in close proximity to their host stars. This Mercury Lander mission concept enables in situ surface measurements that address several fundamental science questions raised by MESSENGER's pioneering exploration of Mercury. Such measurements are needed to understand Mercury's unique mineralogy and geochemistry; to characterize the proportionally massive core's structure; to measure the planet's active and ancient magnetic fields at the surface; to investigate the processes that alter the surface and produce the exosphere; and to provide ground truth for current and future remote datasets.

NASA's Planetary Mission Concept Studies (PMCS) program awarded this study to a multidisciplinary team led by Dr. Carolyn Ernst of the Johns Hopkins Applied Physics Laboratory (APL), to evaluate the feasibility of accomplishing transformative science through a New-Frontiers-class, landed mission to Mercury in the next decade. The resulting mission concept achieves one full Mercury year (~88 Earth days) of surface operations with an ambitious, high-heritage, landed science payload, corresponding well with the New Frontiers mission framework.

The 11-instrument science payload is delivered to a landing site within Mercury's widely distributed low-reflectance material, and addresses science goals and objectives encompassing geochemistry, geophysics, the Mercury space environment, and surface geology. This mission concept is meant to be representative of any scientific landed mission to Mercury; alternate payload implementations and landing locations would be viable and compelling for a future landed Mercury mission.

The study was performed as a Concept Maturity Level 4 preferred point design. The Mercury Lander flight system launches from Cape Canaveral Air Force Station on a fully expendable Falcon Heavy in 2035 with a backup launch period in 2036. The four-stage system uses a solar electric propulsion cruise stage to reach Mercury in 2045. The cruise stage is jettisoned after orbit-matching with Mercury, and the orbital stage uses its bipropellant propulsion system first to bring the remaining three stages into a thermally safe orbit, then to perform apoherm- and periherm-lowering maneuvers to prepare for descent. During the 2.5-month orbital phase, a narrow-angle camera acquires images, at ~1 m pixel scale, for down selecting a low-hazard landing zone. The orbital stage is jettisoned just prior to initiation of the landing sequence by the descent stage, a solid rocket motor (SRM). The SRM begins the braking burn just over two minutes before landing. The descent stage is jettisoned after SRM burnout, and the Lander executes the final landing with a bipropellant liquid propulsion system. Landing uses continuous LIDAR operations to support hazard detection and safely deliver the payload to the surface.

Landing occurs at dusk to meet thermal requirements, permitting ~30 hours of sunlight for initial observations. The RTG-powered Lander continues surface operations through the Mercury night. Direct-to-Earth (DTE) communication is possible for the initial three weeks of the landed mission, followed by a six-week period with no Earth communication. DTE communication resumes for the remaining four weeks of nighttime operations. Thermal conditions exceed Lander operating temperatures shortly after sunrise, ending surface operations. A total of ~11 GB of data are returned to Earth.

The Phase A–D mission cost estimate (50% unencumbered reserves, excluding the launch vehicle) with the 11-instrument payload is $1.2 B (FY25$), comparing favorably with past New Frontiers missions, as well as to the cost cap prescribed in the New Frontiers 4 AO (~$1.1B FY25$). This cost estimate demonstrates that a Mercury Lander mission is feasible and compelling as a New Frontiers-class mission in the coming decade.

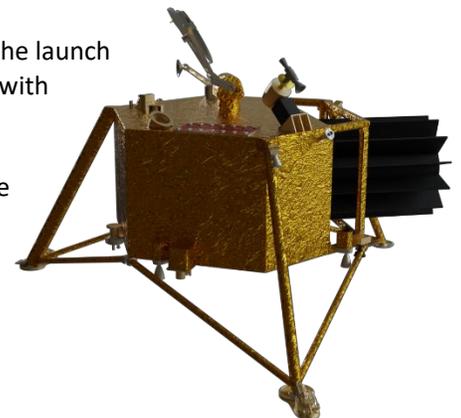

# 1 SCIENTIFIC OBJECTIVES ____________________________

## 1.1. Background & Science Goals

Mariner 10 provided the first close-up reconnaissance of Mercury during its three flybys in 1974 and 1975 [Murray et al. 1974, 1975]. The MErcury Surface, Space ENvironment, GEochemistry, and Ranging (MESSENGER) spacecraft performed three flybys of Mercury in 2008 and 2009 before entering orbit in 2011. MESSENGER's four-year orbital investigation enabled numerous discoveries, several of which led to substantial or complete changes in our fundamental understanding of the planet: the unanticipated, widespread presence of volatile elements such as Na, K, and S [Peplowski et al. 2011; Nittler et al 2011; Evans et al. 2012]; a surface with extremely low iron abundance [Evans et al. 2012; Nittler et al. 2011; Weider et al. 2014] whose darkening agent is likely carbon [Murchie et al. 2015; Peplowski et al. 2016; Klima et al. 2018]; a previously unknown karst-like planetary landform – hollows – that may form by volatile sublimation from within rocks exposed to the harsh conditions on the surface [Blewett et al. 2011; 2016]; expansive volcanic plains [Head et al. 2011] and pyroclastic vents [Kerber et al. 2011] that have shaped Mercury's geology through time; much more radial contraction of the planet than previously thought [Byrne et al. 2014]; an offset of the magnetic equator from that of the planet [Anderson et al. 2011]; crustal magnetization indicating an ancient magnetic field [Johnson et al. 2015; 2018]; unexpected seasonal variability and relationships among exospheric species and processes that generate them [Burger et al. 2014; Cassidy et al. 2015; 2016; Vervack et al. 2016; Merkel et al. 2017; 2018]; an extreme space environment driven by the solar wind [Slavin et al. 2008; 2009; 2014] with unexpectedly energetic heavy planetary ions [Zurbuchen et al. 2008; 2011; Raines et al. 2013; 2014]; and the presence in the permanently shadowed polar terrain of water ice and other volatile materials likely to include complex organic compounds [Lawrence et al. 2013; Neumann et al. 2013; Paige et al. 2013; Chabot et al. 2018].

MESSENGER revolutionized our understanding of Mercury, and the dual-spacecraft ESA–JAXA BepiColombo mission [Benkhoff et al. 2010] promises further revelations in Mercury science. BepiColombo launched in October 2018 and will arrive at Mercury in late 2025, with its nominal one-year orbital mission beginning in spring 2026. Additionally, Earth-based telescopic observations provide a long-term baseline of exosphere and surface observations extending across spacecraft visits, covering Mariner 10 to MESSENGER and continuing into the future (e.g., Sprague et al. [2000]; Mendillo et al. [2001]; Bida & Killen [2017]).

However, remote and orbital investigations have technical limits. Landed, in situ measurements from Mercury's surface are needed to address several fundamental science questions. In particular, MESSENGER revealed that Mercury's highly chemically reduced and unexpectedly volatile-rich composition is unique among terrestrial planets and unlike any predictions of previously proposed hypotheses of the planet's origin. These surprising results have led to a reexamination of the planet's formation and history. In situ measurements from the surface are needed to: (1) understand Mercury's unique mineralogy and geochemistry; (2) characterize the proportionally massive core's structure; (3) measure the planet's active and ancient magnetic fields at the surface; (4) investigate the processes that alter the surface and produce the exosphere; and (5) provide ground truth for current and future remote datasets. Although BepiColombo will further advance our global understanding of Mercury, that mission cannot address the major science questions for which in situ landed measurements are needed, nor will it image Mercury's surface with sufficient resolution [Flamini et al. 2010; Cremonese et al. 2020] to influence the technical approach used to land.

Additionally, unraveling the mysteries about Mercury's origin, evolution, and ongoing processes has implications and expected significance beyond the innermost planet. Mercury is an extreme end-member of planet formation, and its highly reduced nature provides unique clues regarding how planets close to their host stars can form and evolve. Mercury's magnetosphere is also a natural laboratory for understanding the interactions of exoplanets close to their host stars. The acquisition and retention of crustal magnetizations over billion-year timescales has implications for dynamo generation across the major terrestrial bodies.

Understanding the processes that affect the regolith of airless bodies provides key insight into exospheres and space weathering on bodies within our solar system and beyond. **A Mercury Lander would accomplish ground-breaking science, and the results would inform our greater understanding of the formation and evolution of rocky planets in our solar system and those about other stars.**

To guide this Mercury Lander mission concept study, four overarching and fundamental science goals have been identified:

> **Goal 1:** Investigate the highly chemically reduced, unexpectedly volatile-rich mineralogy and chemistry of Mercury's surface, to understand the earliest evolution of this end-member of rocky planet formation;

> **Goal 2:** Investigate Mercury's interior structure and magnetic field, to unravel the planet's differentiation and evolutionary history and to understand the magnetic field at the surface;

> **Goal 3:** Investigate the active processes that produce Mercury's exosphere and alter its regolith, to understand planetary processes on rocky airless bodies, including the Moon;

> **Goal 4:** Characterize the landing site, to understand the processes that have shaped its evolution, to place in situ measurements in context, and to enable ground truth for global interpretations of Mercury.

The scientific motivation driving each of these goals is provided in detail in Appendix B1. To fully evaluate the technical feasibility of a landed mission to Mercury, including the mission design aspects, it is necessary to select a specific landing site on the planet. A landing site in the low-reflectance material (LRM) was chosen for this study. The LRM is believed to be remnants of Mercury's 'exotic' graphite flotation crust [Vander Kaaden & McCubbin 2015] and, hence, represent the earliest solid crustal materials on Mercury, providing a unique window into the planet's earliest differentiation. Additionally, LRM locations are widely distributed across the planet, as shown in Exhibit 1, providing flexibility for a mission concept study with the overall goal of investigating the feasibility of a landed mission to Mercury without being overly limited by a specific choice of landing site.

However, this choice of landing site should not restrict future landed exploration of Mercury. There are compelling scientific cases to be made for a wide range of landing locations, such as the diverse geochemical units on Mercury including the northern smooth plains and the high-Mg region [e.g., Peplowski et al. 2015; Weider et al. 2015; Nittler et al. 2020]; geologically pyroclastic vents [Thomas et al. 2015]; the enigmatic hollows [Blewett et al. 2013; 2016]; and the water-rich, permanently shadowed polar deposits [Chabot et al. 2018]. The overarching science goals remain the same regardless of the ultimate landing site choice (specifics of Goal 1 would necessarily be adapted for a polar deposit lander); specifics of some science objectives and measurements would necessarily be adapted. **The measurements made by the first landed mission to Mercury will be foundational and transformative**, answering high-priority outstanding science questions for Mercury from any location.

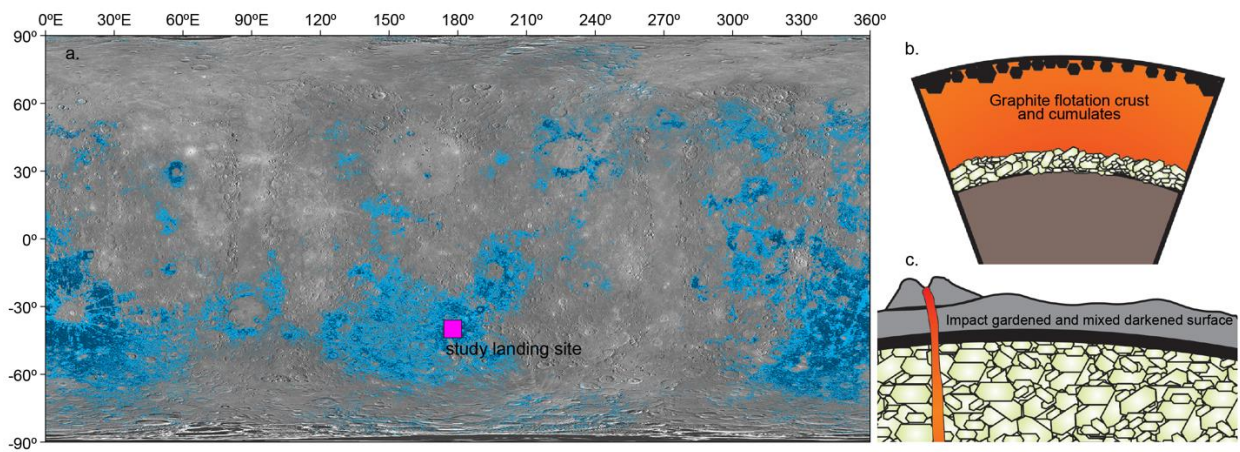

**Exhibit 1.** (a) Mercury's globally distributed low-reflectance material (LRM, shown in blue, Klima et al. [2018]), which may sample ancient, carbon-bearing deposits. (b) A thin, primary graphite flotation crust forms in an early magma ocean. (c) Impacts mix the volcanic secondary crust and the graphite primary crust. (b, c from Vander Kaaden & McCubbin [2019]). In situ geochemical measurements of the LRM will test the graphite crust hypothesis and elucidate the earliest chemical evolution of the planet. The pink square denotes the landing site location (40°S, 178°E) considered for this mission concept study.

## 1.2. Science Objectives & Science Traceability

The purpose of this study was to evaluate the feasibility of a landed mission to Mercury in the next decade that would accomplish ground-breaking science. As discussed here and at length in Appendix B1, there is no shortage of transformative science that can be done by the first landed mission to the innermost planet. Consequently, two overarching philosophies were adopted for this concept study:

1) **Investigate a comprehensive, scientifically robust payload spanning the wide-ranging science measurements that could be made in situ on Mercury's surface.** In particular, this study assured coverage of all four goals in fairly equal detail, rather than choosing to focus on any specific one. The inclusion of a large number of instruments in the concept study provides a more valuable resource for the science community when planning a landed mission to Mercury in the future.

2) **Prioritize landing safely on Mercury.** Consistent with this philosophy, the team decided to focus resources on that fundamental challenge, without which landed science is not possible. As such, the team considered only payload implementations that leveraged previous development efforts, in particular those designed to perform in situ landed measurements on the Moon and Mars. *The first landed measurements on the surface of Mercury are so fundamental that they can be made by existing instrumentation, without the need for major development.*

Exhibit 2 provides traceability from the overarching four science goals to fourteen specific science objectives, the measurements required to fulfill these objectives, and the functional requirements necessary to achieve these measurements. Detailed discussion of the scientific motivation for the science objectives, the rationale for the technical implementation selection, and the instrument performance in relation to the science measurements is given in Appendix B1–2. The functional requirements in Exhibit 2 note constraints placed on the mission concept design by the payload.

An 11-item science payload was chosen for this study, which satisfies the comprehensive science goals and objectives outlined in Exhibit 2. This ambitious instrument suite is just one possible configuration that could accomplish the high-priority science goals. Alternate payload implementations could be designed to return equally compelling science measurements. A future Mercury Lander mission should not be limited to the science payload considered here, but rather should take advantage of technology advancements and use the best instrumentation available at the time of planning such a mission. The comprehensive payload listed in Exhibit 2 is somewhat larger than those of previous New Frontiers missions. It may be advantageous to reduce the payload or to consider foreign contributions. The only absolute requirement to achieve ground-breaking science from a Mercury Lander is to **perform in situ measurements on the surface of Mercury**.

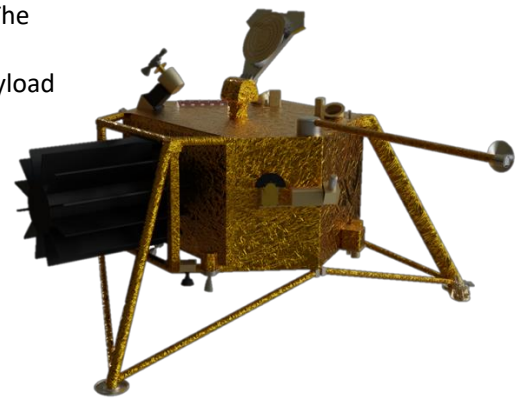

**Exhibit 2.** Science Traceability Matrix.

| SCIENCE GOALS | SCIENCE OBJECTIVES | MEASUREMENT REQUIREMENTS | INSTRUMENT (ANALOGS [TRL]) | FUNCTIONAL REQUIREMENT |
|---|---|---|---|---|
| **Goal 1:** Investigate the highly chemically reduced, unexpectedly volatile-rich mineralogy and chemistry of Mercury's surface, to understand the earliest evolution of this end-member of rocky planet formation. | **1.1** Determine the major- and minor-elemental composition of the LRM, including its C content and volatile-element abundances (e.g., Na, K, S) | Absolute abundances of: C, O, Na, Mg, Si, S, Cl, K, Ca, Fe, Th, U, Cr, Mn, if present at concentrations of >1 wt% | **GRS: Gamma-Ray Spectrometer** (MESSENGER [TRL 9], Psyche [TRL 7], MMX [TRL 7], Dragonfly [TRL 7]) | Continuous operation to avoid instrument degradation; unobstructed FOV of the surface; surface operations ≥72 hrs |
| | **1.2** Determine the mineralogy of the components of the LRM, including any silicate, sulfide, or carbide phases that are present | Identification of silicates, sulfides, carbides, metallic phases, if present at concentrations of >1 wt% | **XRD/XRF: X-Ray Diffractometer/X-Ray Fluorescence Spectrometer** (MSL CheMin [TRL 9], CheMin-V [TRL 6]) | Surface sample must be delivered into the XRD/XRF instrument |
| | **1.3** Investigate the chemical and mineralogical heterogeneity of the landing site | Measurements of Objective 1.2 from two locations at the landing site and from ≥two distinct surface disturbance events | | Ability to collect samples from multiple locations and to produce distinct surface disturbance events |
| **Goal 2:** Investigate Mercury's interior structure and magnetic field, to unravel the planet's differentiation and evolutionary history and to understand the magnetic field at the surface. | **2.1** Investigate the distribution of mass in Mercury's interior, determine the size and state of the core to characterize the solid and liquid portions, and search for seismic activity | Longitude libration amplitudes; obliquity | **RS: Radio Science** (InSight RISE [TRL 9]) | Ka-band communication to enable the most-sensitive science measurements |
| | | Gravitational acceleration change due to solid-body tides; short-period seismic observations | **MAC: Mercury Accelerometer/Short-Period Seismometer** (InSight SEIS-SP [TRL 9]) | Positioned near surface; high data rate from continuous operations needed to detect potential seismic events |
| | **2.2** Measure the magnetic field at the surface to investigate the coupling between the dynamo and external field, the time variation of the field, the strength of the crustal field, and the electrical conductivity structure of the crust and mantle | Measurements of magnetic field at the surface as a function of time, with a precision of 1 nT and at cadence of 20 vector samples per second | **MAG: Magnetometer** (MESSENGER MAG [TRL 9]) | Positioned to minimize contributions from spacecraft-generated fields |
| | **2.3** Investigate the mineralogy of the surface to identify potential magnetic carrier minerals | Covered by Objective 1.2 mineralogical measurements above | | |
| **Goal 3:** Investigate the active processes that produce Mercury's exosphere and alter its regolith, to understand planetary processes on rocky airless bodies, including the Moon. | **3.1** Determine the composition and density of the near-surface neutral exosphere and compare with the surface compositional measurements, to investigate processes releasing materials from the surface | Densities of atomic and molecular species 1–100 amu, M/ΔM ~100, sensitivity ~1 count/sec at density of 10 cm$^{-3}$ | **NMS: Neutral Mass Spectrometer** (BepiColombo STROFIO [TRL 9]) | Unobstructed FOV of space environment, angled 45° toward surface |
| | **3.2** Determine and characterize the incoming and outgoing fluxes of charged particles at Mercury's surface | Identification of low-energy charged particles, 1 eV/e to 20 keV/e, M/ΔM 4–40 over M/q 1–50, angular resolution <20° | **IMS: Ion Mass Spectrometer** (MESSENGER FIPS [TRL 9]) | Unobstructed FOV of space environment, angled 45° away from surface |
| | | Identification of high-energy charged particles, 20 keV to 1 MeV, angular resolution <20° | **EPS: Energetic Particle Spectrometer** (New Horizons PEPSSI [TRL 9]) | Unobstructed FOV of space environment, angled 45° away from surface |
| | **3.3** Determine and characterize the influx of micrometeoroids (dust) at Mercury's surface | Measurements of dust flux with sensitivity to measure 10$^{-15}$ kg m$^{-2}$ s$^{-1}$ | **DD: Dust Detector** (New Horizons SDC [TRL 9]) | Unobstructed FOV of space environment, looking toward zenith |
| | **3.4** Investigate the nature of Mercury's regolith, including particle sizes and heterogeneity | Images of regolith in ≥3 visible colors, pixel scales ≤500 μm @ 1-m distance | **FootCam: Regolith Imagers** (Malin Space Science Systems, ECAM [TRL 9]) | Mounted to resolve 1-mm grains; LED illumination @ 450, 550, 650, 750 nm |
| | **3.5** Investigate the characteristics of space weathering on Mercury | Measurements for Objective 1.2 and 3.4 repeated for ≥two distinct surface disturbances of the same location | | Ability to collect multiple samples from the same location and to produce distinct surface disturbance events |
| **Goal 4:** Characterize the landing site, to understand the processes that have shaped its evolution, to place the in situ measurements in context, and to enable ground truth for global interpretations of Mercury. | **4.1** Connect observations from images acquired by orbiting spacecraft to those from the Lander and determine the geological context of the landing site | Images of landing site acquired during descent, pixel scales 1 cm to 1 m | **DescentCam: Descent Imagers** (Malin Space Science Systems, ECAM [TRL 9]) | Periodic imaging of the surface during descent; two cameras oriented 90° from one another to enable surface imaging despite changing orientation during descent |
| | **4.2** Characterize the geological setting of the landing site, including heterogeneity and landforms, and search for changes over the mission by surface, horizon, and exosphere imaging | Images of the landing site, pixel scale ≤5 cm within 50 m; ≥180° az, 0°– -45° elev | **StaffCam: Panoramic Imager** (MER Pancam [TRL 9], MSL Mastcam [TRL 9]) | Unobstructed access to ≥ 180° of the landing site; articulation to achieve angular coverage |
| | **4.3** Characterize the bulk-element composition of the local landing site and place it into context with the equivalent orbital measurements | Covered by Objective 1.1 elemental measurements above | | |

# 2 HIGH-LEVEL MISSION CONCEPT

## 2.1. Overview

The importance of a landed mission to Mercury was recognized by the 2013–2022 Decadal Survey [National Research Council 2011], and a previous Mercury Lander Mission Concept Study [Hauck et al. 2010] was completed in response, examining the feasibility of a Mercury Lander. The 2010 Mercury Lander Study was conducted prior to MESSENGER's orbital campaign, however, and the study was unable to incorporate MESSENGER's orbital results to inform the science justification for landed in situ measurements. Since that time, MESSENGER's datasets have revolutionized our understanding of Mercury, greatly advancing the scientific case for a landed mission. The degree of scientific thought and consideration brought to bear on this study's Lander payload and scientific measurements is a distinguishing factor from the 2010 study.

The technology landscape has changed substantially in the last decade, with new launch vehicle (LV) availability and increased development and use of solar electric propulsion (SEP) systems. Both are enabling contributions to this design concept, and address the two areas that the 2010 study report recognized as particularly challenging: the impacts of launch energy and ΔV requirements.

In addition, costing guidelines for NASA's New Frontiers missions have changed since 2010. The previous study concluded that the cost of such a mission exceeded the cost cap of a PI-led New Frontiers-class mission. However, LV and Phase E–F mission operation costs are now excluded from the cost cap. Consequently, the costing associated with this new mission concept enables an informed discussion for the 2023 Decadal Survey based on the latest science results and technology capabilities.

This Concept Maturity Level (CML)-4 design concept addresses the primary challenges of a Mercury Lander mission with a four-stage design that launches on an expendable Falcon Heavy (EFH) vehicle. This LV's increased lift capability is mission-enabling. Mass savings are enabled through jettisoning of stages prior to large burns and optimization of propulsion systems for each phase: cruise, orbit, initial descent, and landing.

The cruise stage minimizes the fuel load required with an SEP system for a low-thrust trajectory to Mercury. The reference trajectory launches with a C3 of 14.7 $km^2/s^2$ and includes one Earth, two Venus, and five Mercury gravity assists during a 10-year cruise. Use of SEP for Mercury orbit insertion (MOI), station-keeping, and orbit-lowering maneuvers was evaluated, but results in significantly longer flight times than use of chemical propulsion. Such SEP use would also result in substantial increases to solar array size and mass to accommodate both solar cell degradation and decreased efficiency due to off-pointing needed to maintain solar cell temperatures while in Mercury orbit. A chemical propulsion system is, therefore, chosen for MOI through preparation-for-descent, and the cruise stage is jettisoned after orbit matching with Mercury.

The MOI burn uses the orbital stage's bipropellant system to deliver the remaining three stages to a thermally safe, 100 km x 6000 km polar parking orbit. A narrow angle camera based on the Lunar Reconnaissance Orbiter Camera (LROC) is included on the orbital stage to enable landing site reconnaissance, supplementing BepiColombo's 3 m/pixel imaging capability with 1 m/pixel scale of the landing area from 100-km altitude.

Required landing conditions are met about 2.5 months after MOI. After Mercury's true anomaly position is greater than 130° (MTA130), an apoherm-lowering maneuver to a 100 km x 2000 km orbit and a final periherm-lowering maneuver to a 20 km x 2000 km orbit are executed. These maneuvers are separated by two weeks and prepare the vehicle for descent at aphelion with a near-terminator orbit.

The orbital stage is jettisoned prior to initiation of the landing sequence by the descent stage. Mass efficiency through the braking burn is optimized in the descent stage with a STAR 48GXV SRM. The descent stage is jettisoned after SRM burnout, and the Lander executes the final soft landing with a biprop liquid propulsion system.

Descent and landing timing allows approximately 30 hours of sunlight and three weeks of DTE communication after landing. Landing at dusk is chosen to satisfy vehicle thermal constraints. The radioisotope generator (RTG)-powered Lander continues autonomous operations through the Mercury night. DTE communication is reestablished after six weeks and allows four more weeks of operations and communication prior to sunrise, after which thermal conditions exceed vehicle operating temperatures.

## 2.2. Concept Maturity Level (CML)

The Mercury Lander is CML 4, a preferred design point, and one in which the design and cost return the desired science. The flight system and its major subsystems have been defined with acceptable margins and reserves. Mission-, system-, and subsystem-level trades have been performed with selections resulting in this preferred point design.

## 2.3. Technology Maturity

The Mercury Lander design maximizes the use of high-heritage flight system elements. The sampling system included in this design is TRL-5, and this design concept allows for use of alternate sampling systems. All other components are TRL-6 and above. All instruments are based on flight-proven analogs. Subsystem components have either flight heritage or scheduled flight in the near future, and require limited tailoring to support a Mercury Lander mission, with the exception of the STAR 48GXV solid rocket motor.

The STAR 48GXV is a high performance variation of the flight-proven STAR 48BV. A proof-of-concept STAR 48GXV motor was successfully manufactured and static tested at sea level in December 2013, and the test data indicated that the motor was within 2% of predicted performance.

## 2.4. Key Trades

Multiple solutions were considered for each design decision, with final selections primarily motivated by the prioritization of maximized landed mass. Key trade studies that drove the design are listed in Exhibit 3.

| AREA | TRADE SPACE, RESULT (**BOLD**) | RATIONALE |
|---|---|---|
| Cruise Propulsion | Chemical vs **SEP** | Propellant savings |
| | Aerojet XR5 vs Qinetiq T6 vs **NEXT-C** | Thrust requirements, propellant savings |
| MOI / Orbit Propulsion | **Chemical** vs SEP | Long duration for SEP implementation, impacts to solar array thermal management and cell degradation |
| Braking Burn Propulsion | Bipropellant vs **SRM** | Mass savings, load path efficiency |
| Lander Propulsion | Monopropellant vs **Bipropellant** | Mass savings |
| Landing Area Risk Reduction | Targeted imaging orbits vs **opportunistic imaging** | Thermal constraints in lower orbits |
| Vehicle Stages | 3 vs **4** | Propellant savings through jettison of stages prior to large burns |

**Exhibit 3.** Key trades considered in the study.

# 3 TECHNICAL OVERVIEW

## 3.1. Instrument Payload Description

The Lander's payload was selected to achieve the scientific goals and objectives detailed in Exhibit 2. Analog instruments with their associated TRL numbers are also provided in that exhibit. Exhibit 4 gives a summary of the science payload including current best estimate (CBE), contingency margin allocations (% cont.), and maximum expected value (MEV) mass and power values, and is followed by a brief text and tabular description (Exhibits 5–13) for each instrument. Due to page constraints on the main report, Appendix B2 provides discussion of, or references to, technical details of each instrument (e.g., flight hardware and software) and science operations (e.g., modes, calibration, analysis, data types, and resulting data products). Appendix B2 also provides detail about additional instrumentation considered but ultimately not selected for inclusion in the payload and the rationale for the choices. The concept of surface operations, associated data volume and margin for the landed phase of the mission, and implications for contingency in science data rates and instrument operation options are provided in Section 3.3, Concept of Operations.

| | | **MASS** | | | **AVERAGE POWER** | |
|---|---|---|---|---|---|---|
| | **CBE (KG)** | **% CONT.ᴬ** | **MEV (KG)** | **CBE (W)** | **% CONT.ᴮ** | **MEV (W)** |
| Gamma-Ray Spectrometer (GRS) | 7.7 | 10 | 8.5 | 16 | 30 | 22.9 |
| X-Ray Diffractometer/X-Ray Fluorescence Spectrometer (XRD/XRF) | 4 | 10 | 4.4 | 30 | 30 | 43 |
| Incl. Sampling System (PlanetVac, for two) | 15.43 | 15 | 17.7 | 19.9 | 30 | 28.5 |
| Magnetometer (MAG) | 4.1 | 10 | 4.5 | 5.1 | 30 | 7.3 |
| Accelerometer/Short Period Seismometer (MAC) | 0.76 | 10 | 0.85 | 2 | 30 | 2.86 |
| Neutral Mass Spectrometer (NMS) | 4 | 10 | 4.4 | 7.5 | 30 | 10.7 |
| Ion Mass Spectrometer (IMS) | 1.4 | 10 | 1.5 | 2 | 30 | 2.9 |
| Energetic Particle Spectrometer (EPS) | 1.5 | 10 | 1.7 | 2.5 | 30 | 3.6 |
| Dust Detector (DD) | 1.6 | 10 | 1.8 | 5.1 | 30 | 7.3 |
| Regolith Imagers (FootCam, mass for two, power for one) | 0.7 | 2 | 0.7 | 2.5 | 30 | 3.6 |
| LEDs (for 40 units) | 4 | 2 | 4.1 | 25.6 | 30 | 36.6 |
| Panoramic Imager (StaffCam, includes actuator) | 2.78 | 2 | 2.83 | 2.5 | 30 | 3.6 |
| Descent Imagers (DescentCam, for two) | 0.7 | 2 | 0.7 | 2.5 | 30 | 3.6 |
| **Total Payload Mass** | **48** | | **53** | | | |

**Exhibit 4.** Payload Mass & Power Table.
a. % Contingency based on TRL and APL institutional practice, with 30% total margin included per stage as defined in and required by PMCS Ground Rules
b. % Contingency based on PMCS Ground Rules

**Gamma-Ray Spectrometer (GRS):** Bulk elemental abundances at the landing site are measured by a high-purity, germanium-based GRS with heritage from GRS instruments on MESSENGER [Goldsten et al. 2007], and planned flight on the Psyche [Lawrence et al. 2019a] and Martian Moons eXploration (MMX) [Lawrence et al. 2019b] missions. For this study, the GRS design is simplified, removing the anti-coincidence shield and incorporating a low-power Ricor cryocooler. This simplified design is made possible by the higher signal-to-noise ratio that is achieved via landed measurements as compared to orbital measurements.

**X-ray Diffractometer/ X-ray Fluorescence Spectrometer (XRD/XRF):** A combination XRD/XRF spectrometer provides both mineralogical and geochemical characterization of the regolith at the landing site. For this study, the CheMin-V instrument is incorporated into the payload, drawing heritage from the CheMin instrument on the Curiosity rover [Blake et al. 2012]. CheMin-V will improve upon CheMin by acquiring data more rapidly with improved angular resolution and by collecting quantitative XRF data [Blake et al. 2019], thereby improving the identification of minerals. XRD/XRF analysis requires that the sample be transferred to the instrument; this is accomplished via the **PlanetVac sampling system** [Zacny et al. 2014]. Regolith is collected by the PlanetVac pneumatic samplers, which are mounted on two of the Lander feet. The pneumatic samplers use pressurized gas to loft regolith and pass it through tubes to the XRD/XRF analysis chamber. PlanetVac will fly with NASA's Commercial Lunar Payload Services program in 2023 and on the MMX mission [Zacny et al. 2020] in 2024, and the pneumatic sample transfer system will fly on the Dragonfly mission [Turtle et al. 2019] in 2026.

| ITEM | VALUE | UNITS |
|---|---|---|
| Size/dimensions | 30 x 20 x 20 | cm x cm x cm |
| Mass without contingency (CBE) | 7.7 | kg |
| Average payload power without contingency | 16 | W |
| Average science data rate | 0.5 | kbps |

**Exhibit 5.** Gamma-Ray Spectrometer (GRS) Characteristics.

| ITEM | VALUE | UNITS |
|---|---|---|
| **XRD/XRF** | | |
| Size/dimensions | 30 x 18 x 15 | cm x cm x cm |
| Mass without contingency (CBE) | 4 | kg |
| Average payload power without contingency | 30 | W |
| Average science data rate | 35,000 | kbps |
| **PlanetVac** | | |
| Size/dimensions | 10 x 10 x 10 | cm x cm x cm |
| Mass without contingency (CBE) | 15.43 | kg |
| Average payload power without contingency | 19.9 | W |
| Average science data rate | 32,000 | kbps |

**Exhibit 6.** X-Ray Diffractometer / X-Ray Fluorescence (XRD/XRF) and Sampling System (PlanetVac) Characteristics.

**Magnetometer (MAG):** The MAG makes measurements of the magnetic field at Mercury's surface, and the continuous vector magnetic field observations over the duration of landed surface operations will enable identification of both static and time-varying fields. The MAG is mounted at the end of a boom that is deployed after landing, to minimize contributions from spacecraft-generated fields. The heritage magnetometer instrument assumed for this concept study is the MESSENGER magnetometer [Anderson et al. 2007].

**Accelerometer/ Short Period Seismometer:** The Mercury Accelerometer (MAC) provides direct measurements of the gravitational changes due to tides over the course of the landed mission. High-frequency measurements from the accelerometer allow it to operate as a short-period seismometer, enabling the first observations of the seismicity of Mercury. The InSight SEIS-SP short-period seismometer [Lognonné et al. 2019; Pike et al. 2018] is chosen for this mission concept study and has demonstrated the ability to measure earthquake signals and solid-earth tides [Pike et al. 2018]. Should Mercury have a seismic behavior similar to Mars, it is reasonable to expect several tens of quakes to be detected on Mercury over a roughly 88-day landed mission (see Appendix B2.2.3).

**Neutral Mass Spectrometer (NMS):** The NMS measures the densities of neutral species in the exosphere, including those of both atoms and molecules. The analog used for this mission concept study is STROFIO on the BepiColombo mission [Orsini et al. 2021]. NMS is mounted with an unobstructed field of view (FOV) to the space environment and angled toward the surface.

**Ion Mass Spectrometer (IMS):** Measurements to characterize the low-energy ions at Mercury's surface are made with an IMS. The IMS analog used in this concept study is FIPS onboard MESSENGER [Andrews et al. 2007]. The IMS is mounted with an unobstructed view to the space environment and angled away from the surface.

| ITEM | VALUE | UNITS |
|---|---|---|
| Size/dimensions (Sensor Head) | 8.1 x 4.8 x 4.6 | cm x cm x cm |
| Mass without contingency (CBE) | 4.1 | kg |
| Average payload power without contingency | 5.1 | W |
| Average science data rate | 1.5 | kbps |

Exhibit 7. Magnetometer (MAG) Characteristics.

| ITEM | VALUE | UNITS |
|---|---|---|
| Size/dimensions | 9 x 9 x 8 | cm x cm x cm |
| Mass without contingency (CBE) | 0.76 | kg |
| Average payload power without contingency | 2 | W |
| Minimum science data rate (tide measurements only) | 1.8 | bps |
| Maximum science data rate (short-period seismic data) | 7.2 | kbps |

Exhibit 8. Accelerometer / Short Period Seismometer (MAC) Characteristics.

| ITEM | VALUE | UNITS |
|---|---|---|
| Size/dimensions | 30 x 30 x 15 | cm x cm x cm |
| Mass without contingency (CBE) | 4 | kg |
| Average payload power without contingency | 7.5 | W |
| Average science data rate | 0.23 | kbps |
| Field of view | 24 x 24 | degrees |

Exhibit 9. Neutral Mass Spectrometer (NMS) Characteristics.

| ITEM | VALUE | UNITS |
|---|---|---|
| Size/dimensions | 30 x 20 x 10 | cm x cm x cm |
| Mass without contingency (CBE) | 1.4 | kg |
| Average payload power without contingency | 2 | W |
| Average science data rate | 1.2 | kbps |
| Field of view | 75 x 360 | degrees |

Exhibit 10. Ion Mass Spectrometer (IMS) Characteristics.

| ITEM | VALUE | UNITS |
|---|---|---|
| Size/dimensions | 13 x 22 x 11 | cm x cm x cm |
| Mass without contingency (CBE) | 1.5 | kg |
| Average payload power without contingency | 2.5 | W |
| Average science data rate | 0.23 | kbps |
| Field of view | 20 x 180 | degrees |

Exhibit 11. Energetic Particle Spectrometer (EPS) Characteristics.

**Energetic Particle Spectrometer (EPS):** The EPS measures ions with energies up to 1 MeV per nucleon, as well as energetic electrons. The EPS heritage instrument used for this concept study is PEPSSI onboard New Horizons [McNutt et al. 2008]. The EPS is mounted similarly to the IMS, with an unobstructed view to the space environment and angled away from the surface.

**Dust Detector (DD):** A dust detector, based on New Horizons SDC [Horányi et al. 2009], counts the flux of picogram to nanogram dust particles that are incident on the surface. To have a clear view to space, the dust detector is placed on the top deck of the Lander.

| ITEM | VALUE | UNITS |
|------|-------|-------|
| Size/dimensions (collection area) | 32 x 32 | cm x cm |
| Mass without contingency (CBE) | 1.6 | kg |
| Average payload power without contingency | 5.1 | W |
| Average science data rate | 0.23 | kbps |

**Exhibit 12.** Dust Detector (DD) Characteristics.

**Imagers:** The Lander's regolith imagers (**FootCam**) characterize the regolith and search for local changes to the regolith induced by sampling by the PlanetVac system, which can mobilize regolith surrounding each sampler cone (see Appendix 2.3.4). FootCam consists of two monochrome cameras mounted on two of the three spacecraft landing legs and positioned to observe the corresponding Lander foot and surrounding area. The panoramic imager (**StaffCam**) characterizes the geological setting of the landing site. StaffCam is a monochrome camera collocated with the gimballed high-gain antenna. This mounting location, which extends up from the main body of the Lander, enables panoramic imaging of the landing site. Four-color LED arrays provide illumination to each FootCam imager, with colors attuned to geologically appropriate wavelengths (Appendix B2.3.4). The number and placement of the LEDs will be modeled in more detail pre-flight and placed to optimize nighttime imaging of both FootCam (required) and StaffCam (as possible). Two monochrome descent imagers (**DescentCam**) characterize the landing site, linking it to global maps of the planet via a set of nested images acquired during the descent and landing sequence. The two cameras are oriented with their FOVs 90° from one another, ensuring that the surface remains in view and can be imaged even as the Lander changes orientation relative to the surface during descent. Malin Space Science System ECAM, 5-megapixel CMOS cameras, with heritage from the Mars Exploration Rover (MER) and Mars Science Laboratory (MSL) missions, were selected for all imagers.

| ITEM | | VALUE | UNITS |
|------|---|-------|-------|
| Number of channels – FootCam, StaffCam, DescentCam | | 1 | clear filter |
| LEDs wavelengths | | 450, 550, 650, 750 | nm |
| Size/dimensions – all cameras | Camera | 78 x 58 x 44 | mm x mm x mm |
| | Optics | 75 (l) x 57 (d) | mm x mm |
| Mass without contingency (CBE) – StaffCam with actuator | | 2.786 | kg |
| Mass without contingency (CBE) – FootCam, DescentCam (each) | | 0.35 | kg |
| Average payload power without contingency (each camera) | | 2.5 | W |
| Science data volume per image | | 12.5 | Mbits |
| Field of view – all cameras | | 44 x 35 | degrees |
| Instantaneous field of view – all cameras | | 0.3 | mrad |

**Exhibit 13.** Characteristics of Imagers (FootCam, StaffCam, DescentCam).

## 3.2. Flight System

The Mercury Lander flight system (Exhibits 14 & 22) includes four stages, as described in Section 2.1, High-Level Mission Concept Overview. The cruise stage and descent stage are dedicated propulsion stages, with high-voltage power subsystem components on the cruise stage to support the SEP. The orbital stage is built around its large, biprop propulsion system. To minimize landed mass, the orbital stage also houses those guidance and control (G&C), power, and telecomm components that are not required for landing or surface operations. In addition to the full payload suite, the Lander houses the avionics components that provides command and data handling for the entire flight system, the majority of the telecomm subsystem, a power subsystem including a NextGEN RTG and a small battery, those G&C sensors required for landing, and a smaller bipropellant propulsion system. The allocation of components across the four stages is shown in the high-level system architecture block diagram, Exhibit 15. Mass and low-voltage power budgets include 30% margin, summarized in Exhibits 16 & 17. Mass contingencies listed are allocations from the total 30% margin. The high-voltage power system to support the SEP includes 10% margin per APL design practices and principles. Fifteen power configurations were evaluated; the three configurations that drove solar array (excluding SEP system power support) and battery sizing are shown in Exhibit 16.

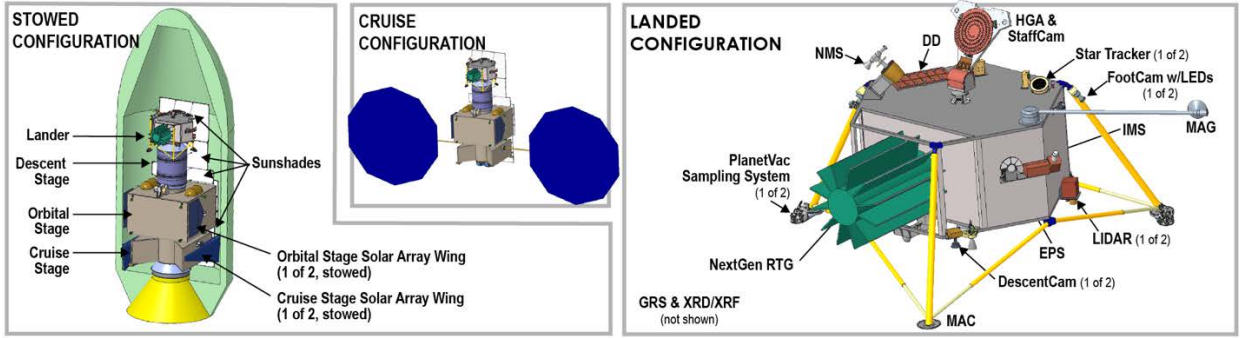

Exhibit 14. Mercury Lander Flight System.

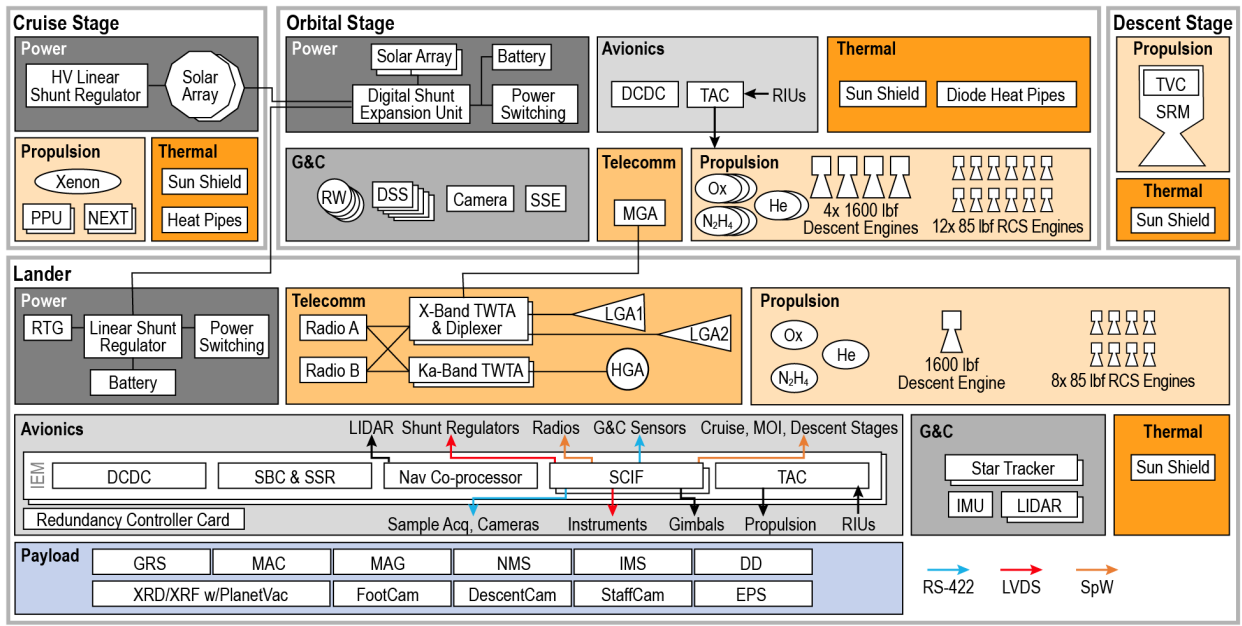

Exhibit 15. High-Level System Architecture Block Diagram.

| | MANEUVER PRE-HEAT | | | | | | SURFACE SCIENCE | | |
| | Cruise | | | On-Orbit | | | | | |
| | CBE (W) | % Cont. | MEV (W) | CBE (W) | % Cont. | MEV (W) | CBE (W) | % Cont. | MEV (W) |
|---|---|---|---|---|---|---|---|---|---|
| **Cruise Stage** | 77 | 30 | 109 | | | | | | |
| Power | 12 | 30 | 16 | | | | | | |
| Thermal | 65 | 30 | 93 | | | | | | |
| **Orbital Stage** | 440 | 30 | 629 | 341 | 30 | 488 | | | |
| Avionics | 12 | 30 | 17 | 12 | 30 | 17 | | | |
| Guidance & Control | 51 | 30 | 74 | 51 | 30 | 74 | | | |
| Power | 32 | 30 | 46 | 32 | 30 | 46 | | | |
| Propulsion | 165 | 30 | 236 | 165 | 30 | 236 | | | |
| Thermal | 180 | 30 | 257 | 81 | 30 | 116 | | | |
| **Descent Stage** | 80 | 30 | 114 | 76 | 30 | 109 | | | |
| Thermal | 80 | 30 | 114 | 76 | 30 | 109 | | | |
| **Lander** | 260 | 30 | 372 | 260 | 30 | 372 | 214 | 30 | 305 |
| Payload | 0 | 30 | 0 | 0 | 30 | 0 | 54 | 30 | 77 |
| Avionics | 21 | 30 | 30 | 21 | 30 | 30 | 21 | 30 | 30 |
| Guidance & Control | 50 | 30 | 72 | 50 | 30 | 72 | 0 | 30 | 0 |
| Power | 37 | 30 | 53 | 37 | 30 | 53 | 21 | 30 | 30 |
| Propulsion | 34 | 30 | 49 | 34 | 30 | 49 | 0 | 30 | 0 |
| Telecomm | 48 | 30 | 69 | 48 | 30 | 69 | 48 | 30 | 69 |
| Thermal | 70 | 30 | 100 | 70 | 30 | 100 | 70 | 30 | 100 |
| **Total** | 857 | | 1225 | 677 | | 968 | 214 | | 305 |

Exhibit 16. Flight Element Power.

### 3.2.1. PROPULSION

The cruise stage is an SEP system carrying two NEXT-C ion engines, one primary and one backup. NEXT was developed by NASA as an improvement to the NSTAR ion engine flown on Dawn and Deep Space 1. NEXT-C is a flight-qualified commercial variant built by Aerojet Rocketdyne; its first flight is slated to be on APL's DART spacecraft, launching in July 2021. The NEXT-C engines are capable of operating at up to 7 kW, delivering more than 4100 s of specific impulse, and have a qualified throughput of 600 kg each.

The custom composite-overwrapped xenon tank has a 650-kg capacity at 2500 psi and is thermally controlled to keep the stored xenon in its supercritical phase above 20°C. The system also features high- and low-pressure xenon flow assemblies, two power processing units (PPU), and a single digital control and interface unit (DCIU). The DCIU controls the feed systems and PPUs, provides the command logic to the engines, and performs error handling for the electric propulsion subsystem. A single biaxial gimbal assembly, modified from the DART gimbal to accommodate two engines, optimizes engine pointing.

| | CBE (KG) | % CONT. | MEV (KG) |
|---|---|---|---|
| **CRUISE STAGE** | | | |
| Harness | 60 | 10 | 66 |
| Structures & Mechanisms | 308 | 14 | 350 |
| Power | 99 | 15 | 113 |
| Propulsion | 232 | 9 | 254 |
| Thermal | 52 | 15 | 60 |
| Unallocated Margin | | | 231 |
| **Total Cruise Stage Dry Bus Mass** | **751** | **30** | **1074** |
| **ORBITAL STAGE** | | | |
| Avionics | 1 | 8 | 1 |
| Guidance & Control | 72 | 13 | 81 |
| Harness | 52 | 10 | 57 |
| Structures & Mechanisms | 477 | 16 | 553 |
| Power | 72 | 12 | 80 |
| Propulsion | 317 | 13 | 360 |
| Telecomm | 3 | 10 | 3 |
| Thermal | 59 | 15 | 68 |
| Unallocated Margin | | | 302 |
| **Total Orbital Stage Dry Bus Mass** | **1053** | **30** | **1507** |
| **DESCENT STAGE** | | | |
| Structures & Mechanisms | 47 | 10 | 52 |
| Motor Assembly | 209 | 10 | 230 |
| Thermal | 28 | 15 | 33 |
| Unallocated Margin | | | 92 |
| **Total Descent Stage Dry Bus Mass** | **264** | **30** | **406** |
| **Lander** | | | |
| Payload | 48 | 13 | 54.2 |
| Avionics | 7 | 8 | 8 |
| Guidance & Control | 41 | 7 | 43 |
| Harness | 21 | 10 | 23 |
| Structures & Mechanisms | 129 | 15 | 148 |
| Power | 99 | 13 | 112 |
| Propulsion | 47 | 9 | 51 |
| Telecomm | 25 | 10 | 27 |
| Thermal | 18 | 15 | 20 |
| Unallocated Margin | | | 134 |
| **Total Lander Dry Bus Mass** | **434** | **30** | **620** |

**Exhibit 17.** Flight Element Mass.

The orbital stage's pressure-regulated bipropellant system consists of four 7000 N (1600 lbf) orbital maneuvering and attitude control (OMAC) main engines and twelve 375 N (85 lbf) Commercial Crew Transportation Capability Reaction Control System (CCtCap RCS) steering thrusters, both developed by Aerojet Rocketdyne for the Boeing CST-100 Starliner, plus components required to control the flow of propellant and monitor system health and performance. Several flight-proven options exist for each subcomponent of the propulsion system, though delta-qualification testing of some components may be required due to the higher operating pressure (650 psi) of the engines. Systems of this size and type are both well-characterized and well-understood, with significant flight history. The propellants are monomethyl hydrazine (MMH) and MON-3 nitrogen tetroxide.

The OMAC and CCtCap RCS engines all use the same fast-acting solenoid pilot valves, which pneumatically open the thruster bipropellant valves using the same pressurized helium source feeding the propellant tanks. For long-duration firings, the pilot valve requires stepping down the source voltage to avoid overheating. The OMAC engine can reach 90% thrust in ≤20 ms. The CCtCap engine can reach 90% thrust in ≤5 ms. The OMAC engine delivers 277 s of specific impulse at steady state, and the CCtCap delivers 286 s. The engines require delta-qualification testing to demonstrate necessary throughput and cycle requirements, including margin.

All orbital stage propellants are stored in identical, custom 364-liter, composite overwrapped titanium tanks. Three tanks are required for MMH and three for MON-3. The tanks contain custom propellant management devices to ensure positioning of gas-free propellant for all maneuvers at the tank outlets. The maximum expected operating pressure (MEOP) for the mission is 650 psi. Helium pressurant is stored at a MEOP of 6000 psi in two custom 198-liter composite-overwrapped titanium pressure vessels. The design uses separate routings of check valves, latch valves, and series-redundant pressure regulators to limit fuel and oxidizer migration to the pressurant tanks. MESSENGER used a similar isolation design in flight. Additional pressurant line routing downstream of the regulators feeds the pneumatic pilot valves for all engines.

The descent stage SRM is a STAR 48GXV, a high-performance variation of the standard STAR 48BV motor that began its development to support Parker Solar Probe (PSP). The original STAR 48GXV stage design for PSP is leveraged for the descent stage, including the avionics assembly and LV adapter assembly, with an isogrid composite lattice structure. The motor is designed to consume 2858 kg of propellant, delivering 308 s of Isp and providing 129 kN (29,032 lbf) of thrust. The motor development plan calls for a full design iteration to incorporate data from a successful 2013 static fire test into the design and to adjust the design to meet the mission performance requirements. Subsequent efforts include two additional test units to complete qualification before delivery of the flight item.

The Lander propulsion subsystem is a smaller version of the orbital stage pressure-regulated bipropellant system. The system consists of one OMAC main engine and eight CCtCap RCS steering thrusters. The main engine provides a deceleration to soft touchdown (< 1 m/s) at the selected landing site on the surface. The eight RCS thrusters provide full, redundant control from the separation of the orbital stage through touchdown. Their orientation provides supplemental thrust to the main engine during the final landing to counteract the Mercury gravity losses, and roll control during the STAR 48GXV burn.

The Lander MMH and MON-3 are stored in identical, custom 50-liter, 650 psi MEOP, composite overwrapped titanium tanks. The two tanks also contain custom propellant management devices to ensure positioning of propellant at the tank outlets. Helium pressurant is stored at 3000 psi in a custom 45-liter composite-overwrapped titanium pressure vessel.

### 3.2.2. MECHANICAL

The full flight stack as well as each individual stage is designed to provide a direct and efficient load path, and to minimize the overall stack height. The overall stack height is minimized to meet the center of gravity (CG) height requirement for the LV and separation system, and to locate the Lander at a height compatible with a LV fairing door to allow RTG integration.

The cruise stage is attached to the LV payload attach fitting on one end and interfaces with the orbital stage at the other end, with the xenon tank and thrusters mounted to a central cylinder of aluminum sheet/string construction. Two aluminum honeycomb panel support structures mount to the exterior of the core cylinder and house the two 30 m$^2$ UltraFlex solar array wings.

The orbital stage consists of a central cylinder that holds the propellant tanks and a support structure for the two 1.5 m$^2$ solar arrays and the orbital stage sunshade. The tanks are mounted to the exterior of the central cylinder in a symmetric pattern to maintain the CG of the stack along the centerline during the orbit insertion burn. The tanks are mounted directly to the cylinder through struts at the top and bottom, in a tank mounting scheme based on the MESSENGER spacecraft structure. Four thrusters are mounted to the bottom of the cylinder, on the inside, and the upper volume of the cylinder accommodates the descent stage SRM.

The descent stage is mounted to the orbital stage and supports the Lander. The adapters at either end are modified from the PSP design to fit the Mercury Lander stack and additional structure is included to hold the descent stage sunshade. The orbital stage adapter is designed to have the entire engine cone nested within the orbital stage, lowering the overall stack height and improving structural integrity.

The Lander configuration accommodates instrument, RTG, and landing stability requirements. The Lander consists of a central aluminum cylinder with a hexagonal structure built around it. Propellant tanks are mounted inside the central cylinder. The majority of the hexagonal structure is closed with aluminum honeycomb panels, and most Lander components are housed within this enclosure. The RTG is located in the open part of the hexagonal structure, mounted to an open titanium frame attached directly to the exterior of the central cylinder. This layout provides an efficient load path, effective heat rejection, and allows RTG installation at the launch pad through a fairing door. The rest of the Lander bus components are mounted opposite the RTG, providing a balanced layout at any fuel level and a low CG, as well as an axisymmetric mounting interface for the landing legs.

The Lander is fitted with three leg assemblies which are configured to provide a stable landing platform. Each leg assembly consists of a telescoping primary strut attached to the top deck and two secondary struts attached to the bottom deck. Crushable honeycomb blocks are incorporated in the primary strut to absorb landing energy and minimize shock. The legs are folded to fit within the LV fairing and behind the Lander sunshade, and deploy to approximately 30 degrees after the Lander sunshade is jettisoned and prior to landing.

### 3.2.3. THERMAL

The thermal design and operation of the cruise, orbital, and descent stages are based on the MESSENGER spacecraft, leveraging heritage from materials and component designs. During the inner cruise and orbital phases, the thermal design relies on four sunshades (one per stage), constructed from high temperature materials similar to those used on MESSENGER, to protect the full vehicle stack from the intense solar environment as the spacecraft stack approaches Mercury perihelion. The sunshades are wrapped around titanium frames and create a benign thermal environment when oriented toward the Sun, allowing for the use of standard electronics and electrical components and thermal blanketing materials. The solar array wings are actively controlled by G&C software, as was done on the MESSENGER and PSP spacecraft, to maintain peak operating temperature and protect the solar cells from damage while providing sufficient power to maintain the SEP engines at the desired power state. Other components that must be Sun-exposed (e.g., the digital Sun sensors (DSS)) utilize thermal designs similar to those developed and successfully used on MESSENGER. These components operate throughout the Mercury year and also during orbits that cross over one of Mercury's "hot poles" that face the Sun at Mercury perihelion. When at spacecraft perihelion, Sun-facing components experience as much as 11 times the solar radiation near Earth. During this time, the sunshade temperatures rise to >300°C.

The cruise stage and orbital stage propulsion components are thermally isolated and use heater power to control the temperature to ~30°C for NEXT-C and ~0°C for the biprop components. The descent stage SRM is completely covered with multi-layer insulation (MLI) and heated to maintain the temperature between 25–30°C uniformly. Because the SRM is always in the umbra created by the sunshade its temperature is maintained by active heater control to meet the maximum temperature (35°C) and gradient (<5°C) requirements imposed by the supplier. Dedicated radiators with embedded fixed conduction heat pipes to uniformly spread and manage the waste heat during NEXT-C operation (DART spacecraft heritage) support the cruise stage PPUs. Electronics in the orbital stage requiring dedicated radiators are attached using diode heat pipes as to effectively attenuate the thermal environments when in Mercury orbit (MESSENGER heritage) to allow for full, unrestricted operation during all parts of the orbital mission phase.

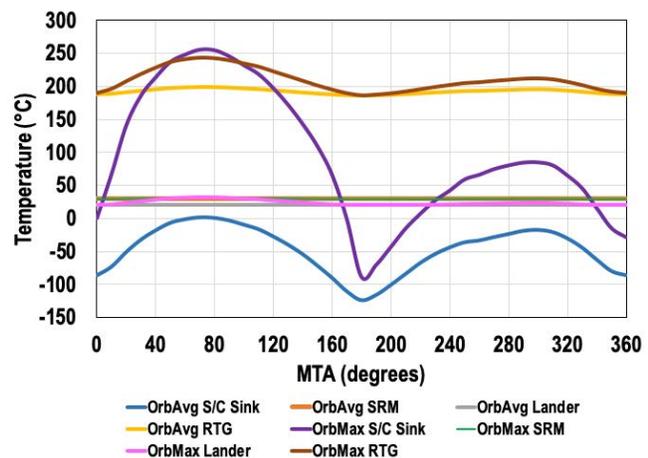

**Exhibit 18.** Maximum and orbit average temperature as a function of MTA for the 100 km x 6000 km orbit. The maximum orbit average spacecraft sink temperature stays below 0°C and the SRM is maintained at 30°C. The peak spacecraft sink occurs at MTA 70 and reaches 255°C. The Lander and RTG maximum temperatures are 30°C and 250°C, respectively.

For most orbits about Mercury, the spacecraft passes between the Sun and the illuminated planet for ~30 minutes. During this period, the sunshades protect the spacecraft from direct solar illumination, but the back of the spacecraft and back side of the sunshades are exposed to the hot Mercury surface. Depending on orbit plane and planet position the once very-cold space sink temperature (< -100°C) behind the sunshades varies from ~0–255°C, occurring at the MTA70 position, during this 30-minute period. A 100 km x 6000 km orbit with an argument of periapsis of 320° is selected, via a comprehensive trade study, as a safe storage orbit that is designed to minimize peak planetary heating and keep the maximum orbit average sink temperature of the stages behind the sunshades roughly at or below 0°C during the Mercury year. This orbit ensures that the SRM is maintained at a temperature much greater than the orbit average spacecraft sink temperature, and thermostatically controlled heaters prevent the risk of over-temperature. Electronics needing dedicated radiators are connected using diode heat pipes to eliminate transient heating effects during the worst-case orbits. Lander RTG temperature excursions from the nominal 180°C during the Mercury year were investigated. As shown in Exhibits 18 and 19, the maximum predicted temperature experienced by the RTG, occurring during the MTA70 orbit, is 250°C, which is consistent with temperatures measured on the New Horizons RTG during the first six months of the mission. This orbit also represents the maximum orbital transient ΔT of 70°C. Prior to landing near Mercury aphelion, the orbital, descent stages, and Lander are transitioned into a 100 km x 2000 km orbit with the same argument of periapsis once the MTA position is greater than MTA130 (Exhibit 20). Once committed to this lower orbit, the landing sequence must be initiated prior to MTA240, about 37 days later.

The Lander thermal control design leverages heritage materials and techniques from both the MESSENGER and New Horizons spacecraft. The Lander is mostly enclosed with MLI, with areas on its surface exposed as radiators to provide cooling for the electronics during hot operation and environmental conditions. Thermostatically controlled heaters are used to protect the electronics during minimum operational cases and cold environmental conditions. Once landed on Mercury, the RTG and sensitive components are in the shadow of the Lander body, protecting them from the intense solar environment, allowing for the use of standard thermal control strategies. The Sun elevation angle is small at landing, which limits Mercury surface temperatures to less than 40°C near Mercury aphelion (MTA180). At Mercury aphelion with a Sun elevation angle of 2°, RTG temperatures are around 200°C (Exhibit 21), lower than temperatures seen on the New Horizons RTG early in flight. After sundown, the Mercury surface reaches low temperatures which are less stressing than those seen in cruise and easily accommodated by the RTG and Lander.

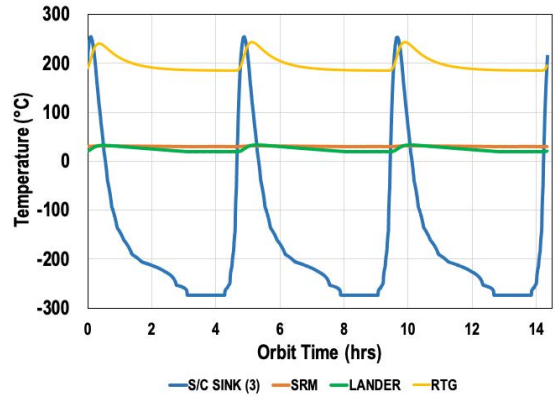

**Exhibit 19.** The time varying temperatures of the spacecraft sink and SRM during the MTA070, 100 km x 6000 km orbit. The maximum spacecraft sink reaches ~260°C whereas the massive and heavily insulated SRM remains around 30°C. Note that the Lander and RTG temperature excursions are also benign.

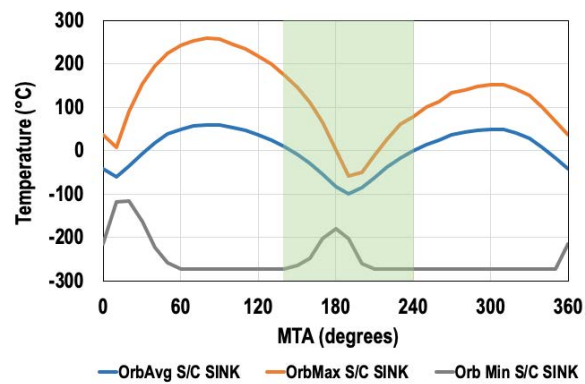

**Exhibit 20.** Maximum and orbit average temperature as a function of MTA for the 100 km x 2000 km orbit. Green rectangle indicates the safe MTA range for the SRM where the orbit average sink (blue) is below 10°C. Landing occurs around aphelion (MTA180). Unlike the 100 km x 6000 km orbit, this orbit cannot safely sustain the SRM for a Mercury year because the orbit average spacecraft sink temperature exceeds 50°C at times.

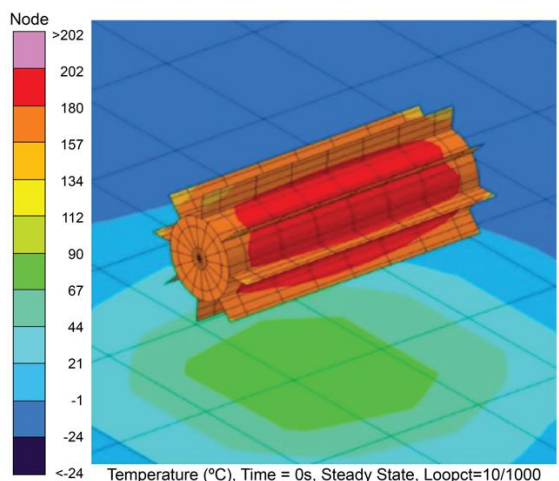

**Exhibit 21.** Lander RTG temperature on surface at Mercury aphelion with 2° Sun elevation.

### 3.2.4. POWER

The electrical power subsystem provides power generation, regulation, distribution, and energy storage for the spacecraft through all mission phases. The high-level system architecture block diagram (Exhibit 15) includes the location of the electrical power subsystem components on each of the stages.

**Power Generation.** A single NextGen RTG provides power in the landed configuration. Solar cell arrays supplement the RTG during all other mission phases. The cruise stage carries two solar array wings (total area ~60 m$^2$) that provide power to the vehicle at two operating voltages: 100 V for ion propulsion, and 30 V to supplement the RTG. Once the cruise stage is jettisoned, two smaller solar cell array wings (3 m$^2$ total), which had been stowed to minimize degradation, are deployed on the orbital stage to supplement the RTG during the orbital phase of the mission.

Both the cruise and orbital stage solar arrays use inverted metamorphic multijunction solar cells with beginning of life (BOL) efficiency of 32% under standard test conditions. The wings for each stage are connected to the vehicle by a single-axis gimbal, and must be rotated off normal (relative to the sunline) to maintain an operating temperature below 165°C. In addition, during the orbital phase the orbital stage wings must be rotated 90° off Sun during sub-solar crossings to limit the peak temperatures. Array degradation estimates apply data from the MESSENGER mission.

**Power Regulation.** Primary power regulation is implemented by two shunt regulators, one each for the 100 V power bus and the 30 V power bus. The topologies are similar, with design heritage from the Van Allen Probes and NEAR missions, but they operate at significantly different power levels. Both regulators implement a hybrid linear/digital topology in which digital stages provide coarse voltage control and linear stages provide fine voltage control. The shunt regulators implement a fault tolerant, 3-stage majority voted control loop to ensure voltage control is maintained under all conditions.

The low-voltage shunt system not only regulates RTG output, but also provides battery charge control. Because two batteries are carried on the system, the low-voltage shunt regulator includes both a float charger to maintain the Lander battery at a fixed voltage during cruise and orbital operations, and switching to allow the battery charge control circuitry to regulate the Lander battery once the orbital stage is jettisoned.

The linear and digital portions of the shunt regulators are packaged separately. The cruise stage components are all located on that stage. However, the digital portion of the low-voltage regulator is located on the orbital stage to reduce landed mass.

**Power Distribution.** Low-voltage power distribution is provided by block redundant power switching units, similar in concept and topology to many previous missions. Power switching for the high-voltage bus is required only for the thruster electronics and is included as part of the high-voltage shunt regulator.

**Energy Storage.** Two lithium-ion batteries provide energy storage, one each located on the Lander and the orbital stage. The former supplements RTG power only during the landing sequence and landed science operations. During the remainder of the mission it is held at a storage state of charge to reduce degradation, and the orbital stage battery provides the supplemental energy required for orbital eclipse or peak load operations. The dual battery implementation reduces landed mass as the orbital energy storage requirements are significantly higher than the landed requirements.

### 3.2.5. GUIDANCE & CONTROL (G&C)

The G&C subsystem contains sensors and/or actuator components on each of the four stages, as well as the software-embedded algorithms required to process data and send commands to the spacecraft. G&C controls the spacecraft orientation, performs trajectory correction maneuvers (TCMs), controls the gimbal assemblies throughout the stages, and processes data from the landing sensors for hazard detection and avoidance.

The G&C software controls the two-axis SEP gimbal on the cruise stage, adjusting it to minimize torque imparted by the engines on the stack and control system momentum while the SEP engines are thrusting. The engines are slightly canted to reduce swirl torque. G&C software also controls the single-axis solar array drive actuators on the cruise and orbital stages to dynamically maintain the solar panels at a thermally safe angle.

A five-head Sun sensor assembly on the orbital stage provides a Sun vector in the body frame that is used for protection in the event of a loss of inertial reference knowledge or a transition to safe mode. Four 100 N-m-s reaction wheels provide attitude control during cruise and Mercury orbit, and are controlled to minimize their power usage. Momentum management and small TCMs are executed using the twelve orbital stage RCS engines in a short, pulsed mode. These attitude control engines are also used to counteract any residual torques created by the four OMAC engines which, following cruise stage separation, are used during large TCMs including the MOI burn and subsequent orbit lowering maneuvers in preparation for descent.

Two orthogonally mounted star trackers on the Lander provide inertial attitude knowledge during cruise through landed operations. An internally redundant four-gyro/four-accelerometer inertial measurement unit (IMU) provides body rate and acceleration data, and propagates attitude when the star trackers are not available. G&C controls the two-axis gimbal on the high-gain antenna (HGA) / StaffCam assembly, and uses data from the two scanning LIDARs for hazard avoidance during landing.

### 3.2.6. AVIONICS

The Lander avionics subsystem controls the full four-stage system and consists of block-redundant, radiation-shielded integrated electronics modules (IEMs), and small, distributed remote interface units (RIUs). The IEMs combine command and data handling (C&DH), navigation processing, G&C functions, and mass memory storage. The IEMs use the 500 mm Al (Van Allen Probe-heritage) 6U chassis, populated with modified PSP-heritage electronics cards. Integrated circuits in the IEMs are designed to tolerate ≥100 krad. Local shielding is used for the few parts that can tolerate only 50–100 krad at a radiation design margin of 2. Each IEM consists of a single board computer (SBC), navigation co-processor (NCP), solid-state recorder (SSR), two spacecraft interface cards (SCIF), thruster/actuator card (TAC) and DC/DC converter card (DCDC). An additional TAC and DCDC are located on the orbital stage to support its propulsion system.

The Parker Solar Probe (PSP)-heritage SBC is modified to host a high-performance, power-efficient, DART-heritage LEON UT700 processor with 32MB SRAM with DMA, and 8 MB nonvolatile MRAM. The processor executes flight software for C&DH functions, a 50-Hz attitude control loop, and on-board autonomy. The FPGA-based NCP supports LIDAR and image processing for hazard avoidance during descent and landing, and is powered off otherwise.

The PSP-heritage TAC provides the necessary power switching for precision actuation of propulsion system thrusters, as well as the interface for the RIUs. The propulsion thruster interface is modified, coupling its 50 Hz resolution pulse-width modulation functionality with a regulator and filter to provide 45 W burst power to open the thruster valve and then 9 W steady-state power for the remainder of the burn.

The redundancy controller card (RCC) distributes inputs from spacecraft resources that do not have redundant ports to each side of the IEM and multiplexes outputs from either IEM side to the appropriate component. The RCC also includes circuitry controlling switches between the two redundant IEMs, which facilitates rapid responses to critical faults.

The remaining components on the Lander IEMs present low design risk to the system due to strong precedent on other programs.

**Exhibit 22.** Flight System Element Characteristics.

| FLIGHT SYSTEM ELEMENT PARAMETERS | | VALUE/ SUMMARY, UNITS |
|---|---|---|
| **GENERAL** | | |
| Design Life | | 10.5 years |
| **STRUCTURE** | | |
| Structures Material (Aluminum, Exotic, Composite, Etc.) | | Central Cylinders & Stage Adapters : Al-7075, Secondary Structures : Al-6061, Honeycomb Panels with Al-6061 Facesheets & Al-5056 Cores, Descent Stage Motor Casing : Composite |
| Number of Deployed/Articulated Structures | Cruise Stage | Solar Array Wings (2), Ion Engine Biaxial Gimbal (1), Stage Separation (1) |
| | Orbital Stage | Solar Array Wings (2), Stage Separation (1) |
| | Descent Stage | Sunshade (1), Stage Separation (1) |
| | Lander | Sunshade (1), Lander Legs (3), Magnetometer Boom (1), HGA (1) |
| **THERMAL CONTROL** | | |
| Type of Thermal Control Used | | MLI, Heat Pipes, Sunshade, Heaters |
| **PROPULSION** | | |
| Estimated ΔV Budget | Cruise Stage | 2 km/s |
| | Orbital Stage | 917 m/s |
| | Descent Stage | 3.3 km/s |
| | Lander | 316 m/s |
| Propulsion Type(s) & Associated Propellant(s)/Oxidizer(s) | Cruise Stage | Solar Electric, Xenon |
| | Orbital Stage | Bipropellant, MMH and MON-3 |
| | Descent Stage | Solid Rocket Motor, TP-H-3340 |
| | Lander | Bipropellant, MMH and MON-3 |
| Number of Thrusters & Tanks | Cruise Stage | 2 NEXT-C Engines (Primary & Redundant), 1 Xenon tank |
| | Orbital Stage | 4 OMAC (1600 lbf), 12 CCtCap (85 lbf) Engines, 3 Oxidizer Tanks, 3 MMH Tanks, 2 GHe Tanks |
| | Descent Stage | 1 STAR 48GXV |
| | Lander | 1 OMAC (1600 lbf); 8 CCtCap (85 lbf) Engines; 1 Oxidizer Tank; 1 MMH Tank; 1 GHe Tank |
| Specific Impulse of Each Propulsion Mode | Cruise Stage | > 4100 s |
| | Orbital Stage | OMAC: 277 s, CCtCap: 286 s |
| | Descent Stage | 308 s |
| | Lander | OMAC: 277 s, CCtCap: 286 s |
| **ATTITUDE CONTROL** | | |
| Control Method | | 3-axis |
| Control Reference | | Inertial, Sun Constrained (Orbit); Velocity, Sun Constrained (Descent); Mercury Nadir, Sun Constrained (Landing) |
| Attitude Control Capability | | < 0.057° (Cruise/Orbit) (3-σ); < 0.75° (Descent) (3-σ); < 0.5° (Landing) (3-σ) |
| Attitude Knowledge Limit, Degrees | | 0.02 (Cruise/Orbit) (3-σ); < 0.1 (3-σ) (Descent/Landing) |
| Agility Requirements, Deg/Sec | | < 1 deg/s(Orbit); < 7.5 deg/s (Descent/Landing) |
| Articulation/#–Axes | | 2x Cruise Solar Array Drive Assembly (SADA) (1 axis); SEP Gimbal (2 axis) ; 2x Orbital SADA (1 axis) STAR48 TVC (2 axis) ; HGA (2 axis) |
| Sensor& Actuator Information | | Star Tracker (x2) < 0.009 deg (3-σ) Transverse, < 0.01 deg (3-σ) Boresight; IMU (4 Gyros X 4 Accelerometers) < 0.05 deg/hour Bias, < 1000 μg ; Sun Sensors (5 heads), 0.5 deg (3-σ); Reaction Wheels (4x 100 N-m-s) up to 0.4 N-m Torque |
| **COMMAND & DATA HANDLING** | | |
| Flight Element Housekeeping Data Rate | | 100 kbps |
| Data Storage Capacity | | 80 Gb |
| Maximum Storage Record Rate | | 1 Mb/s |
| Maximum Storage Playback Rate | | 2 Mb/s |
| **POWER** | | |
| Type of Array Structure | Cruise Stage | UltraFlex Deployed |
| | Orbital Stage | Rigid Deployed |
| Array Size | Cruise Stage | 60 m² |
| | Orbital Stage | 3 m² |
| Solar Cell Type | | Inverted Metamorphic Multijunction |
| RTG Type | | NextGen RTG (16x General Purpose Heat Source) |
| RTG Expected Power Generation | | 373 W Beginning of Life (BOL), 323 W End of Life (EOL) |
| Available Power at Shunt Regulated Voltage | | Cruise Array Propulsion Section: BOL ~9.3 kW @ 0.99 AU, EOL ~8.3 kW @ 0.46 AU Cruise Array Spacecraft Section: BOL ~1.4 kW @ 0.99 AU, EOL ~1.3 kW @ 0.46 AU Orbital Array: BOL 1128 W @ 0.46 AU, EOL 1052 W @ 0.43 AU |
| On-Orbit Average Power Consumption, Fully Margined | | 914 W |
| Battery Type | | Lithium-Ion (both batteries) |
| Battery Storage Capacity | Orbital Stage | 60 Ah BOL |
| | Lander | 4.5 Ah BOL |

### 3.2.7. FLIGHT SOFTWARE

The flight software (FSW) is built upon software successfully flown on multiple APL missions including PSP. The FSW uses a layered architecture to encapsulate functionality into multiple distinct applications, ensuring that functionality is self-contained and readily maintainable.

At the lowest layer of the architecture is the Core Flight Executive (cFE) software. cFE was developed and is maintained by NASA Goddard Space Flight Center. cFE provides lower layer functionality to all applications including a message passing function between applications, an event handling system, and application configuration. Spacecraft control software, including C&DH and G&C software, is layered on top of the cFE. The boot software that starts the FSW is commercial, off-the-shelf software provided by Cobham Gaisler. The boot software runs out of MRAM on power-up or on reset, and configures the hardware for use by application software. Once the hardware is configured, the boot software starts the main application software.

### 3.2.8. TELECOMMUNICATIONS

The telecommunications subsystem includes two X-band low-gain antennas (LGAs), one Ka-band HGA, and two radios on the Lander. Additionally a set of two X-band medium-gain antennas (MGAs) and associated waveguide are housed on the orbital stage for use during cruise and orbit phases. The subsystem is able to close X-band uplink/downlink from Earth via either LGA or MGA (cruise and orbit only), and Ka-band HGA downlink (Lander only). The telecomm subsystem is fully redundant up to the antenna and final antenna feed connections.

A redundant set of X-band diplexers is utilized between the radio and antenna. X-band and Ka-band traveling wave tube amplifiers (TWTAs) are each allocated 33 W input power, which is assumed to result in approximately 11 W output power (33% efficiency, conservatively assumed due to a relatively low input power level).

The two X-band LGAs are a choke circular waveguide horn design, together providing near omnidirectional coverage. The LGAs are used during launch and early orbit phase, TCMs, and emergency operations. The fan-beam MGAs are dual 5-element linear phased array devices, each with an approximately ±45° beam-width on the wide axis and a ±6° 3dB beamwidth on the narrow axis, resulting in a peak gain of 16 dBi. The Ka-band HGA is a 0.6m radial line slot array (RLSA) which is expected to provide a peak gain of approximately 43dBi. The use of an RLSA rather than a conventional parabolic reflector structure streamlines the mechanical packaging of the Lander.

## 3.3. Mission Design & Concept of Operations

Chemical and SEP trajectory options were explored for the cruise and orbital phases of the mission. To design the interplanetary trajectory, an optimal ballistic impulsive ΔV transfer was developed, then the potential propellant savings from converting the impulsive maneuvers to SEP thrust arcs were evaluated.

| EVENT | DATE | V∞ (KM/S) | MASS (KG) |
|---|---|---|---|
| Launch | 3/23/2035 | (C3 = 14.7 km2/s2) | 9406 |
| Earth Flyby | 3/23/2036 | 3.85 | 9406 |
| Venus Flyby 1 | 6/22/2036 | 7.42 | 9406 |
| Venus Flyby 2 | 3/16/2038 | 7.53 | 9397 |
| Mercury Flyby 1 | 5/19/2038 | 5.9 | 9392 |
| Mercury Flyby 2 | 2/14/2039 | 5.17 | 9380 |
| Mercury Flyby 3 | 1/27/2040 | 3.91 | 9330 |
| Mercury Flyby 4 | 1/18/2041 | 3.62 | 9324 |
| Mercury Flyby 5 | 6/26/2042 | 2.51 | 9245 |
| Arrival | 1/13/2045 | 0.00 | 8944 |

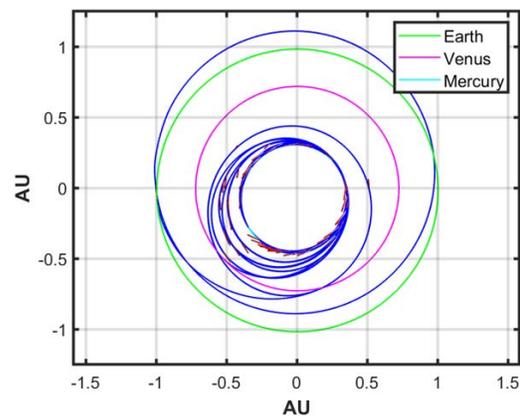

**Exhibit 23.** Reference Cruise Trajectory (blue). SEP Thrust Directions are given in red.

The SEP reference trajectory [Shannon et al. 2020] launches from Cape Canaveral Air Force Station on 23 Mar 2035 with a C3 of 14.7 km²/s². The trajectory consists of one Earth, two Venus, and five Mercury gravity assists (Exhibit 23). The spacecraft arrives at Mercury with $V_\infty$ = 0 km/s on 13 Jan 2045. The primary launch period includes a one-year resonant Earth return in the beginning of the trajectory; the 2036 backup launch period targets the same Mercury arrival date and does not include the Earth flyby. This SEP solution was chosen as the nominal mission trajectory and is shown with the flyby characteristics in Exhibit 23.

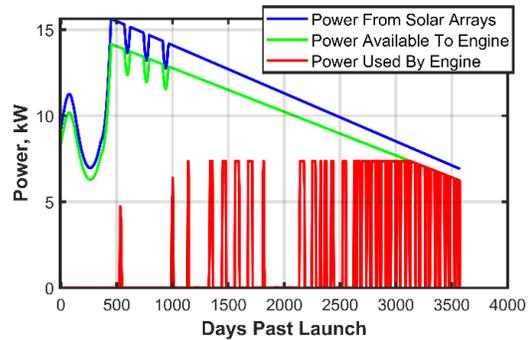

**Exhibit 24.** The cruise stage solar array is sized to provide sufficient power to the engine. As the solar array degrades, the trajectory design maximizes use of power output.

Thermal considerations place a constraint on any available SEP thrust vector pointing. The sunshades are sized to accommodate 15° off-pointing from the Sun-line. Accounting for practical attitude variation, the thrust-Sun angle is constrained to have a ±10° tolerance. Margins held in the SEP model provide robustness to a missed thrust and include 10% available power, 90% thruster duty-cycle, and 6% SEP propellant.

Two NEXT-C engines, one primary and one backup, are baselined. The solar array size was chosen to provide sufficient power to the engine despite the significant time-degradation present at the end of the cruise phase, as shown in Exhibit 24. The end-of-cruise thrust requirements to orbit match with Mercury drive the solar array size.

The possibility of performing portions of the orbit lowering and station-keeping using SEP at Mercury was studied extensively. However, given the large vehicle mass, which would be exacerbated by the larger solar arrays required to support an SEP orbit phase, the specific acceleration provided by the SEP system was too small even at the highest power settings. This results in very-long flight times to transfer to the low-altitude descent orbit, during which the spacecraft is exposed to the surface sub-solar point at low altitudes, violating thermal requirements. Additionally, the spacecraft thrust angle constraint required to maintain thermal shielding in the Sun direction significantly reduces the feasible thrust durations, even at best-case orbit orientations, and adds years to the total mission duration.

| | PARAMETER | VALUE |
|---|---|---|
| **CRUISE PHASE SUMMARY** | | |
| **Primary Launch** | Period | 09 Mar 2035 to 30 Mar 2035 |
| | C3 Min–Max | 12.28–17.57 km²/s² |
| | Duration | 117.5 months |
| | Cruise Propellant Mass (with/without contingency) | 513 / 484 kg |
| | Cruise Propellant contingency | 6% |
| **Secondary Launch** | Period | 09 Mar 2036 to 30 Mar 2036 |
| | C3 Min–Max | 12.56–17.9 km²/s² |
| | Duration | 105.5 months |
| | Cruise Propellant Mass with/without contingency | 545 / 514 kg |
| | Cruise Propellant contingency | 6% |
| | Launch Site | Cape Canaveral Air Force Station |
| | Total Cruise Stage Wet Mass with contingency | 1619 kg |
| | Total Orbital Stage Wet Mass with contingency | 3764 kg |
| | Total Descent Stage Wet Mass with contingency | 3231 kg |
| | Total Lander Wet Mass with contingency (includes instruments) | 723 kg |
| | Launch Adapter Mass with contingency | 73 kg |
| | Primary Total Launch Mass | 9410 kg |
| | Launch Vehicle | Expendable Falcon Heavy |
| | Launch Vehicle Lift Capability | 11255 kg |
| | Launch Vehicle Mass Margin | 1845 kg |
| | Launch Vehicle Mass Margin % | 16% |
| | Maximum Eclipse Period | 20 minutes |
| **ORBIT PHASE SUMMARY** | | |
| | Orbit Parameters, Duration | 100 x 6000 km, 2.5 months<br>100 x 2000 km, 2 weeks |
| | Maximum Eclipse Period | 43.1 minutes |
| | Propellant Mass (with/without contingency) | 2210 kg / 2156 kg |
| | Propellant contingency* | 2.7% |

**Exhibit 25.** Cruise and Orbital Phase Summaries.
*In lieu of a formal navigation analysis, for this study ΔV has been allocated to account for associated errors using engineering judgement. These represent up to 2.7% margin over the nominal trajectory.

An all-chemical approach is implemented for the orbital phase. After orbit matching with Mercury, the cruise stage is jettisoned and the orbital stage delivers the ΔV to place the spacecraft into a thermally safe, 100 km x 6000 km altitude polar parking orbit. The apoherm lowering maneuver occurs after MTA130. Once the predicted longitude of periherm is within 5° of the terminator plane, the periherm lowering maneuver is performed and the orbital stage is jettisoned.

Cruise and orbital phase characteristics are provided in Exhibit 25, and the nominal mission itinerary is presented in Exhibit 26.

| EVENT | DATE | DESCRIPTION |
|---|---|---|
| Launch | 3/23/2035 | Cape Canaveral Air Force Station, $C_3$ = 14.7 km²/s² |
| Arrival | 1/13/2045 | $V_\infty$ = 0 km/s |
| Jettison Cruise Stage | 1/13/2045 | Executed at least 12 hours prior to MOI to allow for contingency operations |
| Mercury Orbit Insertion | 1/13/2045 | Capture into 100 km x 6000 km orbit. Propellant Requirement: 1300 kg |
| Imaging Campaign | 2/11/2045 – 3/22/2045 | Total of Over 76 Hours Over Landing Site at Various Altitudes & Viewing Angles |
| Lower Apoapsis | 3/30/2045 | Lower to 100 km x 2000 km orbit. Propellant Requirement: 712 kg |
| Lower Periapsis | | Lower to 20 km x 2000 km orbit. Propellant Requirement: 38 kg |
| Jettison Orbital Stage & Sunshades / Deploy Lander Legs | 4/12/2045 | 100 seconds prior to start of descent burn |
| Descent | | 75 seconds to burn completion |
| Jettison Descent Stage | | 30 second coast arc |
| Landing | | 34 second final descent |

Exhibit 26. Nominal Mission Itinerary.

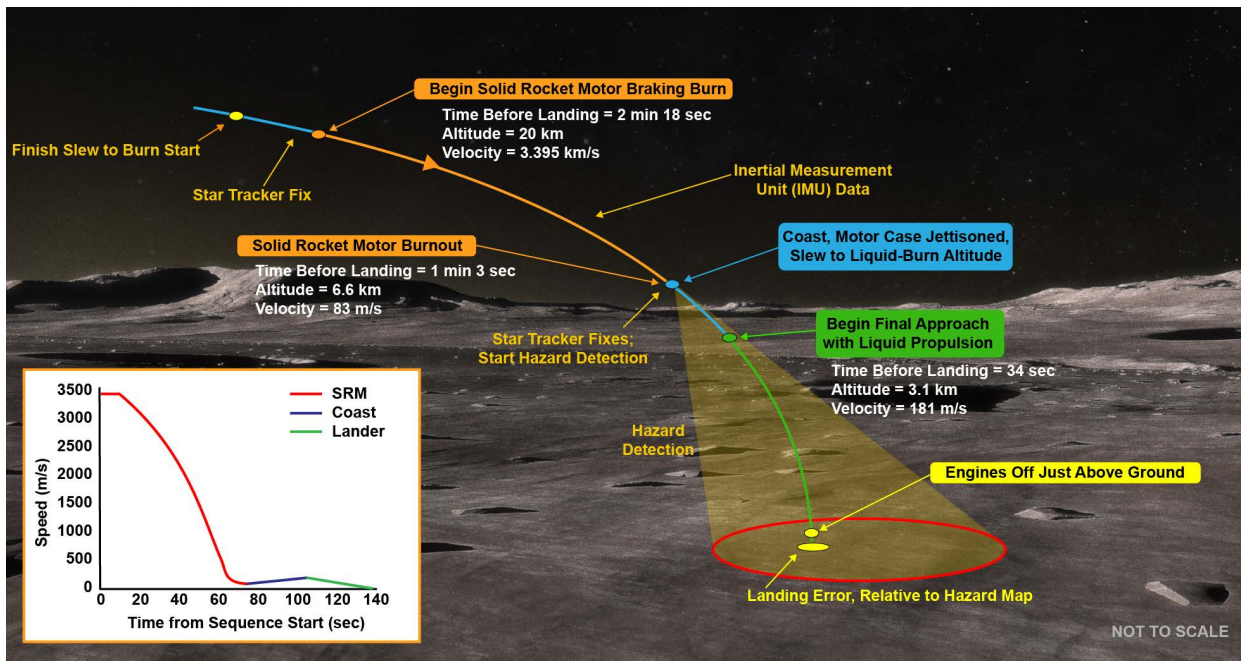

Exhibit 27. The sequence begins with a 75-s braking burn executed by the SRM, which takes out the bulk of the orbital energy, decelerating the Lander from a 3.395 km/s incoming horizontal speed to an 83 m/s, nearly vertical, speed at the end of the burn (ΔV of 3.313 km/s). Landing uses continuous LIDAR operations post SRM burn to support hazard detection.

The Mercury Lander descent and landing sequence is illustrated in Exhibit 27. The braking burn has been designed and simulated using a realistic thrust profile derived from STAR 48GXV SRM qualification test data, coupled with a simple attitude profile that maximizes horizontal speed deceleration while targeting a particular altitude within range to meet landing stage capabilities and hazard detection needs. The G&C software controls pitch and yaw through the thrust vector control (TVC) nozzle gimbals on the STAR 48GXV in 0.1° steps at 10 Hz. Additional attitude control for the SRM burn is provided by the engines in the Lander. The phasing orbit prior to the burn allows for the collection of on-orbit images to down select a landing zone with low hazard content (at the resolution of the on-orbit images), and subsequent tight control of the burn ignition time to target that landing zone, while ensuring full Deep Space Network (DSN) coverage and proximity to the terminator plane. State knowledge during the descent burn is obtained from IMU propagation from the last star tracker fix and the ground estimated state uploaded to the spacecraft before the start of the burn.

The braking burn is followed by a 30-s coast arc, jettison of the descent stage, and determination of whether the nominal landing site is sufficiently clear of hazards using a scanning LIDAR and an on-board hazard map generated from collected data. The near-vertical motion achieved at the end of the braking burn allows for a continuous scan of a single landing area during the coast arc, and an initial landing site is selected at ~3.1 km altitude. Scanning LIDAR performance at this altitude ranges from ~0.53m$^2$/px over 400 m$^2$ to better than 0.1 m$^2$/px over 540 m$^2$, depending on the sampling rate used (10 kHz or 100 kHz, respectively). The guidance system then targets the selected site during the liquid burn while continuing to refine the selection with increased resolution on descent. LIDAR measurements provide translation state corrections in addition to terrain assessment. Follow-on LIDAR measurements focus on the search area and result in cm-level accuracy in hazard detection. This approach builds upon the safe landing strategies developed for lunar applications with the use of multi-resolution algorithms for LIDAR processing, and assumes similar terrain and rock size—frequency distribution as found on the Moon. As with lunar landing, the final 15 m of descent is performed with the IMU only, to avoid any interference of lofted dust with LIDAR, driving the landing accuracy to a few meters from the selected landing site.

The liquid descent was simulated with a 91% throttle assumption to account for variability and margin for lateral control. In particular, using the remaining 9% thrust control can produce more than 3 kN in lateral force, or accelerations larger than 4 m/s$^2$, allowing the Lander to reach any lateral coordinate within the hazard map within the first 15 s of the descent. The short duration of the SRM burn allows state knowledge to be derived from IMU propagation of the star tracker measurements and state estimates at the end of the coast arc. Navigation during this final descent is performed relative to the hazard map generated at the end of the coast arc (3.1 km altitude), and LIDAR measurements are used to provide meter-level navigation accuracy within the map.

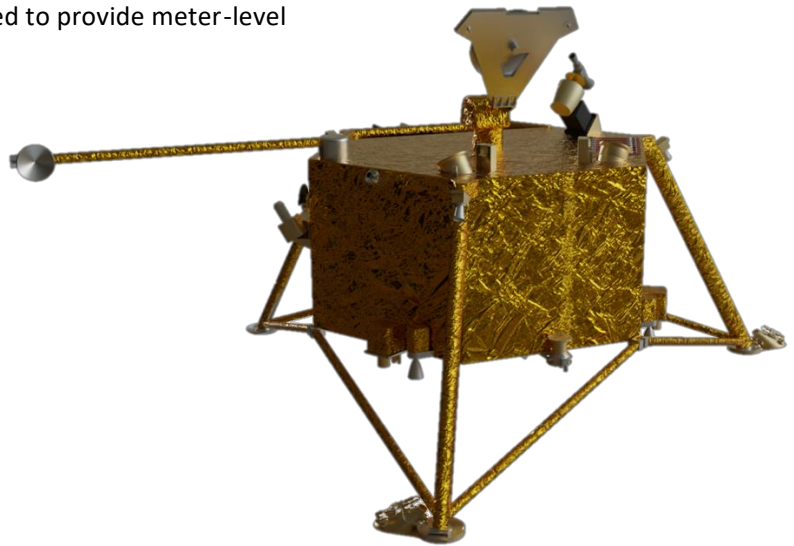

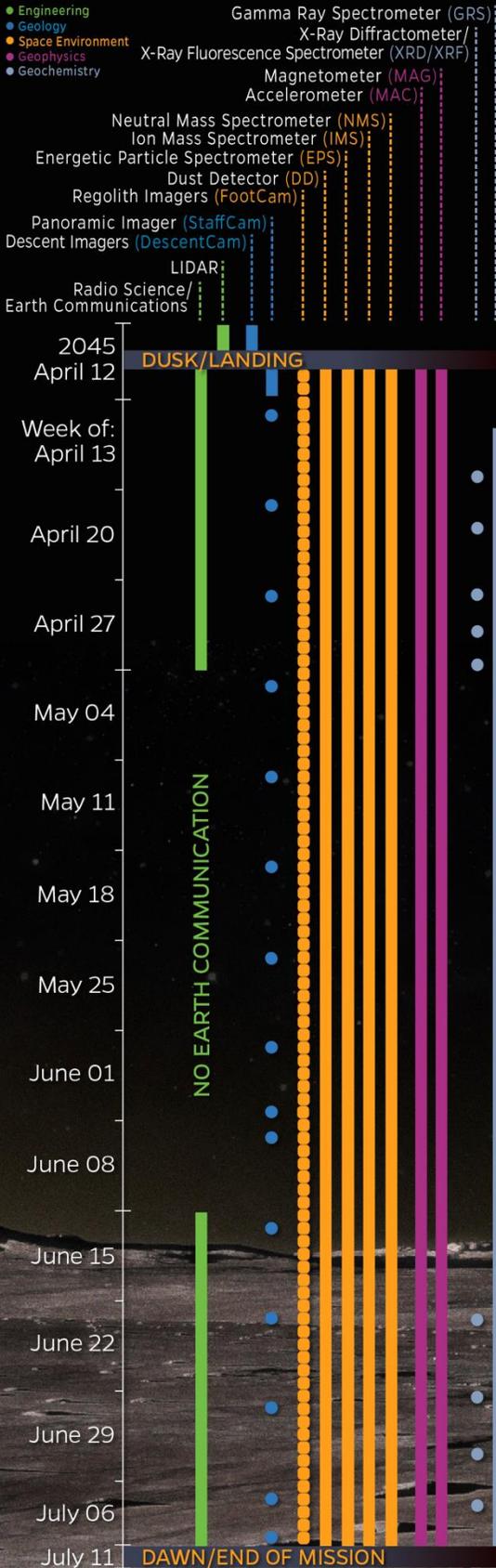

### 3.3.1. LANDED OPERATIONS

The overall concept of landed operations is depicted graphically in Exhibit 28. Additional operations details are given in Appendix B4. The Lander touches down at dusk, about 30 hours before sunset; the exact amount of time in sunlight depends on the local topography of the landing site. During the limited hours of sunlight, StaffCam and FootCam capture images and panoramas of the landing site region (satisfying measurements for Objective 4.2). LEDs are used to enable night-time imaging by FootCam, and will be used to enable night-time imaging by StaffCam as possible (see Section 3.1). The glow from Mercury's Na exosphere may also provide diffuse illumination; whether an adequate signal-to-noise ratio can be attained using this light source will be validated by radiometric modeling before flight. FootCam images are acquired daily, to monitor for change detection, particularly before and after PlanetVac operations. StaffCam acquires a panorama weekly, characterizing the surrounding landscape and any detectable exospheric glow. In particular, on 7 Jun 2045, StaffCam undertakes a dedicated exploratory campaign devoted to imaging the Na exosphere, timed to occur during the maximum seasonal radiance at MTA60 (Exhibit B6). As the Sun rises at the end of the mission, StaffCam and FootCam acquire multiple images and panoramas, and will send them back to Earth until transmissions end when operating temperatures are exceeded.

Many of the instruments operate nearly continuously throughout the landed mission, including NMS, IMS, EPS, DD, MAC, and MAG (once the boom deploys). The radio science investigation also begins, continuing whenever Earth communication is possible. GRS has a cool-down period after landing and then begins continuous operations approximately 36 hours later. The RTG cools concurrently with GRS, allowing it to reach full power output before GRS operations begin. GRS operates truly continuously whereas, owing to power limitations, all other "continuous" instruments momentarily cease operations during the ~one-hour XRD/XRF operations.

XRD/XRF operations occur in coordination with the operation of PlanetVac, which provides regolith samples to the instrument for analysis. During the initial three weeks of operations and with DTE communication, four distinct PlanetVac samples are analyzed. The baseline plan includes one sample from each of the two PlanetVac samplers, and two additional samples from one of the samplers. These four initial samples will provide insight into the nature and amount of compositional diversity that may exist between the locations of the two legs as well as at depth, as PlanetVac excavates deeper into the regolith with each subsequent sample at a given location. Due to the complexities associated with the XRD/XRF instrument and PlanetVac sampling system, no operations are conducted during the period of the landed mission when there is no DTE communication. Once DTE communication is restored for the final 24 days of the mission, the baseline plan includes XRD/XRF analysis of four additional samples; the selection of which PlanetVac(s) to use for these samples is informed by analysis results of the four initial samples.

**Exhibit 28.** Timeline of Landed Operations.

| INSTRUMENT | NMS | IMS | EPS | DD | GRS | XRD /XRF# | MAG | MAC | STAFF CAM | FOOT CAM | DESCENT CAM | TOTAL |
|---|---|---|---|---|---|---|---|---|---|---|---|---|
| **Surface Ops Description** | Cont* | Cont* | Cont* | Cont* | Cont | Specific Collection Times | Cont* | Cont* | Specific Collection Times | Daily | Descent Only | |
| **Total Data Volume Downlinked (MB)** | 228 | 1138 | 228 | 228 | 484 | 93 | 1422 | 4619** | 773 | 1138 | 469 | 10,820 |

**Exhibit 29**. Landed Operations & Associated Data Volume.
    * Continuous except for during XRD/XRF analyses
    # Includes XRD/XRF and PlanetVac data
    ** Only difference between data acquired and data downlinked; MAC acquires 7111 MB of data.

Additional details about the landed science operations and the resulting science data are given in Appendix B4.

Exhibit 29 gives the total data volume acquired for each instrument. Exhibit 30 plots the daily data volume acquired, the downlink available for science data, and the loading on the Lander's SSR, assuming a single 34-m Ka-band capability. The data acquired by all of the instruments are downlinked in full except for data-volume-intensive MAC observations. To record opportunistic seismic activity, MAC records data at a high sampling frequency. During periods with DTE communication, all high-frequency MAC data are downlinked. During the six weeks of no DTE communication, MAC high-frequency data are acquired but only 25% are downlinked. All low-frequency MAC data are downlinked following the completion of the six-week no-DTE-communication period. The science team analyses these low-frequency data in order to select the highest-priority 25% of the high-frequency data for downlinking, sufficient to accommodate one hour of data surrounding over ~250 quakes (an order of magnitude more events than expected given Mars-like activity, see Appendix B2.2.3, B4.2). In Exhibit 30, these high-frequency data produce the increase around days 75–85 in the "daily science data identified for downlink" line. The peak on that same line at day 59 in Exhibit 30 is from the dedicated StaffCam exosphere imaging campaign, and smaller weekly peaks are from the weekly StaffCam images. Overall, about 10.8 GB of science data (nominal data volume) are returned during landed surface operations, and are accommodated by the downlink from a single 34-m Ka-band antenna. The Lander's recorder is cleared of data identified for downlink a few days in advance of sunrise. Data volume contingency and margin are available by adjusting the science downlink plan. In particular, the fully downlinked low-frequency MAC data are used to prioritize the events for which high-frequency data will be downlinked, providing flexibility for data volume contingency scenarios. The addition of any 70-m antenna coverage or other additions would provide additional downlink and margin on the plan.

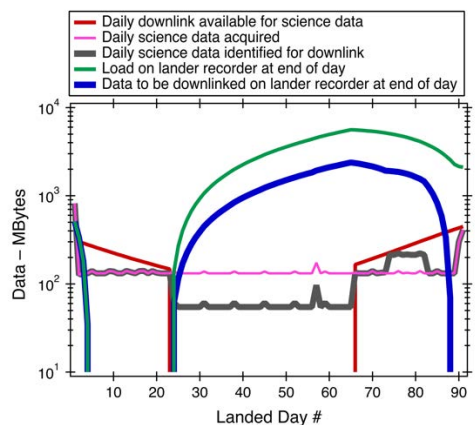

**Exhibit 30.** Daily landed science data volume and downlink plan, using the 34-m Ka-band capability detailed in the text.

### 3.3.2. MISSION OPERATIONS

The NASA DSN is used for communication with the spacecraft during cruise, orbital, and landed phases of the mission. During the ten-year cruise phase and three-month orbital phase, three eight-hour tracks per week using 34-meter antennas are planned for nominal operations. These tracks accommodate housekeeping and science data downlink, uplink of command loads and other necessary activities, and real-time evaluation of spacecraft health and status at regular intervals. Additionally, continuous coverage is scheduled one day before through one day after large maneuvers such as planetary flybys, SEP transitions, and orbit lowering maneuvers through landing. During the landed phase, continuous coverage using 34-meter antennas is required during the time in which there is a line of site from the Earth to the Lander location. The continuous coverage is necessary to downlink the science data volume. Exhibit 31 provides a data system summary.

| | MISSION PHASE: CRUISE | MISSION PHASE: ORBITAL | MISSION PHASE: LANDED |
|---|---|---|---|
| **DOWN LINK INFORMATION** | | | |
| Number of Contacts per Week | 3 | 7 | Continuous |
| Duration of Mission Phase, weeks | 512 | 13 | 8 |
| Downlink Frequency Band, GHz | 8.4 | 8.4 | 32 |
| Telemetry Data Rate(s), kbps | 0.030–0.5 | | 15–55 |
| Transmitting Antenna Type(s) and Gain(s), DBi | MGA, 9-16 dBi | MGA, 9-16 dBi | 0.6m RLSA, 43.8 dBi |
| Transmitter Peak Power, Watts | 33 | 33 | 33 |
| Downlink Receiving Antenna Gain, DBi | 68.2 | 68.2 | 77.5 |
| Transmitting Power Amplifier Output, Watts | 11 | 11 | 11 |
| Total Daily Data Volume, MB/day | 0.045–0.75 | 0.105–1.76 | 170–600 |
| **UPLINK INFORMATION** | | | |
| Number of Uplinks per Day | 3/week | 1 | 1 |
| Uplink Frequency Band, GHz | 7.1–7.2 | 7.1–7.2 | 7.1–7.2 |
| Telecommand Data Rate, kbps | 0.030–0.1 | 0.1–0.4 | 0.027–0.070 |
| Receiving Antenna Type(s) and Gain(s), DBi | MGA, 9-16 dBi | MGA, 9-16 dBi | LGA, -0.1 dBi |

**Exhibit 31.** Mission operations & ground data system summary.

## 3.4. Risk List

The top Mercury Lander mission risks are listed in Exhibit 32. They are classified as either technical (T) or cost/schedule (C/S), and the likelihood (L) and consequence (C) has been assessed for each.

| ID | RISK (TYPE, RATING ( L X C)) | DESCRIPTION / MITIGATION |
|---|---|---|
| A | IF the STAR 48GXV development and testing encounters setbacks, THEN the launch readiness date will be impacted. (C/S, 2 x 3) | • Development of the STAR 48GXV was begun for the Parker Solar Probe mission and included a successful static fire test.<br>• Qualification testing completion with two additional test units. |
| B | If the STAR 48GXV qualification for 10-year flight encounters setbacks, THEN the launch readiness date will be impacted. (C/S, 2 x 3) | • Qualification testing of fuel grain to assure long duration mission.<br>• Qualification testing completion with two additional test units. |
| C | IF analyses determine that a landing area with sufficiently low hazard density cannot be identified prior to descent, THEN additional propellant may be required to enable hazard avoidance over a larger area. (T, 3 x 3) | • Detailed modeling of the hazard environment is informed by MESSENGER and BepiColombo data as well as extrapolations from the lunar environment.<br>• Mercury Lander concept includes landing site imaging in orbital phase, improving surface image pixel scale to 1 m, which will be used for down selection of a safe landing zone. |
| D | IF integration and testing of the four-stage system encounters significant unanticipated complication, THEN additional resources will be required. (C/S, 2 x 3) | • Ample schedule margin is included. |
| E | IF the NextGen RTG does not meet expected performance, THEN an alternate power system design may be required. (T, 2 x 4) | • Lander battery capacity could be increased in conjunction with reduction of payload operating time to accommodate battery recharge.<br>• Descope to an MMRTG implementation may be evaluated. |

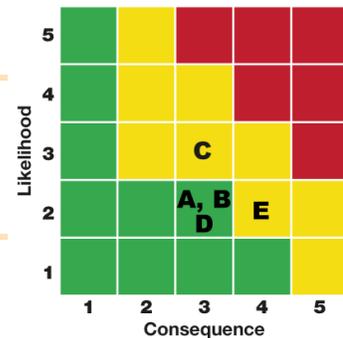

**Exhibit 32.** Top risks identified for the Mercury Lander mission.

# 4 DEVELOPMENT SCHEDULE & SCHEDULE CONSTRAINTS ___

## 4.1. High-Level Mission Schedule

The high-level mission schedule (Exhibit 33) is based on schedules developed for other NASA missions of similar complexity. Key phase durations planned are shown in Exhibit 34 and mission-level milestones are listed in Exhibit 35.

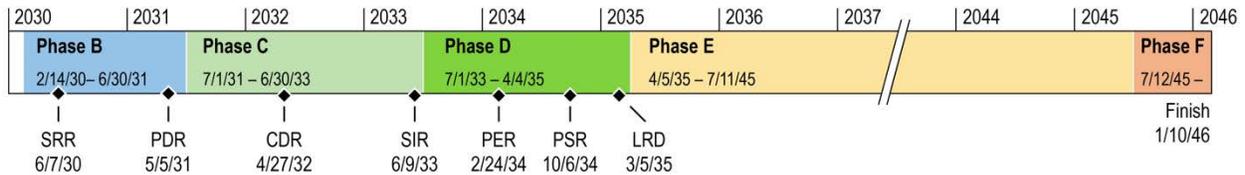

**Exhibit 33.** High-Level Mission Schedule.

| PROJECT PHASE | APPROXIMATE DURATION |
|---|---|
| Phase A – Conceptual Design | 11 mo. |
| Phase B – Preliminary Design | 16 mo. |
| Phase C – Detailed Design | 24 mo. |
| Phase D – Integration & Test | 21 mo. |
| Phase E – Primary Mission Operations | 123 mo. |
| Phase F – Extended Mission Operations | 6 mo. |
| Start of Phase B to PDR | 15 mo. |
| Start of Phase B to CDR | 26 mo. |
| Start of Phase B to Delivery of First Instrument | 37 mo. |
| Start of Phase B to Delivery of Last Instrument | 40 mo. |
| Start of Phase B to Delivery of First S/C Bus Component | 29 mo. |
| Start of Phase B to Delivery of Last S/C Bus Component | 50 mo. |
| System Level Integration & Test | 22 mo. |
| Project Total Funded Schedule Reserve | 8 mo. |
| Total Development Time Phase B–D | 61 mo. |

**Exhibit 34.** Key Mission Phase Durations.

| MISSION LEVEL MILESTONES | DATE |
|---|---|
| System Requirements Review (SRR) | 06/07/2030 |
| Preliminary Design Review (PDR) | 05/05/2031 |
| Integrated Baseline Review (IBR) | 12/09/2031 |
| Critical Design Review (CDR) | 04/27/2032 |
| Mission/Science Operations Review (MOR/SOR) | 08/20/2032 |
| System Integration Review (SIR) | 06/09/2033 |
| Operational Readiness Review (ORR) | 09/30/2033 |
| Pre-Environmental Review (PER) | 02/24/2034 |
| Pre-Ship Review (PSR) | 10/06/2034 |
| Mission Readiness Review (MRR) | 11/20/2034 |
| Safety & Mission Success Review (SMSR) | 01/02/2035 |
| Launch Readiness Review (LRR) | 02/23/2035 |
| Launch Readiness Date (LRD) | 03/05/2035 |

**Exhibit 35.** Mission-Level Milestones.

## 4.2. Technology Development Plan

With the exception of the NextGen RTG, which is being developed under NASA direction by the Radioisotope Power System Program, all technology included in the design is TRL-5 or above. The mission does not require technology development dollars to advance components to TRL 6 as all Mercury Lander mission components will be at or above TRL 6 when required.

## 4.3. Development Schedule and Constraints

The critical path for this mission development is expected to be the descent stage SRM. Launch is constrained to occur no earlier than 2030 per PMCS ground rules for use of the NextGen RTG. The primary launch period is in Mar 2035, and the trajectory is designed to include a backup launch period one year later in Mar 2036. There are 21 launch opportunities in each of the primary and backup periods.

# 5 MISSION LIFE-CYCLE COST

The cost estimate prepared for Mercury Lander is commensurate of a CML-4 mission concept. The payload and spacecraft estimates capture the resources required for a preferred point design and take into account subsystem level mass, power, and risk. The estimate also takes into account the technical and performance characteristics of components. Estimates for science, mission operations, and ground data system elements, whose costs are primarily determined by labor, take into account the Phase A–D schedule and Phase E timeline.

The result is a mission estimate that is comprehensive and representative of expenditures that might be expected if the Mercury Lander mission is executed as described. The Mercury Lander Phase A–F mission cost, including unencumbered reserves of 50% (A–D, excluding LV costs) and 25% (E–F, excluding DSN charges), is $1754.0M in fiscal year 2025 dollars (FY25$). Excluding LV costs, the Mercury Lander Phase A–F mission cost is $1508.0M FY25$ (Exhibit 36). The Phase A–D mission cost estimate (excluding the launch vehicle) is $1191.9M (FY25$), comparing favorably with past New Frontiers missions (Exhibit 37), as well as to the cost cap prescribed in the New Frontiers 4 AO (~$1.1B FY25$). **This cost estimate demonstrates that a Mercury Lander mission is feasible and compelling as a New Frontiers-class mission in the next decade.**

## 5.1. Mission Ground Rules & Assumptions

- Estimating ground rules and assumptions are derived from Revision 4 of the PMCS Ground Rules dated 22 Nov 2019.

- Mission costs are reported using the Level-2 (and Level-3 where appropriate) work breakdown structure (WBS) provided in NPR 7120.5E.

- Responsibility for the mission is spread throughout the NASA community. The cost estimate assumes that APL will lead the Mercury Lander mission and design, develop, manufacture, integrate, and test the four spacecraft stages. APL will also lead mission operations during Phase E. A number of organizations, including APL, will design, develop, and deliver instruments.

- Cost estimates are reported in Fiscal Year 2025 (FY25) dollars.

- The NASA New Start inflation index provided by the PMCS Ground Rules was used to adjust historical cost, price data, and parametric results to FY25 dollars when necessary.

- The mission does not require technology development dollars to advance components to TRL 6 as all Mercury Lander mission components will be at or above TRL 6 when required.

- This estimate assumes no development delays and an on-time launch in Mar 2035.

- Phase A–D cost reserves are calculated as 50% of the estimated costs of all components excluding the LV. Phase E–F cost reserves are calculated as 25% of the estimated costs of all Phase E elements excluding DSN charges.

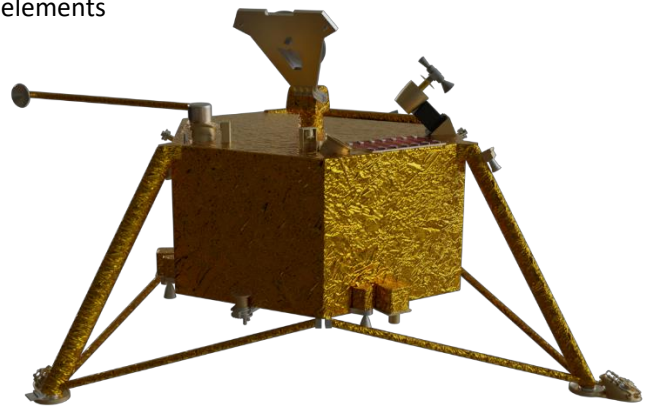

**Exhibit 36.** Estimated Phase A–F Mercury Lander mission costs by Level 3 WBS element (FY25$K).

| | DESCRIPTION | PHASE A–D | PHASE E–F | PRIMARY ESTIMATE TOTAL | PRIMARY ESTIMATE METHODOLOGY | VALIDATION ESTIMATE | REMARKS |
|---|---|---|---|---|---|---|---|
| 1,2,3 | PMSEMA | $102,865 | N/A | $102,865 | See 5.3 | N/A | E–F contained in WBS 7 |
| 4 | Science | $14,927 | $69,463 | $84,390 | See 5.3 | N/A | |
| 5 | Payload | $101,547 | - | $101,547 | N/A | $116,075 | |
| | Payload Management | $7,559 | - | $7,559 | Wrap Factor | $8,797 | 8.2% of hardware based on analysis of VAP, NH, MESSENGER and PSP payload suite data. |
| | GRS | $23,366 | - | $23,366 | ROM BUE, NICM | $25,653 | Primary estimate assumes that GRS will be able to leverage work performed for MMX-MEGANE, Dragonfly-DraGNS, and work done under a MatISSE grant |
| | XRD/XRF | $25,151 | - | $25,151 | MSL-CheMin, NICM | $29,806 | CheMin actuals adjusted for mass difference |
| | Magnetometer | $4,247 | - | $4,247 | MESSENGER-MAG, NICM | $5,247 | |
| | Accelerometer/SP Seismometer | $2,616 | - | $2,616 | InSight SEIS, NICM | $2,957 | Draws heritage from Insight-SEIS-SP but is of much lower complexity |
| | Neutral Mass Spectrometer | $16,879 | - | $16,879 | BepiColombo-STROFIO, NICM | $21,212 | |
| | Ion Mass Spectrometer | $5,959 | - | $5,959 | MESSENGER-FIPS, NICM | $5,231 | |
| | Energetic Particle Detector | $7,999 | - | $7,999 | NH-PEPSSI, NICM | $9,171 | |
| | Dust Detector | $5,132 | - | $5,132 | NH-SDC, NICM | $4,422 | |
| | StaffCam | $266 | - | $266 | Vendor Quote, NICM | $306 | 1 camera. Malin Space Science Systems (MSSS) vendor quote for commercial-off-the-shelf (COTS) ECAM system cameras (N50) |
| | FootCam | $201 | - | $201 | Vendor Quote, NICM | $731 | 2 cameras. MSSS vendor quote for COTS ECAM system cameras (C50). Vendor quote for C50s is lower than N50s. SEER-Space does not differentiate between monochrome and color. |
| | DescentCam | $370 | - | $370 | Vendor Quote, NICM | $429 | 2 cameras. MSSS vendor quote for COTS ECAM system cameras (N50) |
| | DVR8 | $1,802 | - | $1,802 | Vendor Quote, NICM | $2,113 | Shared by all 5 MSSS ECAM Cameras |
| 6 | Spacecraft | $472,500 | - | $472,500 | N/A | $593,014 | |
| | Cruise Stage | $120,353 | - | $120,353 | | $147,180 | |
| | Orbital Stage | $124,528 | - | $124,528 | Vendor Quotes, SEER-H, BUE | $155,977 | $3.1M Orbital Camera |
| | Descent Stage | $41,384 | - | $41,384 | | $56,315 | |
| | Lander | $186,234 | - | $186,234 | | $233,542 | $17.6M PlanetVac, RTGs included as a pass-thru cost to cross-check estimate. $16.3M LIDAR |
| 7 | Mission Operations | $14,920 | $175,580 | $190,501 | See 5.3 | N/A | $19.0M DSN charges |
| 8 | Launch Vehicle & Services | $246,000 | - | $246,000 | See 5.3 | N/A | EFH   + $26.0M RTG |
| 9 | Ground Data Systems | $14,918 | $11,661 | $26,578 | See 5.3 | N/A | |
| 10 | System Integration & Test | $72,904 | - | $72,904 | See 5.3 | N/A | |
| | Subtotal w/ LV | $1,040,582 | $256,704 | $1,297,286 | | | |
| | Subtotal w/o LV | $794,582 | $256,704 | $1,051,286 | | | |
| | Unencumbered Reserves | $397,291 | $59,426 | $456,717 | A–D: 50%, E–F: 25% | | Prescribed by Decadal Ground Rules |
| | Total w/ LV | $1,437,873 | $316,130 | $1,754,003 | | | |
| | Total w/o LV | $1,191,873 | $316,130 | $1,508,003 | | | Comparison point to NF4 Phase A–D cost cap ($1.1B FY25$) |

## 5.2. Cost Benchmarking

The cost and scope of the Mercury Lander concept correspond well to a NASA New-Frontiers-class mission (Exhibit 37). The Phase A–D mission cost estimate, excluding LV, is $1191.9M FY25$. The cost cap prescribed in the New Frontiers 4 AO is $1079.9M FY25$. The team is confident Mercury Lander can be developed below this cost cap with further refinement of the design beyond this CML-4 study.

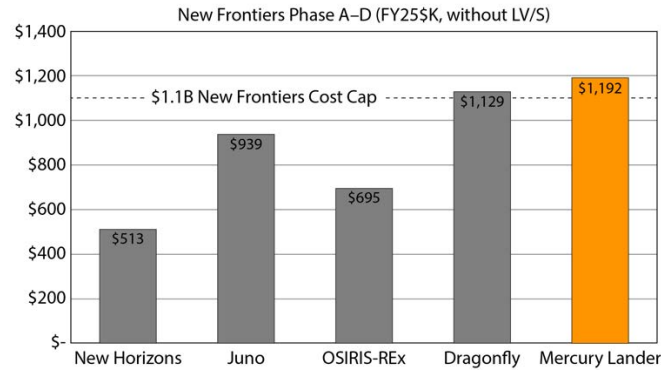

Exhibit 37. Mission-Level Cost Comparison to New Frontiers Missions.

## 5.3. Costing Methodology & Basis of Estimate

The Mercury Lander CML-4 mission cost estimate is a combination of high level parametrics, bottom-up, and analog techniques, and incorporates a wide range of uncertainty in the estimating process. No adjustments were made to remove the historical cost of manifested risk from the heritage data underlying the baseline estimate. Therefore, before reserves are applied, the estimated costs already include a historical average of the cost of risk. This approach is appropriate for capturing risk and uncertainty commensurate with the early stages of a mission. The following paragraphs describe the basis of estimate (BOE) for each element.

**WBS 1, 2, 3 Project Management, Systems Engineering, Mission Assurance (PMSEMA).** Because these functions depend on multiple analysis- and organization-specific characteristics [Hahn 2014], cost analogies to historical missions are preferred over cost model output, which do not take mission characteristics into account. Existing analyses show that hardware costs are a reliable predictor of these critical mission function costs. APL has conducted thorough and rigorous analyses of PMSEMA costs, both for historical APL missions and for analogous missions. The BOE for the Mercury Lander relies on APL's analysis of historical PM, SE, and MA practices on VAP, PSP, and NH. In particular, VAP and PSP are APL's most recent missions that were managed under current NASA requirements (e.g., Earned Value Management System (EVMS, 7120.5E, and 7123) and were delivered on schedule and within budget. Mercury Lander has comparable requirements. The mission PMSEMA cost is 15.9% of the flight system (payload + spacecraft + I&T). This percentage is allowed to vary with hardware costs as part of the mission cost risk analysis, discussed below, to capture uncertainty (e.g., CML-4 level design phase).

**WBS 4 Science.** This element covers the managing, directing, and controlling of the science investigation. It includes the costs of the principal investigator, deputy PI, project scientist, and science team members. This element is largely level of effort. For the Mercury Lander, it is estimated via a rough order of magnitude (ROM) bottom-up estimate (BUE). The cost per year during Phases B–D of $2.92M FY25$ is comparable to MESSENGER, which expended $2.29M FY25$ per year. Similar to Mercury Lander, MESSENGER also carried a large suite of instruments on its payload, resulting in a large and diverse science team. Average costs per year during operations ($6.46M FY25$) is higher than that of MESSENGER ($3.9M FY25$) to reflect an increased science effort for landing and operations at Mercury.

**WBS 5 Payload.** This element includes the lander-hosted instruments (Exhibit 36). All instrument costs underwent an iterative effort among cost, science, and engineering to ensure an estimate that adequately captures the true effort required to develop these instruments. This exercise involved the analysis of analogous costs where appropriate, parametric modeling, and engineering judgement. Additionally, SEER-Space was used as a parametric crosscheck. At the payload level, the project estimate is 15% lower than the sum of the average parametric crosscheck. The detailed BOEs and associated validation estimates can be found in Exhibit 36.

**WBS 6 Spacecraft.** This element includes the cruise stage, orbital stage (including the orbital camera), descent stage, and Lander (including the LIDAR). The BOE relies primarily on a SEER-H parametric estimate. SEER-H was selected as the primary estimating methodology due to the more refined design of the Mercury Lander. The level of detail and design captured in the master equipment list (MEL) allows for specific tailoring of subsystem component technologies and applications. The resulting estimate includes design, fabrication, and subsystem-level testing of all hardware components. The exceptions to this are the propulsion subsystems and SRM, which are estimated via a ROM. Additional details on these BOEs and bookkeeping of electric power subsystem recurring and non-recurring costs are described below. All hardware development costs include supporting engineering models, breadboards, flight parts, ground support equipment, and flight spares. This WBS does not include spacecraft PMSEMA, which is bookkept in WBS 1, 2, 3 consistent with in-house APL mission spacecraft builds. Additionally, SEER-Space was used as a parametric crosscheck. At the spacecraft level, the project estimate is 26% lower than the sum of the average parametric crosscheck (Exhibit 36).

**Propulsion.** Across all four spacecraft stages, there are two bipropellant systems and one electric propulsion system, all with custom tankage. The cost of the cruise stage electric propulsion is based on APL's experience with the development of the SEP on DART. It includes PPU/DCIU development, a custom tank, and expansion on the efforts expended on DART. The two biprop systems are on the orbital stage and the Lander and are estimated via a ROM. This ROM assumes cost savings on the Lander biprop system by including engine/component testing (non-recurring engineering (NRE)) in the orbital stage biprop costs.

**Solid Rocket Motor.** This estimate is a ROM that assumes the Mercury Lander project will resume PSP's development effort. The SRM will need additional development work for space qualification, including associated NRE and ground testing. This testing plan includes two static-fire tests and associated post-fire investigation. The first successful static fire test was performed as part of an early upper stage effort on PSP. The total spent on NRE on the PSP SRM effort before it was discontinued was $26.42M FY25$. It is estimated that the Mercury Lander SRM development effort will take ~three years and $35M FY25$. This includes the cost for two test units and the flight item.

**Power.** The Mercury Lander spacecraft includes shunt regulators on the cruise stage, orbital stage, and Lander. The NRE cost of the shunt regulators is carried in the cruise stage electrical power system costs. The cruise stage shunt regulator is estimated at $21.3M FY25$, which compares favorably to the cost of the shunt regulators developed for VAP. The Mercury Lander spacecraft also includes power switching units (PSUs) on both the orbital stage and Lander. The NRE cost of the PSUs is carried in the orbital stage. The cost of two PSUs on the orbital stage is $27.4M FY25$, which compares favorably to the cost spent on VAP PSUs.

**WBS 7 & 9 Mission Operations (MOps) & Ground Data System (GDS).** The Mercury Lander mission operations estimate includes mission operations planning and development, network security, data processing, and mission management. The pre-launch mission operations estimate is an analogous estimate based off of previous APL efforts on NH, MESSENGER, and PSP. These missions represent typical APL expenditure on pre-launch MOps for projects of comparable scope and complexity. The post-launch mission operations estimate is derived from APL historical costs per month during different operational phases. The GDS estimate is a BUE.

**WBS 8 LV and Services**. The mission requires a LV that does not correspond with any of the options currently described in the Decadal Survey Ground Rules. As such, the figures used in this estimate are based on an evaluation of current best estimates of the cost of the capability that will be required. The price for a LV with expendable Falcon Heavy-type capabilities, based on past pricing to NASA missions of evolved expendable launch vehicles, would be at least $210M for a launch using a standard sized fairing. An additional $10M is included for potential modifications. This does not include National Environmental Policy Act and nuclear launch approval costs which are covered by the cost of the RTGs in WBS 6.

**WBS 10 System Integration & Test (I&T).** This element covers the efforts to assemble the cruise stage, orbital stage, descent stage, and Lander; integrate the four stages into the mission spacecraft; testbeds; and performance of spacecraft environmental testing. The costs are based on a detailed analysis of cost actuals from previous APL missions, including MESSENGER, NH, STEREO, VAP, and PSP. The Mercury Lander I&T effort is estimated at 12.7% of the hardware. Given the use of cost-to-cost factors to estimate I&T, both the cost estimating relationship (CER) and the underlying cost drivers are allowed to vary so that all sources of uncertainty can be quantified. As hardware cost varies, the cost-to-cost factors I&T estimate also varies. This approach allows the estimate to maintain a conservative risk posture given the historical complexity of I&T.

**DSN Charges.** This element provides for access to the DSN infrastructure needed to transmit and receive mission and scientific data. Mission charges are estimated using the Jet Propulsion Laboratory (JPL) DSN Aperture Fee tool. The DSN cost estimate covers pre- and post-contact activity for each linkage.

## 5.4. Confidence & Cost Reserves

Exhibit 38 shows the cost risk ranges used for the probabilistic cost risk analysis.

**PMSEMA.** Given the use of cost-to-cost factors to estimate these functions, both the CER and underlying cost drivers are allowed to range to quantify all sources of uncertainty.

**Science, GDS, & MOps.** These are low-risk cost elements, but subject to cost growth as part of the cost risk analysis.

**Payload.** Given that the point estimate informed by a combination of NICM and historical analogies for each of the 11 instruments, the highest value of the historical analogy, NICM, or the SEER-Space cross-check is used to inform the Mercury Lander payload risk model to capture the uncertainty given the CML-4 level design phase.

**Spacecraft.** Each subsystem is subject to a data-driven risk analysis based on historical APL cost growth. Mass input also varies in the SEER space model consistent with early design programs to 30% over CBE.

**I&T.** I&T as a percentage of the payload and spacecraft from the point estimate is used to inform the risk analysis, allowing I&T to vary with hardware cost.

Per the PMCS Ground Rules, the estimate includes unencumbered cost reserves of 50% of all Phase A–D cost elements (except for the LV) plus 25% of the Phase E–F cost elements (excluding DSN charges). A probabilistic cost risk analysis shows 75% confidence that the Phase A–F mission cost is achievable (Exhibit 39). The high confidence level is driven primarily by the robustness of the required reserves posture for this pre-proposal concept. Given a typical competitive pre-Phase-A NASA environment with 25% reserves on Phase A–D and 10% reserves on Phase E–F, the probabilistic cost risk analysis shows 63% confidence that the Phase A–F mission would be achievable with a less robust reserves posture. A 50th to 70th percentile confidence level is expected and reasonable for a pre-Phase-A concept with this level of reserves.

A coefficient of variation (standard deviation/mean) of approximately 41% indicates appropriate levels of conservatism given the early formulation phase. The model confirms the point estimate and provides a reasonable basis for the Mercury Lander CML-4 study.

| WBS | COST ELEMENT | PROJECT ESTIMATE | 70TH PERCENTILE |
|---|---|---|---|
| 1,2,3 | Mission PMSEMA | $102,865 | $142,882 |
| 4 | Science | $84,390 | $105,488 |
| 5 | Payload | $101,547 | $152,701 |
| 6 | Spacecraft | $472,500 | $644,662 |
| 7 | Mission Ops | $190,501 | $238,126 |
| 9 | GDS | $26,578 | $33,223 |
| 10 | I&T | $72,904 | $101,265 |

**Exhibit 38.** Cost Risk Ranges ($FY25 in thousands).

| DESCRIPTION | VALUE (FY$25K) | CONFIDENCE LEVEL |
|---|---|---|
| Point Estimate | $1,297,286 | 46% |
| Mean | $1,483,719 | |
| Standard Deviation | $601,276 | |
| Cost Reserves | $456,717 | |
| PI-Managed Mission Cost | $1,754,003 | 75% |

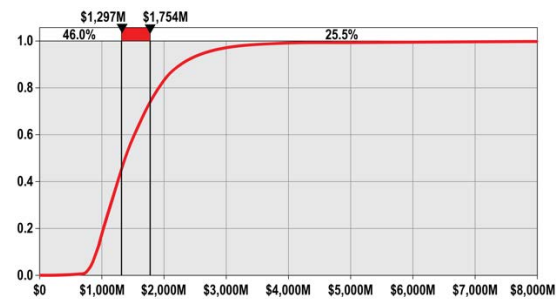

**Exhibit 39.** Cost Risk Analysis & S-Curve Summary ($FY25).

## 5.5. Cost Validation

The cost estimating process for Phases A–F provides a credible basis for generating an accurate forecast of costs associated with Mercury Lander. All elements of the cost estimate are validated using the SEER-Space parametric modeling tool. The validation process has two objectives: (1) assess and confirm the primary estimate and (2) provide project managers, systems engineers, and leads with checks on their costs and insight into the effects of design, schedule, and operational changes on baseline costs. Validation estimates and comparisons to the primary estimate for Mercury Lander are shown in Exhibit 36.

# APPENDIX A: ACRONYMS AND ABBREVIATIONS

| | |
|---|---|
| AMU | Atomic Mass Unit |
| AO | Announcement of Opportunity |
| AU | Astronomical Unit |
| APL | Johns Hopkins Applied Physics Laboratory |
| APXS | Alpha Particle X-Ray Spectrometer |
| BELA | BepiColombo Laser Altimeter |
| BOE | Basis of Estimate |
| BOL | Beginning of Life |
| BUE | Bottom-Up Estimate |
| C&DH | Command and Data Handling |
| CBE | Current Best Estimate |
| CCtCap | Commercial Crew Transportation Capability |
| CDR | Critical Design Review |
| CER | Cost Estimating Relationship |
| cFE | Core Flight Executive |
| CheMin | Chemistry and Mineralogy |
| CG | Center of Gravity |
| CLPS | Commercial Lunar Payload Services |
| CML | Concept Maturity Level |
| CMOS | Complementary Metal Oxide Semiconductor |
| COTS | Commercial Off the Shelf |
| DART | Double Asteroid Redirection Test |
| DCIU | Digital Control and Interface Unit |
| DCDC | DC/DC Converter Card |
| DD | Dust Detector |
| DMA | Direct Memory Access |
| DraGNS | Dragonfly Gamma-Ray and Neutron Spectrometer |
| DSN | Deep Space Network |
| DSS | Digital Sun Sensor |
| DTE | Direct to Earth |

| DTM | Digital Terrain Model |
| DVR | Digital Video Recorder |
| EFH | Expendable Falcon Heavy |
| EOL | End of Life |
| EPS | Energetic Particle Spectrometer |
| ESA | European Space Agency |
| EVMS | Earned Value Management System |
| FIPS | Fast Imaging Plasma Spectrometer |
| FOV | Field of View |
| FPGA | Field-Programmable Gate Array |
| FSW | Flight Software |
| FY | Fiscal Year |
| G&C | Guidance and Control |
| GDS | Ground Data System |
| GHe | Gaseous Helium |
| GRS | Gamma-Ray Spectrometer |
| HGA | High-Gain Antenna |
| I&T | Integration and Test |
| IBR | Integrated Baseline Review |
| IEM | Integrated Electronic Module |
| IMS | Ion Mass Spectrometer |
| IMU | Inertial Measurement Unit |
| IR | Infrared |
| ISAD | Icy Soil Acquisition Device |
| JAXA | Japanese Aerospace Exploration Agency |
| JPL | Jet Propulsion Laboratory |
| JSC | Johnson Space Center |
| KarLE | Potassium (K) Argon (Ar) Laser Experiment |
| LED | Light Emitting Diode |
| LGA | Low-Gain Antenna |
| LIBS | Laser Induced Breakdown Spectrometers |
| LIDAR | Light Detection and Ranging |
| LRD | Launch Readiness Date |
| LRM | Low-Reflectance Material |
| LROC | Lunar Reconnaissance Orbiter Camera |
| LRR | Launch Readiness Review |
| LV | Launch Vehicle |
| MA | Mission Assurance |
| MAC | Mercury Accelerometer |
| MAG | Magnetometer |
| MatISSE | Maturation of Instruments for Solar System Exploration |
| MEL | Master Equipment List |
| MEOP | Maximum Expected Operating Pressure |
| MER | Mars Exploration Rovers |
| MERTIS | MErcury Radiometer and Thermal Imaging Spectrometer |
| MESSENGER | MErcury Surface, Space ENvironment, GEochemistry, and Ranging |
| MEV | Maximum Expected Value |
| MGA | Medium-Gain Antenna |
| MGNS | Mercury Gamma-Ray and Neutron Spectrometer |
| MIV | Micrometeoroid Impact Vaporization |
| MIXS | Mercury Imaging X-ray Spectrometer |
| MLI | Multi-Layer Insulation |
| MMH | Monomethyl Hydrazine |
| MMX | Martian Moons Exploration |
| MOI | Mercury Orbit Insertion |
| MON-3 | Mixed Oxides of Nitrogen (Nitrogen Tetroxide) |
| MOps | Mission Operations |
| MOR | Mission Operations Review |
| MPO | Mercury Planetary Orbiter |
| MRAM | Magnetoresistive Random-Access Memory |
| MRR | Mission Readiness Review |
| MSL | Mars Science Laboratory |
| MSSS | Malin Space Science Systems |
| MTA | Mercury True Anomaly |
| NASA | National Aeronautics and Space Administration |
| NCP | Navigation Co-Processor |
| NEAR | Near Earth Asteroid Rendezvous |
| NEXT-C | NASA Evolutionary Xenon Thruster - Commercial |
| NH | New Horizons |
| NICM | NASA Instrument Cost Model |
| NIR | Near Infrared |
| NMS | Neutral Mass Spectrometer |
| NPR | NASA Procedural Requirement |
| NRE | Non-Recurring Engineering |
| NSTAR | NASA Solar Technology Application Readiness |
| OMAC | Orbital Maneuvering and Attitude Control |
| ORR | Operational Readiness Review |
| OSIRIS-REx | Origins, Spectral Interpretation, Resource Identification, Security-Regolith EXplorer |
| PDR | Preliminary Design Review |
| PEPSSI | Pluto Energetic Particle Spectrometer Science Investigation |
| PER | Pre-Environmental Review |
| PM | Project Management |
| PMCS | Planetary Mission Concept Studies |
| PMSEMA | Project Management, Systems Engineering, Mission Assurance |
| PPU | Power Processing Unit |
| PSD | Photon-Stimulated Desorption |
| PSP | Parker Solar Probe |
| PSR | Pre-Ship Review |
| PSU | Power Switching Unit |
| RCC | Redundancy Controller Card |
| RCS | Reaction Control System |
| RISE | Rotation and Interior Structure Experiment |
| RIU | Remote Interface Unit |
| RLSA | Radial Line Slot Array |
| ROM | Rough Order of Magnitude |
| RS | Radio Science |
| RTG | Radioisotope Thermoelectric Generator |
| RW | Reaction Wheel |
| SADA | Solar Array Drive Assembly |
| SBC | Single Board Computer |
| SCIF | Spacecraft Interface |

| SDC | Student Dust Counter |
|---|---|
| SE | Systems Engineering |
| SEER | System Evaluation and Estimation of Resources |
| SEIS-SP | Seismic Experiment for Internal Structure-Short Period |
| SEP | Solar Electric Propulsion |
| SHERLOC | Scanning Habitable Environments with Raman and Luminescence for Organics and Chemicals |
| SIMBIO-SYS | Spectrometer and Imagers for MPO Bepicolombo Integrated Observatory System |
| SIR | System Integrations Review |
| SMSR | Safety and Mission Success Review |
| SOR | Science Operations Review |
| SPS | Samples Per Second |
| SRAM | Static Random-Access Memory |
| SRM | Solid Rocket Motor |
| SRR | System Requirements Review |
| SSE | Sun Sensor Electronics |
| SSR | Solid-State Recorder |
| STEREO | Solar TErrestrial RElations Observatory |
| STROFIO | STart from a ROtating Field mass spectrOmeter |
| TAC | Thruster/Actuator Card |
| TCM | Trajectory Correction Maneuver |
| TES | Thermal Emission Spectrometer |
| TRL | Technology Readiness Level |
| TWTA | Travelling Wave Tube Amplifier |
| TVC | Thrust Vector Control |
| UV | Ultraviolet |
| VAP | Van Allen Probes |
| VIS | Visible |
| WBS | Work Breakdown Structure |
| XRD | X-Ray Diffractometer |
| XRF | X-Ray Fluorescence |
| XRS | X-Ray Spectrometer |

# APPENDIX B: DESIGN TEAM STUDY REPORT

This appendix provides additional details on the work done by the mission concept study team. In particular, the extensive scientific discussions leading to the science, payload, landing site, and operations selections are captured here.

## B1. Mercury Lander: Science Goals and Objectives

To guide the Mercury Lander mission concept study, four overarching and fundamental science goals were identified, as detailed in the main technical report. Each goal addresses high-priority themes and goals of NASA's 2014 Science Plan and the 2011 Planetary Science Decadal Survey (Exhibit B1). The following sections provide the detailed scientific justification motivating those four science goals. In the sections below, we title each investigation in shorthand terms that correspond to the broad topics they encompass. Goal 1 is "Geochemistry," Goal 2 is "Geophysics," Goal 3 is "Space Environment," and Goal 4 is "Geology." We emphasize, however, that many aspects of our science goals, and the instruments we describe to address those goals, are cross-cutting and these terms are therefore neither exhaustive nor exclusive to each goal.

| DECADAL SURVEY (2011) | NASA SCIENCE PLAN: PSD GOALS (2014) | MERCURY LANDER SCIENCE GOALS |
|---|---|---|
| **Theme 1: Building New Worlds**<br>"What were the initial stages, conditions and processes of solar system formation? What governed the accretion, supply of water, chemistry, and internal differentiation of the inner planets?" | Explore and observe the objects in the solar system to understand how they formed and evolve<br><br>Advance the understanding of how the chemical and physical processes in our solar system operate, interact and evolve | **Goal 1 (Geochemistry):** Investigate the highly chemically reduced, unexpectedly volatile-rich mineralogy and chemistry of Mercury's surface, to understand the earliest evolution of this end-member of rocky planet formation. |
| | | **Goal 2 (Geophysics):** Investigate Mercury's interior structure and magnetic field, to unravel the planet's differentiation and evolutionary history and to understand the magnetic field at the surface. |
| **Theme 3: Workings of Solar Systems**<br>"How have the myriad chemical and physical processes that shaped the solar system operated, interacted, and evolved over time?" | | **Goal 3 (Space Environment):** Investigate the active processes that produce Mercury's exosphere and alter its regolith, to understand planetary processes on rocky airless bodies, including the Moon. |
| | | **Goal 4 (Geology):** Characterize the landing site, to understand the processes that have shaped its evolution, to place the in situ measurements in context, and to enable ground truth for global interpretations of Mercury. |

**Exhibit B1.** The Mercury Lander mission concept science goals address NASA's strategic objectives in Planetary Science.

### B1.1. Goal 1: Geochemistry

Pre-MESSENGER hypotheses for Mercury's origin and extremely large core predicted a variety of silicate compositions for the planet's present make-up, and MESSENGER's compositional measurements were planned to distinguish definitively these competing ideas [Solomon et al. 2001]. MESSENGER orbital measurements from the Gamma-Ray, Neutron, and X-ray spectrometers (GRS, NS, XRS), however, indicated that Mercury's surface is enriched in moderately volatile elements such as K and Na, has high S contents (up to 4 wt%) and low Fe contents (less than 1–2 wt%), and has a surface rich in C (up to 5 wt%) [Evans et al. 2015; Klima et al. 2018; Nittler et al. 2011; 2018; Peplowski et al. 2011; 2012; 2016; Weider et al. 2015; 2016]. These measurements revealed that Mercury's surface did not match predictions from previously proposed hypotheses, which include a giant impact, evaporation models, and direct formation from high-temperature nebular condensates (e.g., Ebel & Stewart [2018]). These surprising results have led to a complete reexamination of the planet's origin and history. Mercury's unique geochemical signatures revealed by MESSENGER are indicative of highly reduced conditions during planetary formation and differentiation [McCubbin et al. 2012; Zolotov et al. 2013; Namur et al. 2016]. The surface composition of Mercury is modeled as Mg-rich silicates (e.g., forsterite, enstatite), oxides, exotic sulfides (e.g., niningerite, oldhamite), and metals [Vander Kaaden & McCubbin 2016; Namur & Charlier 2017]. However, due to the lack of spectral features, in particular in the UV/VIS, MESSENGER instruments were unable to make direct measurements of Mercury's surface mineralogy.

Nevertheless, the measured elemental chemistry and highly reduced conditions have led to new hypotheses for the formation of Mercury that differ from those of all other bodies in our solar system (e.g., Ebel and Stewart [2018]). In particular, the high C content on the surface has been proposed to reflect a primary graphite flotation crust (Exhibit 1) [Vander Kaaden & McCubbin 2015]. Remnants of this exotic graphite flotation crust would represent the earliest solid crustal materials on Mercury, providing a window into the planet's earliest differentiation. After the magma ocean solidified, volcanic eruptions resurfaced the majority of the planet, covering the graphite crust (e.g., Denevi et al. [2018b]). Impacts have since excavated and mixed the graphite with the surface material (e.g., Rivera-Valentin & Barr [2014]). Remnants of the graphite crust are inferred to be concentrated in the LRM exposures, distributed across the surface of Mercury. As materials erupted through this C-rich layer, the melts are hypothesized to have been stripped of their oxygen, producing CO that was lost to space (e.g., through pyroclastic vents [Kerber et al. 2009; Weider et al. 2016]), and resulting in smelting reactions leaving highly reduced metals (e.g., Si-rich metals) on the surface [McCubbin et al. 2017].

**Mercury's surface mineralogy is thus hypothesized to be unlike that of any other solar system terrestrial body, making Mercury the most highly reduced geochemical end-member of the terrestrial planets, and suggesting a unique environment for planetary differentiation and subsequent evolution.** Although MESSENGER's compositional data acquired from orbit challenged previous models of the planet's formation and evolution, only in situ geochemical measurements will enable us to test new hypotheses. BepiColombo is positioned to add to our geochemical knowledge, in particular by improving coverage of elemental compositional measurements over the southern hemisphere and better characterizing silicate mineralogy using orbital thermal infrared imaging spectroscopy. Yet, **direct in situ elemental and mineralogical measurements on Mercury's surface are essential to address the new science questions that have arisen since MESSENGER.**

One crucial measurement is of the major and minor elemental compositions of the LRM at a spatial scale and sensitivity far superior to orbital measurements that were taken by MESSENGER or will be acquired by BepiColombo. In particular, quantifying the LRM's C content, volatile element abundances (e.g., Na, K, S), and minor elements that are not well-resolved from orbit (e.g., Cl, Cr, and Mn) will enable current hypotheses to be tested and provide key constraints to advance petrologic modeling (e.g., Stockstill-Cahill et al. [2012]; Vander Kaaden et al. [2017]) and laboratory experimental studies (e.g., Charlier et al. [2013]; Namur et al. [2016]; Vander Kaaden & McCubbin [2016]). Furthermore, this geochemical information will be instrumental to help answer critical questions about Mercury and our solar system including: What is the composition of the low-reflectance material on Mercury and what does this tell us about the primary processes taking place on the planet? What role does C play in controlling the development of space weathering features on airless surfaces? What do the volatile abundances of Mercury tell us about volatile distribution of the inner Solar System? What can we learn about the composition of the Sun from Mercury's regolith? What can the composition of Mercury tell us about exoplanet formation? Such elemental measurements from the surface of Mercury could also be related directly to MESSENGER and BepiColombo orbital measurements, placing the landed measurements in a global context.

However, the most critical data to be obtained by a Mercury lander from a geochemical standpoint are the mineralogical hosts of the measured elements. Understanding the mineralogy of Mercury's exotic surface materials opens a window into the thermochemical evolution of the planet that does not currently exist. Characterizing Mercury's mineralogy and quantifying the phases present at the 1 wt% level are necessary to interpret the petrologic history, oxidation states, and early processes the planet experienced. Understanding the mineralogical host(s) of Mercury's surprisingly high surface-S content will provide key insights into the planet's differentiation and evolutionary history, and help to constrain the phase that, upon removal, forms Mercury's mysterious hollows, which are closely associated spatially with the LRM (e.g., Blewett et al. [2011]; Thomas et al. [2014]).

Mineralogical measurements acquired from the surface of Mercury will revolutionize our view of the planet, enable the next step in understanding its formation, and advance our understanding of planetary evolution under highly reducing conditions more broadly. Information regarding the mineralogical hosts associated with this landing site are vital to answer outstanding questions about Mercury including: Is the LRM the planet's primary crust? If so, how does the composition of Mercury's primary crust compare with the primary crusts of other planetary bodies? What is the mineralogy associated with the planet's oldest material? Given the mineralogy, what is the oxidation state of the elements on the surface and what insight does this provide into the geochemical evolution of Mercury? Based on the oxidation state of the elements on the surface, did smelting events occur in Mercury's history? Does the mineralogy rectify the oxygen deficit? How can the data from MESSENGER and BepiColombo be refined with new ground-truth data?

### B1.2. Goal 2: Geophysics

Mercury's high bulk density is a critical indicator of the planet's origin and subsequent evolution (e.g., Siegfried & Solomon [1974]; Schubert et al. [1988]; Benz et al. [2007]; Brown & Elkins-Tanton [2009]; Ebel & Stewart [2018]; Hauck et al. [2018]; Margot et al. [2018]). Accurate determination of the interior of Mercury is essential for characterizing the bulk composition of the planet—because each major layer (e.g., crust, mantle, and liquid and solid portions of the core) has a different composition [*Nittler et al.* 2018]—as well as for understanding the conditions of its long-term evolution. **The internal configuration of Mercury is an indicator of how the planet formed and differentiated, and that same structure sets the boundary conditions for how Mercury has evolved.** MESSENGER confirmed the existence of a liquid portion of the metallic core and substantially improved our knowledge of the layering of the interior (e.g., Smith et al. [2012]; Hauck et al. [2013]; Rivoldini & Van Hoolst [2013]; Margot et al. [2018]; Genova et al. [2019]) (Exhibit B2). **However, greater accuracy in determining the thicknesses and densities of these layers, including the solid inner core, is critical for understanding the history of magnetic field generation and global contraction** (e.g., Siegfried & Solomon [1974]; Schubert et al. [1988]; Hauck et al. [2004; 2018]; Christensen [2006]; Tosi et al. [2013]; Cao et al. [2014]; Tian et al. [2015]; Johnson et al. [2018]).

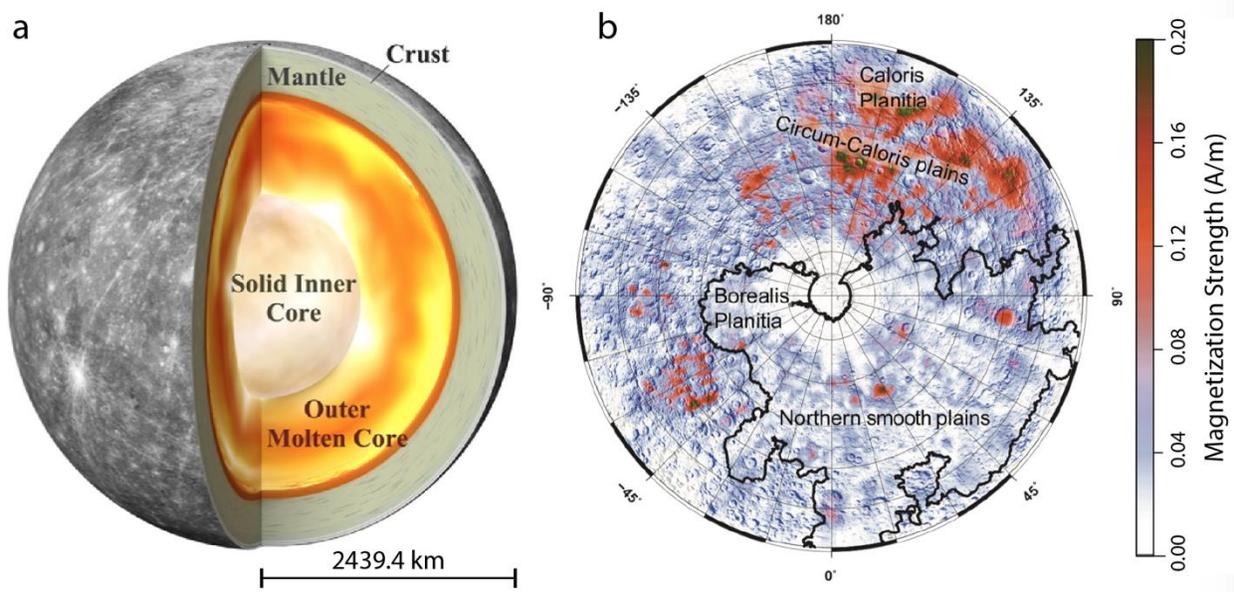

**Exhibit B2.** (a) Mercury's interior structure (from Genova et al. [2019]). (b) Crustal magnetization strength for Mercury's northern hemisphere assuming a 10-km thick magnetized layer (from 30ºN to the pole). The magnetization comprises ancient crustal fields and magnetization induced by the present field (after Hauck & Johnson [2019]). In situ measurements will determine interior structure and directly measure magnetic fields at the surface.

Mercury's rotational dynamics (e.g., libration, obliquity) are sensitive to the interior structure, as well as internal couplings and external forcings, e.g., from internal gravitational coupling among component layers versus perturbations from Jupiter (e.g., Peale [2005]; Margot et al. [2012; 2018]; Stark et al. [2015]). Documenting these internal and external influences is critical for ascertaining Mercury's internal structure, especially the properties of the core (e.g., Dumberry [2011]; Veasy & Dumberry [2011]; Van Hoolst et al. [2012]; Dumberry et al. [2013]; Koning & Dumberry [2013]; Genova et al. [2019]). Direct-to-Earth radio tracking from a stationary lander position over time provides more accurate constraints on planetary rotation of the surface than can be derived from orbital data, by avoiding ambiguities due to spacecraft motion, orbit errors, and aliasing. Currently, analysis based on either gravity [Genova et al. 2019] or altimetry (e.g., Stark et al. [2015]) and Earth-based radar data (e.g., Margot et al. [2012]) yield statistically distinct results for Mercury's average spin-rate. Landed measurements will resolve this discrepancy because they will be a fully independent dataset and approach, which also provides a greater number of precise measurements over a significant portion of a rotation period than Earth-based or orbital data can provide. Further, these data will also provide an extended baseline of observations extending in time from MESSENGER and BepiColombo that is crucial to accurately determine the long-period effects on physical librations, including those forced by Jupiter on timescales of its orbital period around the Sun. Such knowledge is necessary to separate components of the rotational state to determine the internal structure.

The internal evolution of a planet is driven by how heat is generated, transferred from the interior to the surface, and lost. The current thermal state of the interior is an essential constraint for understanding the 4.5 Gyr of planetary evolution. Measurements of the tidal Love numbers (e.g., $k_2$) and phase lag from the tide-raising potential provide constraints on the interior layering and how the crust and mantle deform viscoelastically on tidal periods (e.g., Padovan et al. [2014]; Steinbrügge et al. [2018]). MESSENGER measurements of $k_2$ based on orbital gravity data will be supplemented by those of BepiColombo. However, direct measurements of the tidal changes in the gravity field at the surface are necessary to determine the rheological structure of the interior [Steinbrügge et al. 2018]. Such measurements are capable of characterizing the phase lag of the tidal response, which is sensitive to internal temperatures, and also would provide a direct measurement of the solid body tide. The important interrelationships among the density, thermal, and rheological structures of Mercury's interior present an opportunity to characterize robustly the modern state of the interior and its evolution to the present.

Mercury's magnetic field provides a direct indicator of the dynamics of the interior both in the modern era (via the internal core field) and in the deep past (via crustal magnetization of ancient terranes) (e.g., Johnson et al. [2015; 2018]; Hauck et al. [2018]). The surface magnetic field strength is ~1% that of Earth, and the field is highly symmetric about the planet's rotation axis, but has a magnetic equator that is offset ~480 km north from the geographic equator [Andersen et al. 2011]. The pivotal discovery of the magnetization of ancient portions of Mercury's lithosphere [Johnson et al. 2015] opened new lines of inquiry into how the magnetic field has operated and how the planet evolved [Hood 2016; Hauck et al. 2018; Johnson et al. 2018]. MESSENGER measurements showed that much of the northern hemisphere has a weakly magnetized lithosphere, with some regions having much stronger magnetizations [Hood 2016; Johnson et al. 2018] (Exhibit B2).

Although the weak magnetizations could result from magnetizations induced in the present field, the strong magnetizations are most easily explained as remanent magnetization acquired in an ancient field [Hauck and Johnson 2019]. Furthermore, time-varying fields in Mercury's magnetosphere induce electrical currents in the interior and secondary magnetic fields. These induced fields in MESSENGER data are a probe of interior electrical conductivity structure and have already provided a complementary constraint on the core radius from those offered by geodetic and rotational observations [Johnson et al. 2016]. Orbital mapping by MESSENGER (and, in time, BepiColombo) provides a global picture of these processes. However, small-amplitude crustal fields as well as time-varying fields due to induction in Mercury's mantle, are difficult to detect from orbit because of the strong altitude-dependent decay of the signal. Measurements made on the surface substantially increase the ability to characterize the internal field, in particular crustal fields and time-varying fields induced in the interior. Indeed, measurements of ambient static fields >10 nT above the current models of the core field contribution of the surface would clearly establish crustal contribution to the internal field. Induced fields will be measured by monitoring the slow variation of field strength from dusk to dawn. Field variation of up to order 10 nT will establish field variation.

**Determining the minerals that carry the crustal magnetization is a fundamental issue related to interior composition and is needed to place bounds on the relative contributions of magnetization induced in the present field and acquired in an ancient field.** Understanding the relative contributions is important because the existence of an ancient field at 3.9–3.7 Ga, comparable to or up to 100 times stronger than the present field, places restrictive constraints on models of the thermal history of the core and thus of the interior and evolution of Mercury as a whole [Johnson et al. 2015; 2018; Hauck et al. 2018].

### B1.3. Goal 3: Space Environment

Three primary sources generate exospheres and cause space weathering on airless bodies: solar radiation, charged particles, and micrometeoroids (Exhibit B3; e.g., Killen et al. [2018]). Mercury—under intense solar radiation, with a highly dynamic magnetosphere, and subject to high-speed micrometeoroid bombardment—serves as an excellent laboratory for studying all three sources and the complex interactions among the various processes involved (e.g., Domingue et al. [2014]). As these phenomena affect Mercury's surface, they release neutral atoms and molecules, as well as ions into the exosphere. Remote measurements of the released materials from orbit, via observations of emission, are generally averaged over thousands of kilometers (e.g., Burger et al. [2014]; Merkel et al. [2018]). In contrast, in situ orbital measurements of the exosphere are localized, point measurements but cannot determine where the particles originated (e.g., Raines et al. [2013]). Thus, orbital observations provide only an overall sense of the outputs of each process. Similarly, although the input flux of charged particles impinging on Mercury's surface has been estimated through space-based observations from MESSENGER (e.g., Raines et al. [2014]; Winslow et al. [2014]), there is still considerable ambiguity regarding these particles' contributions. Many factors that cannot be measured from orbit, such as unexpected magnetic field configurations and small-scale plasma processes, could substantially alter the flux and energy distribution of particles that actually reach the surface. Our current understanding of the micrometeoroid (dust grain) influx at Mercury relies primarily on models (e.g., Christou et al. [2015]). The majority of dust grains may also be charged [Mann et al. 2004] and thus subject to the same unknown factors that affect the flux of charged particles. **Only in situ measurements from the surface can make the precise, small-scale measurements that connect all these pieces together into a complete picture of the processes at work on the surface of Mercury.**

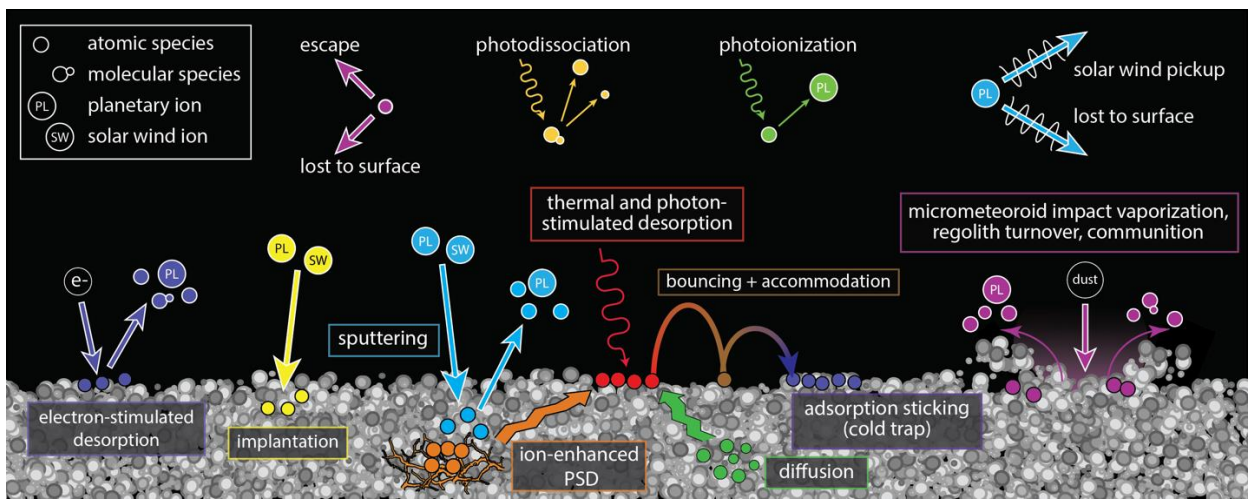

**Exhibit B3.** Processes that act on Mercury's surface to generate and maintain the exosphere and contribute to space weathering of the regolith. Exospheric sources and space weathering processes are illustrated at the surface itself; intermediate and loss processes in the exosphere are illustrated at the top. In situ measurements will directly resolve the contributions of each processes.

Landed observations on Mercury will enable the quantification of release processes in detail through concurrent, local-scale measurements of both neutrals and ions released from the surface, incident fluxes of charged particles and micrometeoroids, and detailed measurements of the surface mineralogy. Furthermore, a lander that experiences both twilight and night conditions is given the opportunity to distinguish the direct effects of sunlight from those of charged particles and dust by observing trends in the inputs and outputs with time. Surface measurements are also necessary for addressing other factors, including: relationships among the incoming sunlight, charged particles, micrometeoroids, and the released neutral and ionized species; temporal variability of the incoming and outgoing fluxes; whether the stoichiometry of the surface minerals is reflected in the released material; how the fluxes inform both recycling to the surface and loss to space; and to what extent physical regolith parameters (e.g., binding energies) play a role in these interactions.

Equally integral to a complete understanding of how material is released from Mercury's surface are measurements of the surface itself. Solar-wind irradiation and micrometeoroid impacts contribute to space weathering of the surface, which occurs on all airless bodies. The effects of these processes on the microstructure, chemistry, and optical properties of material at Mercury are poorly understood (e.g., Hapke [2001]; Bennett et al. [2013]; Pieters & Noble [2016]). Beyond compositional (i.e., elemental and mineralogical) measurements, it is important to understand the physical parameters of the regolith (e.g., particle size, strength, porosity) that also affect how the release processes operate. In particular, the physical properties of the regolith dictate the depth to which electromagnetic radiation or charged particles can penetrate, controlling how quickly the products from these interactions can diffuse back to the surface and be released to the exosphere, and governing the rate at which gardening of the regolith brings fresher materials to the surface. Investigating the character of the regolith in the near subsurface in color at pixel scales ≤500 µm (to resolve mm-sized grains) would enable further understanding of space weathering on Mercury. **Landed in situ regolith measurements can address key questions, including how the effects of the release processes change with the regolith's physical parameters, the nature of gardening on the surface, and the space weathering environment at Mercury.**

### B1.4. Goal 4: Geology

Although some MESSENGER images resolved surface features as small as a few meters across (e.g., Blewett et al. [2018]), the vast majority of the surface was observed at much lower resolution (yielding global image mosaics at 166 m/pixel, e.g., Denevi et al. [2018a]). BepiColombo will provide important new images of the innermost planet, acquiring global coverage at 50 m/pixel and local coverage ranging from ~10 m/pixel soon after beginning orbital operations, to 2–3 m/pixel locally later in the mission [Flamini et al. 2010; Cremonese et al. 2020]. But even with several-meter-scale images, there remains a gap between orbital observations and in situ, lander-scale observations that must be bridged to tie a landing site to our global framework for Mercury. Connecting observations from orbit to touchdown by acquiring nested images during the descent and tracking prominent landforms would help to obtain positional data for the lander [Grotzinger et al. 2012]. Nested descent images would also enable characterization of the site itself in the context of a continuum of scales across the descent sequence, such as the size–frequency distribution of boulders and craters and the visibility of different landforms at different scales. Such data would provide key information for placing the landing site in particular, and inferences of Mercury's surface more generally, into the context of orbital observations.

Mercury's local geological characteristics are currently unknown, particularly at outcrop scales. Yet vital insights have been gained by landers and rovers operating at that scale on other planetary bodies (e.g., Squyres et al. [2006]; Smith et al. [2009]; Eppes et al. [2015]), affording us a view of planetary processes impossible to achieve from orbit (Exhibit B4). Imaging a substantial fraction of the lander surroundings (≥180º azimuth, coverage from the horizon down to the near-field surface) with the ability to resolve 10-cm features within 50-m of the lander will return invaluable geomorphological, textural, and structural information with which to ask key outstanding questions of the landscape itself [Bell et al. 2003; Grotzinger et al. 2012]. Characterizing the landing site is necessary for identifying, for instance, local small-scale volcanic and tectonic features, as well as evidence for hollows at the LRM-rich landing site. Landed images will determine local landing site regolith and rock heterogeneity, and if there is evidence for processes that have altered and continue to alter the surface. These images could reveal textures, landforms, etc., at the surface that have not been recognized from orbit.

**Comprehensive assessment of the landing site will provide context with which to better understand global observations.** Documenting the chemical composition of a landing site will provide "ground truth" for the compositional and geochemical data for Mercury returned by MESSENGER [McCoy et al. 2018; Nittler et al. 2018] and planned from BepiColombo's orbital mission. Measurements of elements and minerals not detected from orbit will provide information about the entire mineral assemblage, allowing for a comprehensive interpretation of LRM deposits globally distributed across the planet.

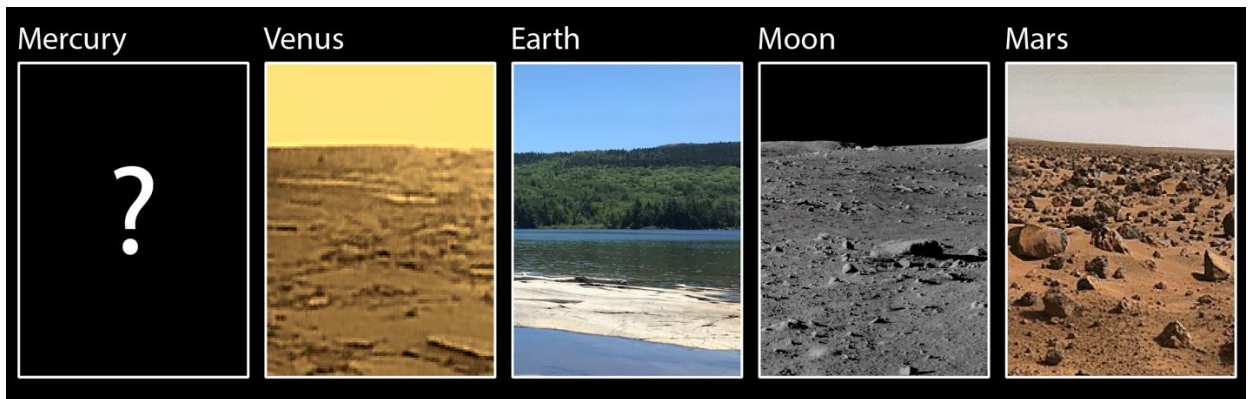

**Exhibit B4.** Mercury is the only major terrestrial body for which in situ surface data are lacking, yet the planet holds unique value in understanding how planets form and evolve.

## B2. Concept Study Science Payload

In this concept study, we considered an 11-instrument science payload, as detailed in the Science Traceability Matrix of the main report (Exhibit 2), to address our extensive set of science goals and objectives. This mission concept is meant to be representative of any scientific landed mission to Mercury; alternate payload implementations and landing locations would be viable and compelling for a future landed Mercury. The next sections detail our payload choices, describing the rationale and scientific measurements for the instruments selected for the concept study.

### B2.1. Geochemistry Payload

The instrument payload we selected for this mission concept study to address Goal 1 (Geochemistry) includes a gamma-ray spectrometer (GRS) and an X-ray diffractometer/X-ray fluorescence spectrometer (XRD/XRF). The XRD/XRF instrument requires delivery of surface samples into the instrument, so in this concept study, XRD/XRF is coupled with the PlanetVac sampling system. The payload items and their capabilities are discussed below.

### B2.1.1. Gamma-Ray Spectrometer (GRS)

Data from the GRS will be used to determine elemental compositions of the materials at the landing site to a depth of tens of centimeters. High-resolution, in situ data yield tighter constraints to understand Mercury's unique geochemistry and provide insights into the volatile-rich nature of the planet, and its thermal and magmatic evolutionary history. The GRS data will also provide a crucial ground truth for the orbital elemental measurements made by the MESSENGER and BepiColombo GRS and XRS instruments. The GRS is a high-purity, germanium-based sensor that makes high-energy-resolution measurements of gamma-ray emissions from Mercury's surface. The instrument selected for this concept study is based on the MESSENGER GRS [Goldsten et al. 2007], with updates from ongoing GRS instrument development for the upcoming Psyche [Lawrence et al. 2019a] and Martian Moons eXplorer (MMX) [Lawrence et al. 2019b] missions. For a Mercury lander, the GRS is simplified, removing the anti-coincidence shield and incorporating a low-power Ricor cryocooler. This simplified design is made possible by the higher signal-to-noise that can be achieved via landed measurements.

The GRS will measure gamma-ray emissions from Mercury's surface that result from cosmic-ray bombardment of near-surface materials. The cosmic rays liberate neutrons, which interact with atomic nuclei to produce element-specific gamma-ray emissions. The GRS will measure gamma-ray emissions from major and minor elements (O, Mg, Si, Al, Ca, Fe, C, Na, S, Ti, Mn) and naturally radioactive elements (K, Th, and U). These gamma-ray emissions will be used to characterize the elemental composition of Mercury's surface, in a ~1 $m^3$ volume beneath the lander, following procedures developed for the analysis of GRS data from the Near Earth Asteroid Rendezvous (NEAR) [Peplowski et al. 2015; Peplowski 2016] and MESSENGER [Peplowski et al. 2011; 2012; 2014, 2015] missions. Landed, in situ measurements will improve statistical uncertainties of many elements measured in MESSENGER data. For example, measurements of the concentrations of Na, Mg, Si, S, Cl, Fe, Cr, Mn, if present at concentrations of >0.5 wt%, will be completed with better than 10% statistical uncertainties. Measurements of the concentrations of C, O, and Ca, if present at concentrations of >1 wt%, will be completed with better than 20% statistical uncertainties. Measurements of the concentration of K, if present at >100 ppm concentration, will be collected with better than 10% statistical uncertainty. Measurements of the concentrations of Th and U, if present at >10 ppb concentrations, will be done with better than 20% statistical uncertainty. By improving on the statistical uncertainties and characterizing the composition of the local region, the GRS will provide valuable context for the measurements made by the XRD/XRF instrument, which samples smaller regions.

**B2.1.2. X-ray Diffractometer/X-ray Fluorescence (XRD/XRF) & PlanetVac**

A combination XRD/XRF spectrometer can provide both mineralogical and elemental characterization of the regolith at the landing site [Blake et al. 2019]. Powder XRD is a powerful crystallographic technique that, in combination with Rietveld refinements and full-pattern fitting methods, can be used to quantify crystalline and X-ray amorphous components, crystallite size and strain, and unit-cell parameters (e.g., Bish & Howard [1988]; Chipera & Bish [2002]). The refined unit-cell parameters of minerals can be used to infer crystal chemistry because ionic substitution within a lattice affects the unit-cell lengths and angles (e.g., Morrison et al. [2018]). XRF spectroscopy is a geochemical technique used to quantify major, minor, and trace elemental abundances within a sample.

The combination of XRD and XRF measurements would directly address the goal to investigate the highly chemically reduced, unexpectedly volatile-rich mineralogy and chemistry of Mercury's surface. XRD data will be used to identify and quantify the Mg-rich silicates, oxides, sulfides, and metals predicted to be on the surface of Mercury [Vander Kaaden & McCubbin 2016; Namur & Charlier 2017], as well as other minerals on the surface, to a detection limit of ~1 wt%. The refined unit-cell parameters of minerals will be used to identify crystal chemistry (e.g., Morrison et al. [2018]), which is important for constraining magmatic evolution on Mercury. Minor and trace elements derived from XRF measurements will inform elemental substitutions within minerals. Even if there is an unknown mineral on the Mercurian surface, XRD patterns could be used to solve the crystal structure.

The CheMin-V instrument was adopted for this study, drawing heritage from the CheMin instrument on the MSL Curiosity rover [Blake et al. 2012; 2019]. MSL-CheMin is a combination XRD/XRF that operates in transmission (i.e., Debye-Scherrer) geometry. MSL-CheMin accepts a few tens of mg of drilled rock powders and soils, sieved to <150 µm, or drilled rock powders delivered directly from the drill bit to the instrument [Blake et al. 2012; Rampe et al. 2020]. CheMin-V will improve upon CheMin by acquiring data more rapidly with improved angular resolution and by collecting quantitative XRF data [Blake et al. 2019], thereby improving the identification of minerals. The instrument will have two reusable sample cells that can be analyzed simultaneously. Diffraction data will be collected on charge-coupled devices (CCDs) and XRF data will be collected on silicon drift detectors. Full XRD/XRF analyses will be completed in 1 hour.

To accomplish the geochemical analyses, the surface sample of Mercury must be acquired and transferred to the XRD/XRF spectrometer. The PlanetVac sampling system [Zacny et al. 2014] was selected for this task. PlanetVac is the sampling instrument of choice for other planetary sampling firsts, such as sample return from Phobos by the MMX mission [Zacny et al. 2020], analysis of the lunar surface via NASA's Commercial Lunar Payload Services (CLPS) program, and pneumatic sample transfer on the surface of Titan on NASA's Dragonfly mission [Turtle et al. 2019].

PlanetVac will collect the surface regolith and transport the material into the XRD/XRF (CheMin-V) cell through two phases, sample acquisition and sample transfer. Two PlanetVac sample acquisition systems will be mounted on two different lander legs. This configuration will allow sampling from two distinct surface locations for characterizing differences and similarities between the two sample sites and for providing redundancy and robustness. Nozzles directing the compressed gas flow from the sampler cone loosen and loft surface material into the pneumatic sample transfer lines. Here, the sample is transferred by the pressure differential caused by the released compressed gas and the environmental vacuum at the transfer lines exhaust. Providing this pressure differential is an onboard compressed gas canister sized to accommodate eight sample collection operations, four per sampling system, with margin. At the sample ingest position, a deflector plate is used to syphon the transferred sample into the XRD/XRF analysis cell. The cell is self-metering and will accept ~100 mg of regolith. Once filled, any additional transferred particles will flow naturally around the deflector plate and out to an exhaust. This approach is well suited for instruments that require extremely small and known sample volume [Zacny et al. 2012]. The XRD/XRF instrument can confirm successful sample transfer by analysis, further simplifying the sampling system by removing the need for onboard sensors to image sample transfer.

The architecture of the PlanetVac sampling system allows for multiple sites to be sampled. By carrying two independently operated sample acquisition systems on separate lander legs, a lateral sampling distribution will be obtained from two distinct surfaces. A vertical sampling profile will also be obtained by allowing the sampler to "burrow" into the regolith, achieved with a longer duration compressed gas release at the nozzle. Cross contamination is mitigated by flushing the pneumatic lines between sampling events with gas, to clear residual material. The reusability of the XRD/XRF cells allows for multiple samples from each PlanetVac cone to be analyzed. Measurements from distinct PlanetVac cones will be used to determine the mineralogy and geochemistry of the regolith from two different landing site locations. Multiple samples from a single PlanetVac will be used to investigate compositional changes with depth (see also Section B2.3.4). Section B4. discusses the full operations planned for the geochemical instruments.

### B2.1.3. Geochemistry Payloads Considered, But Not Included

To meet the geochemistry goals of this mission concept study, numerous geochemical instruments were considered beyond the final selected payload. These considerations included: an age-dating mass spectrometer similar to the KArLE instrument (e.g., Cohen et al. [2014]), an Alpha Particle X-Ray Spectrometer (APXS) comparable to the one on the Mars Exploration Rovers (MER) and Mars Science Laboratory (MSL) (e.g., Rieder et al. [2003]; Gellert et al. [2006]), visible and near-infrared (VIS/NIR) spectrometers such as ultra-compact imaging spectrometers [Van Gorp et al. 2014], mid-IR spectrometers analogous to mini-TES on MER (e.g., Christensen et al. [2003]), Raman spectrometers similar to SHERLOC on Mars2020 [Beegle et al. 2016], laser induced breakdown spectrometers (LIBS) comparable to the ChemCam instrument on MSL [Wiens et al. 2012], and Mössbauer instruments as seen on MER (e.g., Klingelhofer et al. [2002]). Given the uncertainty of the chemical composition of Mercury's surface, and in particular the ancient age associated with the LRM, an age-dating mass spectrometer was not deemed a top scientific priority for this mission and this landing site. The featureless VIS/NIR spectra from MESSENGER that indicated the presence of low-FeO silicates [Izenberg et al. 2014], along with the low-light landing conditions, reduce the science return expected from a landed VIS/NIR spectrometer. Although an APXS-like instrument could provide elemental measurements needed to fulfill the geochemical objectives in this study, a GRS is able to make similar measurements without requiring specific instrument positioning relative to the surface, decreasing the complexity of surface operations. Given the difficulty in definitively verifying fine-grained materials and opaque materials, as well as the higher priority assigned to obtaining definitive in situ mineral identifications, mid-IR and Raman spectrometers were not ultimately chosen for the payload in this concept study. In addition, although a Mössbauer instrument would be extremely useful to understand the valence state of the limited iron available on Mercury's surface, this instrument requires a radioactive source, which would substantially complicate the lander design. Further, given the low-Fe content at the surface, the integration time required for useful measurements with a Mössbauer instrument might exceed the lifetime of the mission, and so this instrument was not included in our study.

As illustrated by numerous planetary surface missions (e.g., Viking, Mars Phoenix, MSL Curiosity, Venera, Luna, etc.), sample acquisition and delivery can be one of the most difficult aspects of a mission. Traditional means of gathering a sample with a planetary lander utilize a robotic arm and scoop. Robotic arms, however, require considerable volume, mass, power, electronics, and control software. Scoops, on the other hand, use gravitational force to move the sample down the gravity vector. However, if material is cohesive, the cohesion forces could dominate, causing failure of gravity-driven sample transfer. This behavior was observed during the Mars Phoenix mission: the soil sample inside the Phoenix scoop, called the Icy Soil Acquisition Device (ISAD), failed to fall out of the scoop when the scoop was placed above the instrument's inlet port. The ISAD percussive system had to be used to move the sample—but there is no guarantee that even percussion will be enough to dislodge sticky material. Hence for this concept study, to minimize similar issues, the PlanetVac sampling system was selected. We note that even with PlanetVac, additional sampling depth could be achieved if a drill were coupled with this sampling system, allowing for the drill cuttings to be transferred by the pneumatics [Zacny et al. 2015] to the instrument payload.

Overall, with the advancement in technology over the coming years, it is possible that those instruments or sampling systems not selected, or other more technologically advanced instrumentation, will be made available for consideration when a Mercury lander mission is ultimately designed.

### B2.2 Geophysics Payload

The instrument payload we selected for this mission concept study to address Goal 2 (Geophysics) includes a radio science (RS) investigation, a magnetometer (MAG), and an accelerometer (which we abbreviate as MAC for the Mercury Accelerometer); these items are discussed below.

### B2.2.1. Radio Science (RS)

Radio science utilizes the on-board telecommunications system to establish a coherent two-way link between stations on Earth and the lander. This link allows the measurement of distances between the stations and lander (ranging data), or, by using the Doppler effect, the measurement of the line-of-sight velocity between stations and lander (Doppler data). Radio science investigations are a common part of planetary missions, as they are needed for deep-space navigation. In addition, radio science data have been used to infer the gravitational fields of objects throughout the solar system, providing critical constraints on the interior structure of these objects. Although most of these investigations have involved an orbiting spacecraft, the Viking, Pathfinder, and InSight missions to Mars have provided a consistent set of data with which to refine Mars' orientation and infer its interior structure [Yoder & Standish 1997; Folkner et al. 1997; Yoder et al. 2003; Konopliv et al. 2006; 2011; Kuchynka et al. 2014; Jacobson et al. 2018; Folkner et al. 2018].

This lander concept study has the capability for both X-band (8–12 GHz) and Ka-band (26–42 GHz) measurements, with the latter providing data with less noise and data that are less susceptible to interference from solar plasma (e.g., Bertotti et al. [1993]; Asmar et al. [2005; 2019]; Iess et al. [2012]). The use of Ka-band data with proper calibration for effects from, for example, the Earth's atmosphere and solar plasma, will allow the determination of the line-of-sight velocity with a precision close to 0.01 mm/s at 60-s integration time (e.g., Iess et al. [2012; 2018]; Asmar et al. [2019]). For comparison, InSight currently uses an X-band system [Folkner et al. 2018], which in general has a precision close to 0.1 mm/s at 60-s integration time. Data will be collected on a daily basis during contact periods, with data sessions planned for continuous 24-hour communications. As described in Section 3.3.1, there are two distinct communication periods in the mission lifetime, which will allow for the collection of measurements at different times with the planet in a different orbital phase.

From these Doppler measurements from a fixed location on Mercury, changes in the planet's orientation can be measured precisely. Using Viking ranging data, Yoder and Standish [1997] were able to determine Mars' precession rate, which combined with the planet's gravitational flattening ($J_2$) coefficient results in a measurement of the planet's moment of inertia, providing a measure of radial density distribution within the planet. Using additional Pathfinder measurements, Folkner et al. [1997] were able to improve the error on the moment of inertia by a factor of 10. For Mercury, the moment of inertia has been determined by virtue of the planet being in an equilibrium state called the Cassini State I (e.g., Peale [1976]; Margot et al. [2012; 2018]; Genova et al. [2019]). By determining the planet's longitudinal librations, currently determined from Earth-based radar observations [Margot et al. 2007] and gravitational harmonics [Smith et al. 2012; Genova et al. 2019], a measurement of the planet's moment of inertia can be obtained. Precise measurements from a fixed position on the planet can be used to improve longitude librations and thus further improve our knowledge of the moment of inertia of Mercury, which in turn can be used to provide additional constraints on Mercury's interior structure and current thermal state.

### B2.2.2. Magnetometer (MAG)

Surface measurements of the vector magnetic field can be used to address questions regarding Mercury's magnetic field, help elucidate interior structure, and improve our understanding of exospheric processes such as surface precipitation. Continuous vector magnetic field observations over the roughly 88-day duration of surface operations will allow both static and time-varying fields to be identified. Magnetospheric processes occur over a wide range of time scales, from sub-second to a Mercury solar day, and the magnetometer will thus measure the magnetic field at 20 samples-per-second (sps), returning this full data stream to Earth. Onboard down-sampling to e.g., 1 sps, together with calculation of a 1-sps data stream that captures the root-mean-square fluctuation at frequencies above 1 Hz, can yield a lower data volume for times of more limited downlink capability, if needed. Suitable heritage instruments include the MESSENGER magnetometer [Anderson et al. 2007] and the InSight magnetometer [Banfield et al. 2019], which is being adapted for use on the Europa Clipper mission. Magnetometer placement should be at the end of a boom that is deployed after landing and has a boom length approximately that of the lander itself, to minimize contributions from spacecraft-generated fields. A star camera, co-located at the end of the boom, will provide magnetometer orientation information, and a dedicated small sunshield and heater are needed to maintain the magnetometer temperature within the instrument operating range (approximately –50°C to + 50°C). In addition, pre-launch spacecraft magnetic cleanliness and magnetic characterization protocols (e.g., Banfield et al. [2019]) should be employed.

During the period of surface operations, local time, heliocentric distance, and solar wind conditions will all affect the magnetic field recorded at the landing site. Superposed on this background time-varying field will be the effects of crustal magnetization. Crustal field models based on MESSENGER data suggest contributions to the surface field strength from crustal magnetization may be on the order of tens of nanoteslas (nT). However, these models cannot capture the shortest wavelengths in the field and experience from the InSight mission indicates that the surface crustal field can be considerably greater than satellite predictions, providing important constraints on local magnetization [Johnson et al. 2020]. The MAG will measure Mercury's magnetic field with a precision of 1 nT. MESSENGER provided no constraints on the southern hemisphere crustal field; however, BepiColombo may do so toward the end of the Mercury Planetary Orbiter (MPO) lifetime, if periapsis at that time is in the southern hemisphere. Such measurements would provide helpful regional context for crustal fields in the vicinity of the landing site.

### B2.2.3. Accelerometer (MAC)

Accelerometers, depending upon design characteristics, are used to measure vibrations, motions, and changes in gravity, and hence can function as seismometers and/or gravimeters. Potential analog instruments considered for this payload include the Rover Inertial Measurement Units from MSL, which are Northrup Grumman LN-200S units (e.g., Lewis et al. [2019]) and the InSight SEIS-SP short-period seismometer [Lognonné et al. 2019; Pike et al. 2018]. The latter instrument has the demonstrated ability to measure quake signals and solid-earth tides [Pike et al. 2018], and represents an appropriate analog capable of multi-purpose operation for both short-period seismic observations and measurement of the gravitational acceleration change from solid-body tides. The instrument will measure accelerations in three axes with low noise (SEIS-SP can operate at 0.25 ng Hz$^{-1/2}$) and collect continuous data at up to 100 sps. Measurements of acceleration changes in the vertical direction will also resolve the changes in the vertical acceleration of gravity at the landing site surface due to tidal variations.

The accelerometer will thus provide direct measurements of the gravitational changes due to tides over the course of the landed mission. These point measurements, when combined with radio science tracking data for this mission and the global orbital data from MESSENGER and BepiColombo, will constrain the $k_2$ tidal Love number, which is a reflection of the rheological behavior of the interior (e.g., Padovan et al. [2014]; Steinbrügge et al. [2018]). Further, when combined with data for the moment of inertia from radio science and the value of $h_2$ tidal Love number that describes actual surface displacements because of tides, a precise $k_2$ value can be used to tightly bound estimates of the size of an inner core [Steinbrügge et al. 2018].

High-frequency measurements from the accelerometer will detect seismic events. Operating as a short-period seismometer, the accelerometer will be able to provide the first information on the seismicity of Mercury. As a one-plate planet, Mercury's seismicity may be comparable to Mars, which is similar to intraplate regions on Earth [Giardini et al. 2020]. Mercury is rife with tectonic features related to the planet's global contraction [Byrne et al. 2014; Watters et al. 2016], including geologically young scarps <50 Myr indicative of ongoing shortening tectonics. Further, similar to the Moon, tidally induced quakes are also likely. Indeed, the largest expected tidal displacements on Mercury at ~2.4 m [Steinbrügge et al. 2018] are at least an order of magnitude larger than they are on the Moon at ~0.1 m [Williams & Boggs 2015]. Should Mercury have a seismic behavior similar to Mars (where InSight found 174 quakes in its first 10 months), it would be reasonable to expect several tens of quakes to be detected on Mercury over a roughly 88-day landed mission. Those quakes would contain information about the nature of the crust and tectonics on Mercury and have the potential to constrain the depth of the core due to the strength of the ScS core-mantle boundary reflected waves from the shallow core [Stähler et al. 2017]. Such data are necessary to constrain the planet's structure and evolution, including to provide insight on whether Mercury continues to contract today.

### B2.2.4. Geophysics Payloads Considered But Not Included
To further characterize the thermal and rheological structure of the interior, which are important for understanding the evolution of the planet as a whole, a heat-flow probe was also considered. Determination of the heat flux would be important for understanding the cooling history of Mercury and the temperatures at depth that control the rheology of the interior, and would be an important complement to GRS measurements of heat producing elements. Autonomous heat-flow probes are challenging to operate on planetary surfaces, however, as it is typically necessary to take measurements below the diurnal thermal wave (of order 1 m on Mercury, Vasavada et al. [1999]), and the success of self-penetrating devices depends on (poorly understood) regolith material properties. Ultimately, owing to mass and complexity constraints, a heat-flow probe was not included in the payload considered for this mission concept study. However, future advancements in the architecture and deployment of such instruments, as well as whether remote radiometer instruments would be sufficiently capable over the limited operational lifetime of a Mercury lander, should be considered in future mission opportunities.

### B2.3. Space Environment Payload
The instrument payload we selected in this mission concept study to address science Goal 3 (Space Environment) includes a neutral mass spectrometer (NMS), an ion mass spectrometer (IMS), an energetic particle spectrometer (EPS), a Dust Detector (DD), and regolith imagers (FootCam). The primary measurements needed to achieve Goal 3 are focused on the neutral species and ions released from the surface by the processes acting on it, the charged particles and micrometeoroids that impact the surface, and the character of the regolith itself. Measurements of the surface-sourced exospheric neutrals and ions would ideally be carried out with an instrument with sensitivity to a large particle-mass range and high mass resolution; however, the resource limitations of a lander preclude such an instrument. Instead, these measurements must be obtained with three smaller, but still highly capable instruments: a NMS, an IMS, and an EPS. The DD and FootCam, coupled with the ability to disturb the surface with PlanetVac, complete the nominal science payload adapted for this mission concept to investigate Goal 3, as discussed below.

### B2.3.1. Neutral Mass Spectrometer (NMS)

The NMS will measure the densities of neutral species in the exosphere, including those of both atoms and molecules. Atomic species are important to quantify because they represent those that have been observed remotely both from the ground and/or by MESSENGER. They are thus the "ground truth" for tying remote measurements of the exosphere to Mercury's surface. To understand the remote observations fully, however, in situ measurements of the release rates at the surface are needed as crucial inputs to the models used to interpret remote observations. Although laboratory measurements can give some insight on potential release rates, it is extremely difficult to reproduce the conditions at the Mercury surface, providing further rationale for a lander-based investigation. For example, it is thought that atoms returning to the nightside surface will adsorb and be re-released as the surface rotates into the light at dawn. Accommodation of such atoms on the surface is poorly understood, and landed measurements are the only means by which we can begin to address the issue. With in situ measurements of the near-surface exospheric atoms, the true atomic source rates owing to *all* the actual processes can be obtained.

Molecules are an important aspect of the neutral measurements because both ground-based and MESSENGER measurements suggest that some higher-energy exospheric atoms, particularly Ca, achieve those energies via the photodissociation of a molecule released during micrometeoroid impact vaporization (MIV; e.g., Killen et al. [2005]; Burger et al. [2014]). Laboratory measurements and theoretical studies have strongly suggested that MIV processes at Mercury do, in fact, deliver not only atoms but also molecules and ions directly to the exosphere (e.g., Berezhnoy [2018]). Because molecules generally emit radiation less efficiently than atoms, their detection around Mercury is difficult, and no neutral molecule has yet been detected.

Because of the key but unexplored role that molecules play in sourcing the exosphere, an NMS instrument is needed to make these measurements. Although it would be desirable to have a mass range out to ~150 atomic mass units (amu) to cover Xe (as the noble gases are important tracers of planetary evolution), a NMS with a range of ~100 amu is sufficient to cover the majority of atoms (amu $\leq$ Ni) and molecules (e.g., MgS, CaS) that would likely be detected. Similarly, whereas a high mass resolution would enable some potential ambiguities to be resolved if the molecules happened to overlap in mass-space, a mass resolution (M/$\Delta$M) of ~100 is reasonable for a small NMS. Finally, the sensitivity of the NMS needs to be high, of order 1 count/sec at density of 10 $cm^{-3}$ [Orsini et al. 2021]. A potential analog for such an instrument is STROFIO on the BepiColombo spacecraft.

### B2.3.2. Ion Mass Spectrometer (IMS) & Energetic Particle Spectrometer (EPS)

The second objective for Goal 3, as given in the science traceability matrix, is to measure ions both incidental to the surface, from the magnetosphere, and released from the surface by MIV and/or sputtering. Measurements of the ion species at Mercury's surface could be accomplished with an IMS to characterize the low-energy ions, and an EPS to measure the higher-energy ions and electrons.

Mercury's magnetosphere is surface-bounded, like the exosphere, because the collisionless exosphere does nothing to impede the flow of plasma. At mid-latitudes on the nightside, magnetospheric plasma, mostly protons and electrons, is expected to impact the surface after being accelerated planetward by magnetic reconnection in the magnetotail. The flow of protons toward the surface has been characterized, with energies up to the 13 keV maximum that could be measured by MESSENGER [Dewey et al. 2018]. Planetary ions present in the magnetotail (e.g., $Na^+$, $O^+$, $He^+$) are also likely accelerated toward the surface, potentially at much higher energies owing to their increased mass. For example, $Na^+$ ions would be up to 23 times higher in energy than the corresponding protons in the flow. Modeling of the Mercury magnetosphere shows that these energies may go as high as 100 keV [Delcourt et al. 2003]. Although a minor component of the magnetotail (<10% by mass), these high-energy species are important to measure because they are much more efficient at ion sputtering on the surface. MESSENGER measurements of X-ray fluorescence on the surface are likely direct evidence of electron impacts, probably in the 1–5 keV range (Lindsay et al. 2016). Electrons were measured in Mercury's magnetotail at energies ranging from 10–300 keV [Ho et al. 2012; 2016; Lawrence et al. 2015, Baker et al. 2016; Dewey et al. 2017], whereas MESSENGER was not capable of measuring electrons with energies <10 keV.

Ions coming in from the magnetosphere have two main roles in the generation of the exosphere. The first is direct sputtering of atoms (and possibly molecules) from the surface regolith. Sputtering yields at Mercury are estimated from a combination of theory and laboratory measurements and can be widely disparate (e.g., Leblanc & Johnson [2003]; Mura et al. [2007]). Without systematic measurements of the incoming ion fluxes—and the outgoing neutrals and ions—at the surface over time, exospheric models are hampered in their ability to predict both the regular time-varying roles of the solar wind and planetary ions in sputtering, as well as the effects of increased sputtering as caused, perhaps, by the passage of a coronal mass ejection. The second role is in the indirect enhancement of neutral species via a process known as ion-enhanced photon-stimulated desorption (PSD). In this process, the incoming ions break bonds in the surface material, damaging the structure of the minerals, which leads to less energy required for the primary volatile release process of PSD to act and/or an enhanced diffusion of new volatile material to the surface. The effects of ion-enhanced PSD have been noted for Na at Mercury (e.g., Burger et al. [2010]), but a quantification of the process requires a much greater knowledge of the near-surface ion fluxes than we currently have or can get from in situ observations in orbit.

The complex, time-varying dynamics of Mercury's magnetosphere make extrapolating plasma measurements taken from orbit down to the surface complicated except in certain regions (e.g., at magnetospheric cusps). The location and intensity of magnetotail plasma precipitation on Mercury's nightside surface depends on the state of the magnetosphere, which can vary on time scales of seconds to hours owing to changing inputs from the solar wind. Mapping orbital measurements down to the surface requires detailed knowledge of the magnetic field topology at these same time scales. This knowledge is well beyond the current state of the art, and is likely to remain so indefinitely as it would require many simultaneous spacecraft measurements.

Typical planetary IMS instruments measure ions in the 1 eV/e to 60 keV/e range, but these instruments often have a limited field-of-view (FOV) when not on a spinning spacecraft. Ideally, the IMS would have a large simultaneous FOV to allow for concurrent measurements of ions incident to, as well as upwelling from, the surface, along with angular resolution to distinguish the two sources (<20° angular resolution). The vast majority of the ions will likely be <20 keV/e, so an instrument with a large simultaneous FOV but a limited energy range would be more suitable. The IMS should have sufficient mass resolution to distinguish among the known and expected planetary ions (e.g., $Mg^+$ and $Na^+$; $K^+$ and $Ca^+$), and a M/ΔM of >4 for singly charged ions. The EPS can cover the higher energy range for ions, up to 1 MeV per nucleon, as well as energetic electrons. Measuring only particles incident on the surface, the EPS does not need such a broad FOV as the IMS; however, angular resolution of <20° would be helpful in understanding the processes. Ions released directly from the surface through MIV and/or sputtering are also important to measure, as they allow for a complete accounting of the species driven from the surface by all the processes combined. These ions are likely in the 1–10 eV energy range and could be measured by either the IMS or, potentially, the NMS if it has a low-energy-ion mode. Analogs for the IMS and EPS instruments are FIPS onboard MESSENGER [Andrews et al. 2007] and PEPSSI onboard New Horizons [McNutt et al. 2008].

### B2.3.3. Dust Detector (DD)

The third objective of Goal 3 is to characterize the incoming dust flux at the surface. Knowing the population of incoming dust is critical for completing the inventory of exospheric source processes, for understanding the space weathering environment at Mercury, and for the development of models of regolith production and gardening. MIV has been shown to be the primary process behind the production of both Ca [Burger et al. 2012; 2014] and Mg [Merkel et al. 2017] in the Mercury exosphere. Furthermore, the Ca source rate has shown a strong correlation with the passage of Mercury through the dust trail of comet 2P/Encke [Killen & Hahn 2015]. Measurement of the dust flux at the surface is therefore a crucial, missing piece in our knowledge of the MIV source of exospheric material. BepiColombo does have a dust instrument [Nogami et al. 2010] and will provide valuable observations. However, the importance and advantage of measurements taken by a landed instrument is that the incoming dust flux will be obtained simultaneously in both space and time with the other exospheric measurements, allowing for the establishment of correlations between dust and species release and thus a better understanding of the direct effects of MIV on the surface—critical observations for understanding the nature of space weathering at Mercury in general. Although a measurement of particle sizes and impact energies would be valuable, the primary goal is to measure the flux, with sensitivity to measure $10^{-15}$ kg m$^{-2}$ s$^{-1}$. Thus, the dust detector is envisioned as a simple sampling array akin to the Student Dust Counter on New Horizons [Horányi et al. 2009].

### B2.3.4. Regolith Imager (FootCam)

The Mercury Lander regolith imaging suite (FootCam) consists of two monochrome cameras mounted on two of the three spacecraft landing struts. Each camera will be supported by a four-color LED array, with those colors attuned to geologically appropriate wavelengths (nominally 450, 550, 650, and 750 nm). The number and placement of the LEDs will be modeled in more detail pre-flight and placed to optimize nighttime imaging of both FootCam (required) and StaffCam (as possible). The cameras are positioned such that each camera can observe both the contact of the lander foot with the surface and the associated PlanetVac sampling system. (The third spacecraft leg and landing strut assembly does not have a designated regolith imager.) The cameras used for this mission concept study have heritage from navigation and hazard cameras on MER and MSL (i.e., MSSS ECAM 5-megapixel CMOS cameras or equivalent, 44° x 35° FOV and 0.3 mrad iFOV).

FootCam will return images of the landing pads to provide insight into the regolith properties encountered by the lander, as well as information regarding space-weathering effects. These images, with a pixel scale of approximately 360 μm, will help to inform geological studies of the landing site, including characterization of the cohesion, mechanical strength, and texture of the "soil". Repeated imaging during the entire duration of operations will also allow for changes in the local surface materials to be detected. FootCam images will give information on the texture and size distribution of particles in the regolith, which will provide important insights into the maturation of the regolith on sub-millimeter to centimeter scales.

FootCam will image the immediate area surrounding each corresponding lander foot before and after each sampling event to search for any local changes to the regolith induced by the PlanetVac system or ongoing surface processes. The gas efficiency of the PlanetVac sampling process was found to be ~5500:1 (5500 grams of regolith lofted by 1 gram of gas) when tested on a lunar simulant in a vacuum chamber (~130 Pa) during a parabolic flight simulating lunar gravity conditions, with particle velocities approaching 10 m/s [Zacny et al. 2010]. Terrestrial vacuum-chamber testing for both martian and lunar regolith simulants under Mars-like atmospheric pressures (800 Pa) yielded 1000:1 gas efficiency [Zacny et al. 2014]. The efficiency of the sampling process is related to the ratio of the pressure of the gas to the ambient pressure on the surface of the planet [Zacny et al. 2010]. Given the low pressure of the Mercurian surface, these tests suggest that each PlanetVac sampling event will likely mobilize sufficient amounts of regolith to produce observable changes to the surface surrounding each sampler cone. Importantly, the extent to which material moves during each sample event can be used to infer the mechanical properties of the regolith (e.g., cohesion). Color imaging will help reveal whether fresher, unweathered (i.e., brighter) material is disturbed from immediately below the surface and exposed during sampling. Thus, in addition to providing information on the success of each sampling event, FootCam images will return valuable information on the nature of Mercury's regolith and the depth and degree of space weathering on the surface. These results will, in turn, inform models of space weathering and regolith gardening on Mercury in particular, and on airless bodies more generally.

Additionally, the elemental and mineralogical data on the regolith composition obtained from the GRS and XRD/XRF (Goal 1) will provide key insights into the nature of the effects of space weathering on Mercury. By monitoring the input fluxes of two major agents of space weathering (micrometeoroids with the DD, and solar-wind and energetic particles with the IMS and EPS), it will be possible to improve our understanding of the contribution of each agent to the space weathering of the soil. The ability of the PlanetVac to collect multiple samples for analysis in the XRD/XRF will provide some measure of a depth profile, whereby compositional and mineralogical changes, which may be attributed to space weathering processes and/or exospheric interactions, can be investigated by the FootCam.

### B2.3.5. Space Environment Payloads Considered, But Not Included

Although other exosphere-related measurements are possible with a lander, the resource limitations did not allow for them (i.e., science return judged less valuable than that from other, included instruments). Measurements of the near-surface altitude distribution of exospheric emission (tens of km above the surface) with an ultraviolet and visible spectrometer were considered, as it is nearly impossible to make such low-altitude measurements remotely with high spatial resolution. Thus, small contributions to the exosphere from processes such as thermal desorption have eluded detection thus far. However, although the NMS will not provide such altitude distributions near the surface, it will return accurate measurements of several source rates such that processes "missing" in current remote observations can be identified by mismatches in the incoming and outgoing fluxes.

An instrument combining most, if not all, of the aspects of the NMS, IMS, and EPS could be preferable to three different instruments, and may allow for a greater range in certain measurements. However, such instruments are generally quite large and utilize significant power, making them unlikely to fit within the lander resources. Similarly, a dust detector with the capability for measuring particle size and impact velocity would be desirable, but was ultimately deemed to be prohibitive owing to resource limitations.

A small all-sky imager with appropriately chosen filters would be desirable for observing the exospheric emission from a variety of species over large portions of the sky on short timescales. Although such emission can be measured remotely, both from ground-based telescopes on Earth and from instruments in Mercury orbit, the former suffer from interference caused by the terrestrial atmosphere and a lack of spatial resolution, whereas the latter are generally limited to small regions at a time. Thus, an all-sky imager would provide a highly unique perspective on Mercury's exosphere; however, it would necessarily require that a large data volume be captured and transmitted for the imager itself to be of the greatest utility. Given that such a volume would be difficult with the limited lander resources and that remote observations can provide some level of similar science, an all-sky imager was excluded. A small set of images covering part of the sky will be acquired with StaffCam (see Section B2.4.2) and thus provide some small measure of the exospheric emission variability on short timescales.

A robotic arm was also considered for the payload, potentially with attached VIS/NIR spectrometers or imaging cameras. This arm would have served to disturb the uppermost layers of surface regolith, enabling observation of changes in texture (and/or composition) with depth, providing insight into the nature of space weathering as it affects the mineralogy and morphology of regolith on the surface. However, mass constraints prevented its inclusion on the lander. Similarly, the featureless nature of Mercury's surface in the VIS/NIR indicated that data collected from an arm-mounted spectrometer may not reveal much about the nature of the regolith material. However, the inclusion of the PlanetVac sampling system on the payload, and by the landing process itself, will disturb surface material at the landing site and will enable the investigation and measurement of many of these parameters, without the inclusion of a robotic arm.

## B2.4. Geology Payload

The instrument payload we selected in this mission concept study to address Goal 4 (Geology) includes cameras to resolve the surface during the lander's descent and a panoramic camera mounted atop the lander to survey the landscape.

### B2.4.1. Descent Imagers (DescentCam)

The lander payload for this mission concept study includes two decent cameras ("DescentCam") to characterize the landing site, linking the local landing site to the global maps of the planet. The two cameras will be oriented with their FOVs 90° from one another, so that the surface can be imaged even as the lander changes it orientation relative to the surface during descent. For this mission concept study, the MSSS ECAM 5-megapixel CMOS cameras (44° x 35° FOV and 0.3 mrad iFOV) were adopted, drawing on heritage from the MER and MSL missions (as for the FootCam for Goal 3). The descent cameras will return a set of nested images of the landing site locale and region at different scales—with pixel scales of 6 m at 20-km altitude, 30 cm at 1-km altitude, and 0.6 mm at 2-m altitude—playing a critical role in bridging the gap between orbital-scale MESSENGER and BepiColombo data and lander-scale observations from the ground. These descent images will also assist in determining precisely the landing site itself.

### B2.4.2. Panoramic Imager (StaffCam)

The lander payload includes a panoramic imager ("StaffCam"), which is a multispectral camera located on the top of the gimballed high-gain antenna. This mounting location, which extends up from the main body of the lander, enables access to a full 360° view of the landing site (sun keep-out-zones will prevent the full 360° from being taken before dusk). For this concept study, StaffCam was selected to be the same as cameras with heritage from the MER Pancam and MSL Mastcam, both of which comprise a Malin Space Science System (MSSS) ECAM, 5-megapixel CMOS camera or equivalent. The concept of operations for StaffCam assumes a ~2650 × 1944 pixel sensor imaging with a 44° x 35° FOV and 0.3 mrad iFOV. Some of the LEDs required for FootCam will be mounted in such a way to enable illumination of the surface around the lander through the Mercury night, as possible. StaffCam will survey the landing site from the near field to the horizon, imaging the morphology of the landing site and the surrounding landscape at pixel scales of 1.5 cm from 50 m. StaffCam will acquire a set of complete 360° panoramas (when able) via a series of nine overlapping images; areas of overlap will permit the generation of stereo scenes.

The panoramas, coupled with data from the descent imager, place the landing site-scale morphology and geology within the context offered by orbital imaging from previous and future missions. Photometric analysis of areas viewed under differing illumination will offer clues to the soil texture, porosity, etc., as well as the scattering properties of regolith particles. StaffCam images of the horizon under post-dusk and pre-dawn illumination conditions may help identify and characterize dust particles from electrostatic levitation or thermal lofting. Mid-Mercury-night imaging of the sky will attempt to detect the faint sodium emission of the Mercury nightside exosphere (i.e., the "tail" region) near its maximum likely extent and intensity. This exosphere imaging campaign is considered opportunistic; whether an adequate signal-to-noise ratio can be attained using this light source will be validated by radiometric modeling before flight. Repeated StaffCam images of the near- and far field will also allow detection of possible changes in the vicinity of the landing site, perhaps due to activity of the lander (e.g., resulting from PlanetVac operations) or other events/processes such as mass wasting or localized tectonic activity from thermal loading.

### B2.4.4. Geology Payloads Considered, But Not Included

A radiometer was also considered during the concept study, and would have been used to characterize the thermal properties of the surface in the immediate vicinity of the lander, perhaps at a similar wavelength range as the MErcury Radiometer and Thermal Infrared Spectrometer (MERTIS) instrument onboard BepiColombo [Hiesinger et al. 2010]. For this concept study, the radiometer was not selected because it was deemed a lower science priority than the selected payload listed. A micro-imager was also discussed during early phases of the concept study. However, the need for exact positioning of such a micro-imager on an articulated platform such as an arm was deemed to increase greatly the complexity of the surface operations and resources required for the mission, so it was not included in this payload.

## B3. Landing Site Knowledge

A driving scientific goal of this study is to perform in situ measurements of Mercury's surface, with a particular interest in measuring the LRM to test the hypothesis that this material reflects the planet's ancient, graphite-rich flotation crust. Prospective landing sites are therefore necessarily restricted to areas of the planet where LRM is exposed (Exhibit 1). In this section, we discuss the current best characterization of this landing site based on MESSENGER data, and then briefly review the data BepiColombo are positioned to acquire that will further our understanding of the location. The exact landing site for this mission study was driven by science, i.e., to be in a region with extensive LRM deposits. We further refined the site selection based on thermal considerations and direct-to-Earth communication opportunities during the timeframe of the landed operations (as detailed in the main technical report). The site ultimately selected for this study is situated at approximately 40°S, 178°E.

### B3.1 MESSENGER Characterization of the Landing Site

This landing site region contains substantial exposures of LRM and is part of the Mercury terrain type characterized as "intercrater plains." These plains have a rough, hummocky texture and a high spatial density of small, superposed craters <15 km in diameter [Trask & Guest 1975; Strom 1977; Schaber & McCauley 1980; Leake 1981]; intercrater plains are substantially more rugged in appearance than Mercury's smooth plains units. Intercrater plains are likely dominantly volcanic in nature, with sustained impact bombardment responsible for their present texture [Whitten et al. 2014; Byrne et al. 2018]. The region in which our targeted site lies is among the oldest on Mercury [Denevi et al. 2018a].

The region we selected encompasses several named features, including Liang K'ai, Dowland, and Dostoevskij basins (Exhibit B5). The region also includes numerous sites where pyroclastic volcanic activity occurred [Byrne et al. 2018; Jozwiak et al. 2018], and several rayed craters, the most prominent of which is Bashō. Smooth plains units are present, although almost all instances are hosted within earlier impact craters and basins (Exhibit B5b).

No images are currently available that show our nominal landing site at the scale of the lander itself, and so it is possible only to broadly estimate the morphology of the site at present. Even the highest-resolution images returned by MESSENGER during its low-altitude campaign are ~2–3 m/px, which are insufficient to resolve any features at the scale of the lander, including boulders or similar morphological textures that might pose a hazard to safely landing on the surface.

We can, however, draw *some* inferences from those high-resolution MESSENGER images. The surface of an area of intercrater plains imaged by MESSENGER at ~2 m/px (Exhibit B5c) shows craters from several meters to several hundred meters in diameter; craters at still smaller diameters, as well as similarly sized boulders, are likely present but cannot be resolved in the MESSENGER images. Even so, the terrain between the larger craters appears relatively level, with little evidence for short-wavelength (i.e., tens of meters), positive-relief topography casting shadows in the scene. Therefore, it is possible, and perhaps likely, that much of Mercury's intercrater plains, between craters hundreds of meters in diameter and thus resolvable with MESSENGER and BepiColombo data, may be sufficiently level to permit the safe, autonomous landing of a robotic spacecraft. Further quantitative investigations into this topic are highly worthwhile for future Mercury lander mission planning.

By way of analogue, the Apollo landing sites on the Moon might offer some useful insights. Although unlikely to be as topographically benign as where the Apollo 11 astronauts ultimately touched down, Mercury's intercrater plains may more closely resemble the Apollo 17 site—where a small tectonic scarp and several impact-related massifs bracket a relatively smooth area landing area (Exhibit B5d).

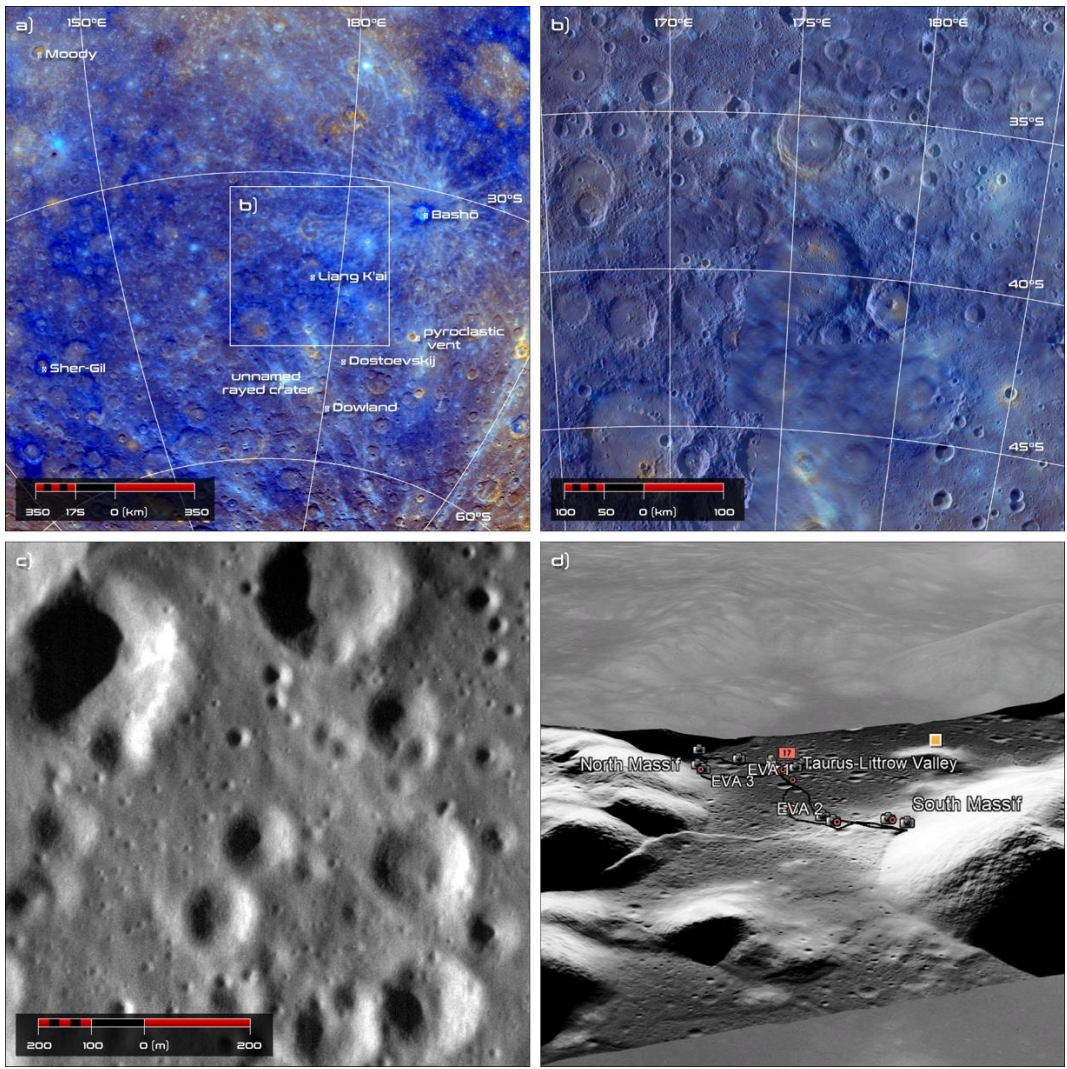

**Exhibit B5.** (a) The region of Mercury where our nominal landing site is situated, shown at a view scale of 1:10M and with enhanced color. Prominent named features are labeled, as is an example rayed crater and pyroclastic vent. Note the expanse of blue, low-reflectance plains material (LRM) throughout the region. The approximate location of panel (b) is marked by the white outline. (b) The same region at a view scale of 1:3M, with the enhanced color global mosaic superposed on the global monochrome mosaic. (c) A MESSENGER MDIS image from that mission's low-altitude operations phase, showing an exemplar region of the intercrater plains at a resolution of ~2 m/px. (d) An oblique view of the Apollo 17 landing site in Google Earth; note the massifs that flank the landing site, and the Lee Lincoln thrust fault scarp that strikes left–right.

**B3.2. BepiColombo & Landing Site Characterization**

The BepiColombo mission is on its way to Mercury and will start collecting scientific measurements of Mercury as early as 2022. During the six flybys of the planet, only a limited number of instruments will be able to observe Mercury's surface. Improvements to the MESSENGER observations of the landing site will therefore only come from the orbital phase of the mission, planned to begin in 2025. The MPO spacecraft will be positioned to characterize Mercury's surface at scales and wavelengths not achieved by MESSENGER. The Spectrometer and Imaging for MPO BepiColombo Integrated Observatory SYStem (SIMBIO-SYS) instrument suite [Flamini et al. 2010; Cremonese et al. 2020] will acquire stereo images to support the creation of targeted digital terrain models (DTMs), high-resolution images, and near-infrared spectra up to 2400 nm, and the MERTIS instrument [Hiesinger et al. 2010] will acquire observations in the infrared domain and will measure the surface temperature.

The MPO spacecraft will have an elliptical orbit that evolves during the mission. Its periapsis will drift rapidly to the south, and the altitude of the spacecraft will decrease. Although the nominal MPO mission is for one Earth year [Benkhoff et al. 2010], it is expected that by the end of a likely one-year extension, the spacecraft argument of periherm will be located at 40°S, with an estimated altitude of 270 km. These orbital properties will position the spacecraft for optimal observations of the concept study landing site shown in Exhibit B5 and located at 40°S. In particular, at this phase of the BepiColombo mission:

- SIMBIO-SYS will acquire high-resolution images of order 2–3 m/px and produce DTMs of resolution ~100 m/px.
- MERTIS will obtain unprecedented surface temperature measurements with a resolution of roughly one kilometer per pixel.
- The Mercury Imaging X-Ray Spectrometer (MIXS) [Fraser et al. 2010] and Mercury Gamma and Neutron Spectrometer (MGNS) [Mitrofanov et al. 2010] will provide observations of the southern hemisphere at scales that were not feasible with MESSENGER.
- The BepiColombo Laser Altimeter (BELA) [Gunderson et al. 2010] will obtain topographic measurements of the southern hemisphere.

These observations will provide important, new measurements of Mercury's surface, new insights on Mercury's southern hemisphere, and new information to help characterize the overall scientific setting of the landing site. However, given that any landing site would have to be characterized to a sub-meter (i.e., lander) scale to fully identify and mitigate potential landing hazards, BepiColombo will not provide the necessary spatial resolution to pin-point a hazard-free landing site. Should BepiColombo successfully operate beyond its primary one-year mission, the altitude and periapsis will continue to diminish and drift to southern latitudes. Although this change in orbital position might not be sufficient to obtain the required resolutions for the landing site selected for this study, dedicated and well-prepared MPO observations may be able to offer measurements that could be useful for characterizing highly localized regions farther in Mercury's southern hemisphere, potentially at a sub-meter scale to identify hazards. (It is also likely, however, that lighting condition limitations would restrict the utility of these data). Overall, the risk posed by the uncertainty regarding the Mercury surface at lander scale is not likely to be retired by BepiColombo measurements, and will thus remain one of the most substantive challenges facing any landed mission to Mercury.

# B4. Landed Science Operations

The overall concept of landed operations is outlined in the main report (see Section 3.3.1, Exhibit 28). The sections below provide more detailed discussions of the landed operations of each instrument in the mission concept payload needed to achieve the science goals outlined in the study.

**B4.1. Descent Science Observations**

DescentCam will commence full-frame imaging at a frequency of 0.5 Hz before the firing of the solid rocket motor (SRM) on the descent stage, pausing during the firing, and resuming during the landing phase after the descent stage burn-out and jettison; this jettison occurs at an altitude of ~7 km, around one minute prior to landing. The final landing approach with liquid propulsion will begin at an altitude of approximately 3 km, some 30 s prior to landing, for which DescentCam will acquire an uninterrupted, nested imaging sequence (for a total of about 60 images at progressively higher resolutions across smaller spatial scales).

A light detection and ranging (LIDAR) instrument on the lander is classified as an engineering instrument to aid with landing. Although the primary objective of the LIDAR is not to return science data, but rather to enable hazard avoidance and landing safely on Mercury, the LIDAR data would also provide opportunistic science and will be fully transmitted to Earth. In addition to characterizing the topography of the local landing site, the LIDAR data can provide information regarding shape, surface roughness, and reflectivity of the Mercury surface at a variety of scales, augmenting the science return of the descent camera and other imaging instruments. The LIDAR will operate continuously after the completion of the SRM burn.

**B4.2. Continuously Operating Landed Measurements**

Many of the instruments operate nearly continuously throughout the landed mission, including NMS, IMS, EPS, DD, MAC and, once the boom deploys, MAG. The Radio Science investigation also begins after landing, continuing whenever Earth communications are possible. GRS has a cool-down period after landing and then begins continuous operations about 36 hours later. GRS operates truly continuously; owing to power limitations, all other continuous instruments momentarily cease operations during the approximately one-hour XRD/XRF operations that occur periodically, which are described in more detail in Section B4.4. The continuous operation of the GRS enables the highest sensitivity elemental measurements for the mission and minimizes the potential for degradation of the instrument during landed operations. Measurements beyond the required 72 hours will lead to increased precision as well as enable insights into regolith density and depth-dependent concentration variations.

Continuous MAG observations will span local times and heliocentric distances from ~6:00 pm at aphelion just after landing to midnight at perihelion some 44 days later, to just after 6:00 am at aphelion at the end of the mission. Global magnetospheric models based on MESSENGER data (e.g., Johnson et al. [2012]; Korth et al. [2015; 2017]) suggest that the large-scale internal and external fields will result in a change in surface field strength from ~195 nT at dusk/dawn to ~235 nT at midnight [Johnson et al. 2018], on top of which will be a superposed time-varying signal of –5 nT at aphelion to +5 nT at perihelion from changes in the internal dipole moment induced by variations in the solar wind [Johnson et al. 2016]. These variations, together with shorter time scale changes, could be used to probe the electrical conductivity structure of the silicate portion of the planet. Transient changes in the magnetic field on time scales of fractions of a second to minutes will contribute to understanding the plasma data and magnetospheric processes.

Seismic and tidal potential observations will be continuous over the course the full course of the landed mission. Quake activity can occur at any time throughout the mission and complete monitoring is necessary to quantify the scale and frequency of tectonic activity, particularly tidally driven activity as a result of Mercury's eccentric orbit. Operating as a seismometer, MAC will operate at a 100-sample-per-second rate that generates more than 7 GB of data, a considerable fraction of the total available downlink. However, downlinking the full data sets during the periods of direct communication will provide an important baseline on seismicity and thermomechanical behavior of the surface and lander during the post-sunset and pre-dawn periods. Lower-resolution data (1 sample per second), plus statistical information from the 100-sample-per-second data that characterizes its variability, will be downlinked upon reconnection from the period without Earth communications. Those data will be analyzed on Earth to prioritize time periods with seismic activity for full resolution downlinking. Prioritizing 250 hours of data from that period will permit as many as 125 events during that 6-week period to have two hours of full-resolution data downlinked for each of those events. This prioritization process can be extended to the second communications period to further manage both stored and downlinked data volumes.

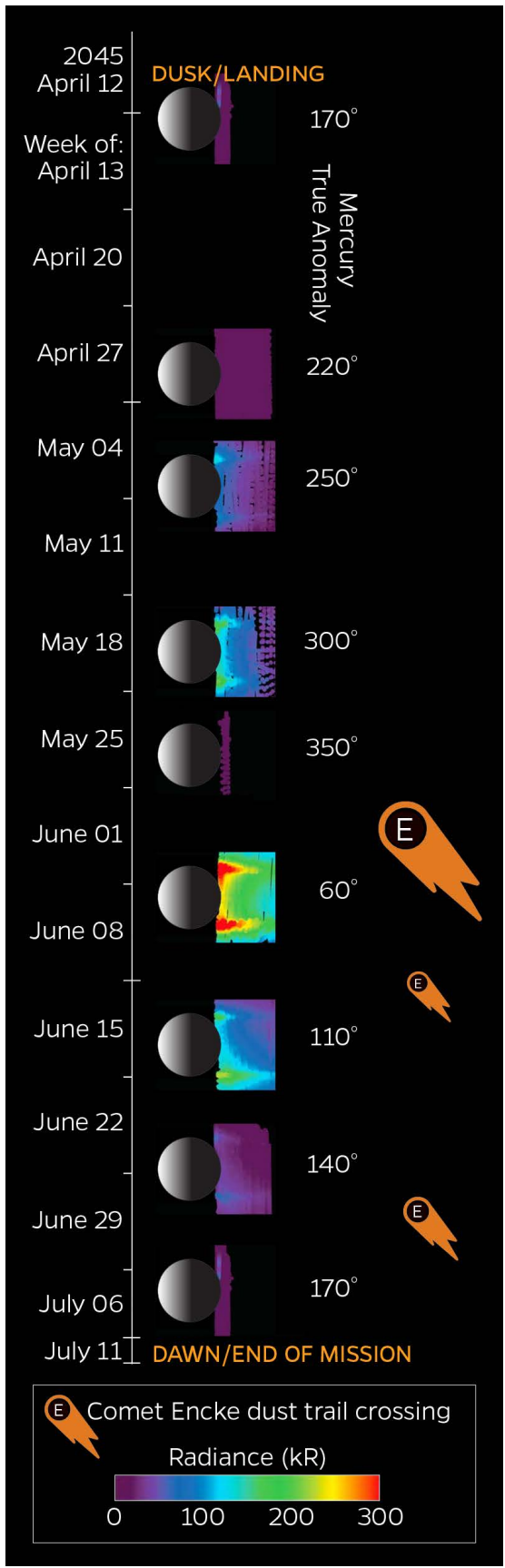

Mercury's exosphere is continuously produced and maintained at all times, and understanding the time-varying nature of the sources and their relationships to the exosphere requires that as many observations as possible are obtained. The timescales for variations observed in the exosphere and magnetosphere have been as small as minutes to even seconds, and it is impossible to know when the shortest-timescale phenomena are going to be observed. Hence, continuous operations of NMS, IMS, EPS, and DD are planned during the landed mission, within available power constraints. As a balance between the space environment observational needs and those of the rest of the lander, a sampling rate of once every 10 seconds for all the instruments has been baselined. This rate is enough to capture the shorter timescales relevant to the exosphere, without exceeding or dominating the total data rate of the lander. If data volume limitations arise, the sampling rate could be reduced to once every 100 seconds for these instruments, but for this mission concept study this rate serves as a contingency.

With the combined Mercury space environment landed measurements of the NMS, IMS, EPS, and DD, the source processes for the exosphere can be examined as a truly complete system for the first time. Correlations between the incoming and outgoing material will be established on multiple scales. Because the lander observations will span one Mercury year, the correlations will provide unparalleled insight into the seasonal variations of the source rates, as shown in Exhibit B6.

Exhibit B6. Seasonal coverage from one full Mercury year of landed science operations. The color maps are generated from the Na emission in the tail region of the exosphere observed by MESSENGER and averaged over the orbital phase of the mission (adapted from Cassidy et al. [2016]). The size of the comet Encke symbols qualitatively indicates the amount of dust likely to impact Mercury (bigger equals more dust).

Because operations would start near the dusk terminator and end near the dawn, the well-known dawn/dusk asymmetries in Mercury's exosphere can be investigated—albeit not simultaneously with the lander, but perhaps with supporting ground-based data from the opposite side. By operating at a continuous cadence on the order of 10–100 seconds throughout the lander phase, the solar-wind-driven time-variability and relationships among the space environment measurements will be characterized in great detail. Should there be unusual circumstances at the surface—perhaps owing to solar energetic particle events, coronal mass ejections, magnetospheric substorms, or a larger meteoroid impact—the lander will provide important knowledge of how the processes change.

### B4.3. Imaging Operations

During the ~30 hours of sunlight from landing at dusk until nightfall, the StaffCam and FootCam will capture images and panoramas of the landing site region. StaffCam's initial operation will consist of a set of panoramas of the entire landing site (save in the sun keep-out zone) immediately after landing. Imaging obtained with the StaffCam and FootCam will allow comparison of more heavily scoured material directly under the landing site with material that is less disturbed farther afield. A full panorama comprises nine frames, where a single image is estimated at 12.5 Mbit with compression, times two for stereo (with overlap), at two different tilt angles to capture the near-field and horizon around the lander. Nighttime panoramas will be acquired in all four LED colors. Immediately after landing, the FootCam will image the landing pads to acquire information regarding the mechanical and textural properties of the regolith. As for StaffCam, a single FootCam image is estimated at 12.5 Mbit (compressed).

To enable night-time operations, LEDs will be used, although the glow from Mercury's Na exosphere may also provide diffuse illumination, particularly when the tail is extended (Exhibit B6) and total emission intensities rival those of a moderate aurora on Earth. StaffCam acquires a panorama weekly, characterizing the surrounding landscape and the exospheric "glow". Panoramas without LED illumination will also be acquired, to search for the possible detection of lofted particles and to characterize the exospheric glow. In particular, on 7 June 2045, which corresponds to Mercury true anomaly of 60°, StaffCam will undertake a dedicated exploratory imaging campaign with multiple panoramas devoted to imaging the Na exosphere, timed to occur during the maximum seasonal radiance (> 300 kR), as seen in Exhibit B6.

FootCam images are acquired daily to monitor for change detection, especially before and after PlanetVac operations. FootCam images will be obtained in colors utilizing the dedicated LEDs for such imaging. These daily images will provide before-and-after views of each PlanetVac sampler site when that instrument is active. Additionally, FootCam images taken at the 7 June 2045 anticipated peak of the Na exosphere extent and intensity will provide useful insight into illumination conditions of the Mercury soil during this phenomenon.

As the Sun rises at the end of the mission, StaffCam and FootCam acquire multiple images and panoramas, streaming them back to Earth until the lander fails and the transmissions end.

### B4.4. XRD/XRF & PlanetVac Operations

XRD/XRF operations occur in coordination with the operation of PlanetVac, to provide regolith samples to the instrument for analysis. Before initial operations begin, images obtained by DescentCam, StaffCam, and FootCam from the sunlit dusk operations will be analyzed by the science team to characterize the regolith properties, such as grain size and morphology, and the placement of the PlanetVac samplers on the surface. Based on comparison with lunar landings and test data from Earth (e.g., Metzger et al. [2011]), the area under the lander will probably experience several cm of scouring, depending on the regolith texture and exhaust pressure. For the purpose of sampling, this scouring may allow access to somewhat less space-weathered material for mineralogical analysis. These images will inform the subsequent operations plans for PlanetVac and XRD/XRF. The baseline plan for this study's landed mission operations include eight XRD/XRF analyses, four from each PlanetVac sampler. This plan would allow for both lateral and vertical heterogeneities of the landing site to be assessed, but could also be adjusted based on the specifics of the local landing site.

Of all the instrument operations on this Mercury Lander concept study, the combined XRD/XRF and PlanetVac operations are the most iterative, with the science team actively making decisions to inform future sampling and analysis plans based on the data collected previously. Consequently, given the complexities associated with the XRD/XRF and PlanetVac operations, no operations are planned for the six-week period when there will be no communication with Earth, i.e., from 4 May to 16 June 2045. During the periods of direct-to-Earth communication, all XRD/XRF and PlanetVac data can be fully downlinked to Earth within a few hours of being acquired.

During the first three weeks of nighttime operations with Earth communication, after the initial wellness checks are completed, four distinct PlanetVac samples will be analyzed. The baseline plan includes one sample from each of the two PlanetVac samplers, and two additional samples from one of the samplers. These four initial samples will provide insight into the compositional diversity between the locations of the two legs as well as at depth, as PlanetVac will excavate deeper into the regolith with each subsequent sample at a given location. The daily FootCam images will capture the regolith before and after each PlanetVac sampling. These images will provide information on morphological changes occurring with each collection, as well as information regarding depth of sampling.

The baseline plan includes obtaining samples and analysis from each of the two PlanetVac samplers during the first ten days of the landed operations. These initial XRD/XRF analyses are planned to last one hour each. To assess the efficacy of this measurement time and evaluate the impact of longer analysis time on data quality, several days later these same two samples, which will still be in the XRD/XRF cells, will be reanalyzed for a four-hour duration. Approximately two weeks into the landed mission, both previously acquired samples will be dumped, and empty-cell XRD/XRF analyses conducted, to ensure cells are empty before accepting new samples. During the third week of landed operations, two additional samples will be collected with one of the PlanetVac samplers, with XRD/XRF analysis of each sample followed by subsequent cell dumping and empty-cell analyses conducted a day afterward.

Once Earth communication is regained for the final twenty-four days of the landed mission, wellness checks will be carried out for both the PlanetVac system and XRD/XRF payload. The baseline plan includes analysis of four additional samples, with empty-cell analyses between those sample measurements; which of the two PlanetVac instruments will be used for those four subsequent samples will be decided on the basis of the previous results. The nominal operational plan will result in eight PlanetVac samples total, four from each PlanetVac system, which will allow for horizontal and vertical heterogeneities of the landing site to be assessed.

# APPENDIX C: REFERENCES